\documentclass[12pt]{article}
\usepackage{graphics}
\usepackage{epsfig}
\usepackage{amsmath}
\usepackage{amssymb}

\usepackage{geometry}
\newcommand{\be}{\begin{eqnarray}}
\newcommand{\ee}{\end{eqnarray}}
\newcommand{\nn}{\nonumber}
\newcommand{\gev}{{\rm GeV}}
\newcommand{\mev}{{\rm MeV}}

\newcommand{\<}{\langle}
\renewcommand{\>}{\rangle}

\def\rmii{a}
\def\rmiii{b}
\def\infniii{c}

\begin{document}

\begin{flushright}
\begin{tabular}{l}
{\tt RM3-TH/08-15}\\
{\tt ROM2F/2008/24}\\
\end{tabular}
\end{flushright}

\vspace{1cm}

\begin{center}
\LARGE{Electromagnetic form factor of the pion\\ from twisted-mass lattice QCD 
at $N_f = 2$}

\vskip 0.5cm
\begin{figure}[h]
  \begin{center}
    \includegraphics[draft=false]{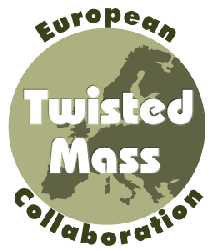}
  \end{center}
\end{figure}

\large{ \textbf{
     R.~Frezzotti$^{(\rmii)}$,
     V.~Lubicz$^{(\rmiii)}$,
     S.~Simula$^{(\infniii)}$
}}

\vspace{0.5cm}

\footnotesize{

$^{(\rmii)}$ Dip. di Fisica, Universit{\`a} di Roma Tor Vergata and INFN, Sez. di Roma Tor Vergata,
\\ Via della Ricerca Scientifica, I-00133 Roma, Italy
\vspace{0.2cm}

$^{(\rmiii)}$ Dipartimento di Fisica, Universit{\`a} di Roma Tre and INFN, Sez. di Roma Tre,
\\ Via della Vasca Navale 84, I-00146 Roma, Italy
\vspace{0.2cm}

$^{(\infniii)}$ Istituto Nazionale di Fisica Nucleare, Sezione di Roma Tre,
\\ Via della Vasca Navale 84, I-00146 Roma, Italy
\vspace{0.2cm}

}

\end{center}

\normalsize

\vspace{0.5cm}

\begin{abstract}

\noindent We present a lattice calculation of the electromagnetic form factor 
of the pion obtained using the tree-level Symanzik improved gauge action with 
two flavors of dynamical twisted Wilson quarks.
The simulated pion masses range approximately from $260$ to $580~\mev$ and the 
lattice box sizes are chosen in order to guarantee that $M_\pi L \gtrsim 4$. 
Accurate results for the form factor are obtained using all-to-all quark 
propagators evaluated by a stochastic procedure. 
The momentum dependence of the pion form factor is investigated up to values of 
the squared four-momentum transfer $Q^2 \simeq 0.8~\gev^2$ and, thanks to the use 
of twisted boundary conditions, down to $Q^2 \simeq 0.05~\gev^2$.
Volume and discretization effects on the form factor appear to be within the 
statistical errors.
Our results for the pion mass, decay constant and form factor are analyzed using 
(continuum) Chiral Perturbation Theory at next-to-next-to-leading order. 
The extrapolated value of the pion charge radius is $\langle r^2 \rangle^{phys} 
= 0.456 \pm 0.030_{\mbox{stat.}} \pm 0.024_{\mbox{syst.}}$ in nice agreement 
with the experimental result.
The extrapolated values of the pion form factor agree very well with the experimental 
data up to $Q^2 \simeq 0.8~\gev^2$ within uncertainties which become competitive 
with the experimental errors for $Q^2 \gtrsim 0.3~\gev^2$.
The relevant low-energy constants appearing in the chiral expansion of the pion 
form factor are extracted from our lattice data, which come essentially from a 
single lattice spacing, adding the experimental value of the pion scalar 
radius in the fitting procedure. 
Our findings are in nice agreement with the available results of ChPT analyses 
of $\pi - \pi$ scattering data as well as with other analyses of our 
collaboration.

\end{abstract}

\newpage

\pagestyle{plain}

\section{Introduction}

The investigation of the physical properties of the pion, which is the lightest 
bound state in Quantum Chromodynamics (QCD), can provide crucial information on
the way low-energy dynamics is governed by the quark and gluon degrees of 
freedom.
In this respect for space-like values of the squared four-momentum transfer, 
$Q^2 \equiv - q^2 \geq 0$, the electromagnetic (e.m.) form factor of the pion, 
$F_\pi(Q^2)$, provides important insights on the distribution of its charged 
constituents, namely valence and sea light quarks. 
At momentum transfer below the scale of chiral symmetry breaking ($Q^2 \lesssim 
1~\gev^2$) the pion form factor represents therefore an important test of 
non-perturbative QCD.

The current experimental situation is as follows. For values of $Q^2 \lesssim 
0.2~\gev^2$ the pion form factor has been determined quite precisely at CERN 
SPS \cite{CERN} by measuring directly the scattering of high-energy pions 
off atomic electrons in a fixed target. 
At higher values of $Q^2$ the pion form factor is extracted from cross section 
measurements of the reaction $^1H(e, e^\prime \pi^+)n$, that is from electron 
quasi-elastic scattering off virtual pions in a proton. 
The separation of the longitudinal and transverse response functions as well as 
the extrapolation of the observed scattering from virtual pions to the one 
corresponding to on-shell pions have to be carefully considered for 
estimating the systematic uncertainties.
Using the electroproduction technique the pion form factor has been determined 
for $Q^2$ values in the range $0.4 \div 9.8~\gev^2$ at CEA/Cornell \cite{CEA}, 
for $Q^2 = 0.35$ and $0.70~\gev^2$ at DESY \cite{DESY1,DESY2} and, more 
recently, for $Q^2$ in the range $0.6 \div 1.6~\gev^2$ \cite{JLAB1} and 
for $Q^2 = 1.60, 2.15$ and $2.45~\gev^2$ \cite{JLAB2} at the Thomas 
Jefferson National Acceleration Facility (JLab).
A careful reanalysis of the systematic uncertainties for the data of 
Refs.~\cite{DESY1,DESY2,JLAB1} has been carried out in Refs.~\cite{JLAB3,JLAB4}.

It is well known that at small values of $Q^2$ the pion form factor can be 
reproduced qualitatively by a simple monopole ansatz inspired by the Vector 
Meson Dominance (VMD) model with the contribution from the lightest vector 
meson ($M_\rho \simeq 0.77~\gev$) only. 
This is not too surprising in view of the fact that in the time-like region 
the pion form factor is dominated by the $\rho$-meson resonance 
\cite{CERN_timelike}.

More interesting is the quark mass dependence of the pion form factor, which can 
be addressed by QCD simulations on the lattice and by Chiral Perturbation Theory 
(ChPT). 
The latter, which is known at next-to-leading (NLO) \cite{GL} and 
next-to-next-to-leading (NNLO) order \cite{BCT} for the pion form 
factor, can be used as a guide to extrapolate the lattice results 
from the simulated pion masses down to the physical point, 
obtaining at the same time an estimate of the relevant 
low-energy constants (LEC's) of the effective theory.

Initial studies of the pion form factor using lattice QCD dates back to the late 
80's \cite{MS88,Draper} giving strong support to the vector-meson dominance 
hypothesis at low $Q^2$. 
Within the quenched approximation, which neglects the effects of the sea quarks, 
several lattice investigations have been carried out using Wilson \cite{LHP}, 
Sheikholeslami-Wohlert \cite{Clover_q}, twisted Wilson \cite{tm_q} and 
Ginsparg-Wilson \cite{GW_q} fermions.
The effects of the quenched approximation might be limited because, thanks to 
charge-conjugation and isospin symmetries, the e.m.~pion form factor receives 
no contribution from the so-called disconnected diagrams in which the vector 
current is attached directly to a non-valence quark (see Ref.~\cite{Draper}). 
However there are effects from sea quarks which do not interact directly with 
the external current, and they can be taken into account only by performing 
unquenched gauge simulations.

There are few results for two flavors of dynamical fermions from JLQCD \cite{Clover1} 
and QCDSF/UKQCD \cite{Clover2} collaborations adopting Clover fermions and again from 
JLQCD \cite{overlap} using overlap quarks. Finally only two studies with three flavors 
of dynamical quarks are available to date, namely from Ref.~\cite{LHP}, where 
domain-wall valence quarks and Asqtad sea quarks are mixed, and from Ref.~\cite{DWF}, 
where the domain-wall formulation is used for both sea and valence quarks.

As far as the lattice results for the (squared) pion charge radius at the physical 
point are concerned, the present situation is a bit puzzling.
Some collaborations \cite{LHP,Clover1} have found that their extrapolations 
underestimate significantly (up to $\simeq 30\%$) the well-known experimental 
value $\langle r^2 \rangle^{exp.} = 0.452 \pm 0.011~\mbox{fm}^2$ \cite{PDG}, 
while other collaborations \cite{Clover2,overlap,DWF} have obtained values 
consistent with experiment within the errors.

The European Twisted Mass (ETM) collaboration has recently produced a large number 
of gauge configurations with two flavors of dynamical quarks \cite{ETMC1,ETMC2,ETMC_tm} 
using the Wilson twisted-mass fermionic action \cite{twisted-mass} and the tree-level 
Symanzik improved (tlSym) gauge action \cite{gauge}.
In order to obtain (almost) automatic $\mathcal{O}(a)$ improvement the Wilson 
twisted-mass fermions have been tuned to maximal twists \cite{improvement}.
An intensive, systematic program of calculations of three-point correlation 
functions relevant for the determination of meson form factors both in the 
light and in the heavy sectors has then been started. 
Preliminary results, concerning the vector and scalar form factors of the pion, the 
Isgur-Wise universal function and the transition form factors relevant in $K_{\ell 3}$ 
and $D \to \pi (K)$ semileptonic decays have been presented in Ref.~\cite{ETMC_pion}.

In this paper we concentrate on the e.m.~form factor of the pion and we present 
the results of several measurements performed with pion masses in the range 
from $\simeq 260~\mev$ to $\simeq 580~\mev$, using six values of the quark 
mass at a lattice spacing of $\simeq 0.09~\mbox{fm}$ and two values of 
the quark mass at a lattice spacing of $\simeq 0.07~\mbox{fm}$.
The lattice box sizes are chosen in order to guarantee that $M_\pi L \gtrsim 4$ 
for minimizing as much as possible finite size effects.
Thanks to the use of all-to-all propagators evaluated by the stochastic procedure 
of Ref.~\cite{stoc} (see also \cite{ETMC2}) the statistical precision of the 
extracted form factor is quite impressive.
The momentum dependence of the pion form factor is investigated up to values of the 
squared four-momentum transfer $Q^2 \simeq 0.8~\gev^2$ and, thanks to the use of 
twisted boundary conditions (BC's) \cite{Bedaque,Tantalo}, down to $Q^2 \simeq 
0.05~\gev^2$. 
The $Q^2$-shape at the simulated pion masses is well reproduced by a single monopole 
ansatz with a pole mass lighter by $\approx 10\% \div 15\%$ than the lightest 
vector-meson mass.
Volume and discretization effects on the form factor are estimated using few 
simulations at different volumes and lattice spacings, and they turn out to 
be within the statistical errors.

The extrapolation of our results for the pion mass, decay constant and form factor 
to the physical point is carried out using (continuum) ChPT at NNLO \cite{BCT}. 
The extrapolated value of the (squared) pion charge radius is $\langle r^2 
\rangle^{phys} = 0.456 \pm 0.030_{\mbox{stat.}} \pm 0.024_{\mbox{syst.}}$ in 
nice agreement with the experimental result $\langle r^2 \rangle^{exp.} = 
0.452 \pm 0.011~\mbox{fm}^2$ \cite{PDG}.
The extrapolated values of the pion form factor agree very well with the experimental 
data up to $Q^2 \simeq 0.8~\gev^2$ within uncertainties which become competitive 
with the experimental errors for $Q^2 \gtrsim 0.3~\gev^2$.
The relevant low-energy constants (LEC's) appearing in the chiral expansion of the 
pion form factor are extracted from our lattice data adding in the fitting procedure 
the experimental value of the pion scalar radius \cite{BCT}. 
The latter helps constraining one the LEC's ($\bar{\ell}_4$), which in turn is 
beneficial for reducing the uncertainties of the other LEC's.
We get: $\bar{\ell}_1 = -0.4 \pm 1.3 \pm 0.6$, $\bar{\ell}_2 = 4.3 \pm 1.1 
\pm 0.4$, $\bar{\ell}_3 = 3.2 \pm 0.8 \pm 0.2$, $\bar{\ell}_4 = 4.4 \pm 0.2 
\pm 0.1$, $\bar{\ell}_6 = 14.9 \pm 1.2 \pm 0.7$, where the first error is 
statistical and the second one systematic.
Our findings are in nice agreement with the results of ChPT analyses of $\pi - \pi$ 
scattering data \cite{CGL}. 
The values found for $\bar{\ell}_3$ and $\bar{\ell}_4$ agree very well both with 
earlier ETMC results from Refs.~\cite{ETMC1,ETMC2} and with the recent ETMC 
determination of Ref.~\cite{ETMC_scaling}. 
This is quite reassuring because different kinds of systematic uncertainties may 
affect the two analyses: the present one being a NNLO analysis limited mainly 
to data from a single lattice spacing, and that of Ref.~\cite{ETMC_scaling} 
having two values of the lattice spacing but limited mainly to NLO in ChPT.

The plan of the paper is as follows.
In the next Section we briefly discuss the implementation of twisted BC's for the 
quark fields. 
In Section~\ref{sec:correlators} we present the calculation of two- and three-point 
correlation functions performed in the Breit reference frame, where the values of 
four-momentum transfer $Q^2$ are independent of the simulated pion mass. 
We also briefly show the stochastic procedure used for our unbiased estimate of 
the all-to-all propagators employed in this work.

In Section~\ref{sec:results} we firstly illustrate the very precise results obtained 
for the renormalization constant of the vector current and then we compare them 
with other determinations. 
Our accurate results for the momentum dependence of the pion form factor for the 
various simulated pion masses are presented and both volume and discretization 
effects are investigated. 

In Section~\ref{sec:radius}, using a single monopole ansatz to fit the 
momentum dependence of the form factor, the charge radius and the curvature 
are calculated at the simulated pion masses and analyzed both in terms of 
the ChPT expansion at NNLO from Ref.~\cite{BCT} and adopting a simple 
polynomial fit.

The mass and momentum dependencies of our lattice points for the pion form factor 
are analyzed in Section~\ref{sec:ffpion} without any model assumption, but 
using only the functional forms dictated by ChPT at NNLO.

In Section~\ref{sec:LECs} the final values of the relevant LEC's, including the 
estimate of the systematic errors, are presented and it is shown that the 
extrapolated form factor at the physical point agrees very well with the 
experimental data in the whole range of values of $Q^2$ considered.
Finally Section~\ref{sec:conclusions} is devoted to our conclusions.

\section{Lattice all-to-all quark propagators with twisted boundary conditions 
\label{sec:propagators}}

In lattice QCD simulations the spatial components of the hadronic momenta $p_j$ 
($j = 1, 2, 3$) are quantized.
The specific quantized values depend on the choice of the BC's applied to the 
quark fields. 
The most common choice is the use of periodic BC's in the spatial directions
 \be 
    \psi(x + \hat{e}_j L) = \psi(x) ~ ,
    \label{eq:periodic}
 \ee
that leads to 
 \be
    p_j = n_j \frac{2 \pi}{L} ~ , 
    \label{eq:pj}
  \ee
where the $n_j$'s are integer numbers. 
Thus the smallest non-vanishing value of $p_j$ is given by $2 \pi / L$, which 
depends on the spatial size of the (cubic) lattice ($V = L^3$).
For instance a current available lattice may have $L = 32 ~ a$, where $a$ is the 
lattice spacing, and $a^{-1} \simeq 2.5~\gev$ leading to $2 \pi / L \simeq 
0.5~\gev$. 
Such a value may represent a strong limitation of the kinematical regions accessible 
for the investigation of momentum dependent quantities, like e.g.~form factors.

In Ref.~\cite{Bedaque} it was proposed to use twisted BC's for the quark 
fields
 \be
    \widetilde{\psi}(x + \hat{e}_j L) = e^{2 \pi i \theta_j} ~ \widetilde{\psi}(x)
    \label{eq:twisted}
 \ee
which allows to shift the quantized values of $p_j$ by an arbitrary amount equal 
to $2 \pi \theta_j / L$, namely
 \be
    \widetilde{p}_j = p_j + \theta_j \frac{2 \pi}{L}  = n_j \frac{2 \pi}{L} + 
    \theta_j \frac{2 \pi}{L} ~ .
    \label{eq:pj_twisted}
  \ee

The twisted BC's (\ref{eq:twisted}) can be shown \cite{Bedaque} to be equivalent 
to the introduction of a U(1) background gauge field coupled to the baryon 
number and applied to quark fields satisfying usual periodic BC's (the Aharonov-Bohm 
effect). 
In Ref.~\cite{Tantalo} the twisted BC's were firstly implemented in a lattice QCD 
simulation of two-point correlation functions of pseudo-scalar mesons. 
The energy-momentum dispersion relation was checked confirming that the momentum 
shift $2 \pi \theta_j / L$ is a physical one.
In Ref.~\cite{GMS06} the twisted BC's were firstly applied to the calculation of 
the vector and scalar form factors relevant to the $K \to \pi$ semileptonic decay. 
It was shown that the momentum shift produced by the twisted BC's does not introduce 
any additional noise and easily allows to determine the form factors with good 
accuracy at quite small values of $Q^2$, which are not accessible when periodic 
BC's are considered\footnote{We mention that a new application of twisted BC's 
to the evaluation of the vector form factor at zero-momentum transfer has been 
proposed in Ref.~\cite{DWF}.}.

On the lattice, for a given flavor, the all-to-all quark propagator $S(x, y) \equiv 
\< \psi(x) ~ \overline{\psi}(y) \>$, where $\< \dots \>$ indicates the average over 
gauge field configurations weighted by the lattice QCD action, satisfies the 
following equation
 \be
    \sum_z D(x, z) ~ S(z, y) = \delta_{x, y} 
    \label{eq:Snormal}
  \ee
where $D(x, z)$ is the Dirac operator whose explicit form depends on the choice 
of the lattice QCD action\footnote{We omit in this Section color and Dirac indices 
for simplicity.}. 
In what follows we work with the fermionic twisted-mass Lattice QCD (tmLQCD) action 
with two flavors of mass-degenerate quarks given in Ref.~\cite{ETMC1}, tuned at 
{\em maximal} twist in the way described in full details in Ref.~\cite{ETMC2}. 
Therefore, in the so-called physical basis the operator $D(x, z)$ is given explicitly 
by 
 \be
    \label{eq:Dirac}
    D(x, z) & = & K(x, z) + i \gamma_5 \tau_3 ~ W(x, z) + a m~ \delta_{x, z} ~ , \\
    \label{eq:kinetic}
    K(x, z) & = & \frac{1}{2} \sum_{\mu = 1}^4 \gamma_\mu ~ \left\{ \delta_{x, z - 
                  a \hat{\mu}} ~ U_\mu^{\phantom\dagger}(x) - \delta_{x, z + 
                  a \hat{\mu}} ~ U_\mu^\dagger(z) \right\} ~ , \\
    \label{eq:Wilson}
    W(x, z) & = & \left( 4r + a m_{\rm crit} \right) \delta_{x, z} - \frac{r}{2} 
                  \sum_{\mu = 1}^4 \left\{ \delta_{x, z - a \hat{\mu}} ~ 
                  U_\mu^{\phantom\dagger}(x) + \delta_{x, z + a \hat{\mu}} ~ 
                  U_\mu^\dagger(z) \right\} ~ ,
 \ee
where $U_\mu(x)$ is the gauge link, $m$ is the bare twisted quark mass, 
$m_{\rm crit}$ is the critical value of the untwisted quark mass (needed 
to achieve maximal twist), $\tau_3$ is the third Pauli matrix acting in 
flavor space, and $r$ is the Wilson parameter, which is set to $r = 1$ 
in our simulations. 

We now want to consider the case in which a valence quark field satisfies the 
twisted BC's (\ref{eq:twisted}) in the spatial directions and is anti-periodic 
in time.
This is at variance with what has been done in the production of the ETMC gauge 
configurations, which include two sea quarks with periodic BC's in space and 
anti-periodic ones in time \cite{ETMC2}. 
However it has been recently shown \cite{SV05} that for many physical quantities, 
which do not involve final state interactions (like, e.g., meson masses, decay 
constants, semileptonic form factors and e.m.~transitions), the use of different 
BC's on valence and sea quarks produce finite-volume effects which remain 
exponentially small. 
In this way there is no need for producing new gauge configurations for each 
quark momentum, and this is quite relevant in the case of gauge configurations 
with dynamical fermions.

The corresponding quark propagator $\widetilde{S}(x, y) \equiv \< \widetilde{\psi}(x) ~ 
\overline{\widetilde{\psi}}(y) \>$ still satisfies Eq.~(\ref{eq:Snormal}) with the 
same Dirac operator $D(x, z)$ but with different BC's: 
 \be
    \sum_z D(x, z) ~ \widetilde{S}(z, y) = \delta_{x, y}
    \label{eq:Stwisted}
  \ee
Technically in order to work always with fields satisfying periodic BC's in space 
and time we follow Refs.~\cite{Bedaque,Tantalo} by introducing a new quark field 
as 
 \be
    \psi_{\widetilde{\theta}}(x) = e^{- 2 \pi i \widetilde{\theta} \cdot x / L } 
    \widetilde{\psi}(x) 
    \label{eq:psi_new}
 \ee
where the four-vector $\widetilde{\theta}$ is given by ($L/2T, \vec{\theta}$). 
In such a way the new quark propagator $S^{\widetilde{\theta}}(x, y) \equiv \< 
\psi_{\widetilde{\theta}}(x) ~ \overline{\psi}_{\widetilde{\theta}}(y) \> $ 
satisfies the equation 
 \be
    \sum_z D^{\widetilde{\theta}}(x, z) ~ S^{\widetilde{\theta}}(z, y) = \delta_{x, y}
    \label{eq:Stheta}
  \ee
with a modified Dirac operator $D^{\widetilde{\theta}}(x, z)$ but periodic BC's in 
both space and time.
The new Dirac operator is related to Eq.~(\ref{eq:Dirac}) by a simple re-phasing 
of the gauge links
 \be
    U_\mu(x) \to U_\mu^{\widetilde{\theta}}(x) \equiv e^{2 \pi i a 
    \widetilde{\theta}_\mu / L} ~ U_\mu(x) ~ .
    \label{eq:rephase}
 \ee

In terms of $S^{\widetilde{\theta}}(x, y)$, related to the quark fields 
$\psi_{\widetilde{\theta}}(x)$ with periodic BC's, the all-to-all quark propagator 
$\widetilde{S}(x, y)$, corresponding to the quark fields $\widetilde{\psi}(x)$ 
with twisted BC's, is simply given by 
 \be
    \widetilde{S}(x, y) = e^{2 \pi i \widetilde{\theta} \cdot (x - y) / L} ~ 
    S^{\widetilde{\theta}}(x, y) ~ . 
    \label{eq:Stilde}
 \ee

\section{Two- and three-point correlation functions\label{sec:correlators}}

We are interested in the calculation of the vector form factor of a charged pion 
defined through the relation
 \be
    \langle \pi^+(p^\prime) | \widehat{V}_{\mu}(0) | \pi^+(p)  \rangle = F_\pi(q^2) ~ 
    (p + p^{\prime})_\mu ~ ,
    \label{eq:ff}
 \ee
where $p$ ($p^\prime$) is the initial (final) pion four-momentum, $q^2 = (p - p^\prime)^2$ 
is the squared four-momentum transfer and $\widehat{V}_{\mu}$ is a conserved e.m.~current 
on the lattice. 
Splitting $\widehat{V}_{\mu}$ into an isovector and an isoscalar part, it is easy to show 
that the matrix elements of the isoscalar component between pion states is vanishing in 
the continuum limit, thanks to charge conjugation and isospin symmetries\footnote{Note 
that in tmLQCD the charge conjugation symmetry is preserved, while the isospin one 
is broken at finite lattice spacings.}.

Thus, up to discretization effects we take $\widehat{V}_{\mu}$ at a generic (Euclidean) 
space-time point $x = (t_x, \vec x)$ in the following form
 \be
     \widehat{V}_{\mu}(x) & = & Z_V ~ V_{\mu} (x)~ , \\
     V_{\mu}(x) & = & \frac{1}{2} \left[ \bar{u}(x) \gamma_{\mu} u(x) - 
     \bar{d}(x) \gamma_{\mu} d(x) \right] 
     \label{eq:vector}
 \ee
with $Z_V$ being the renormalization constant of the isovector part of the vector 
current at maximal twist (cf.~Ref.~\cite{improvement}).

The insertion of the current (\ref{eq:vector}) generates two types of Feynmann 
diagrams, the so-called connected and disconnected diagrams\footnote{The terms 
``connected" and ``disconnected" refer to fermionic lines only.}. 
In the former the external current is attached to the valence quarks, whereas in 
the latter the current interacts with the sea quarks. 
However, in the continuum limit the vanishing of the pion-to-pion matrix element 
of the isoscalar current ($\bar{u} \gamma_\mu u + \bar{d} \gamma_\mu d$) implies 
that the connected diagrams stemming from the u- and d-quark terms of the 
current (\ref{eq:vector}) are equal in absolute value and opposite in 
sign, while their disconnected counterparts are vanishing for each 
quark flavor.
Thus, in tmLQCD the disconnected diagrams for the e.m.~pion form factor represent 
a pure discretization effect, which turns out to be of order ${\cal{O}}(a^2)$ 
(see later on).
Therefore, up to discretization effects, it is enough to consider only the 
connected insertion of one single flavor of Eq.~(\ref{eq:vector}).

From Eq.~(\ref{eq:ff}) the pion form factor can be extracted from both the time 
and the spatial components of the vector current. 
However for reasons which will become clear during this Section, we work in the 
Breit reference frame where $\vec{p}^{\,\prime} = - \vec{p}$, so that the spatial 
components of the vector current are vanishing identically.
Therefore we limit ourselves to consider the following two- and three-point 
correlation functions
\be
    \label{eq:c2pt}
    C^\pi (t, \vec p ) & = & \sum_{x, z} \langle O_\pi(x) ~ O_\pi^\dagger(z) 
    \rangle ~ \delta_{t, t_x - t_z} ~ e^{-i \vec p \cdot (\vec x - \vec z)} ~ , \\
    \label{eq:c3pt}
    C_0^{\pi \pi} (t, t^\prime, \vec p, \vec{p}^{\,\prime}) & = & \sum_{x, y, z} 
    \langle O_\pi(y) ~ V_0(x) ~ O_\pi^\dagger(z) \rangle ~ \delta_{t, t_x - t_z} 
    ~ \delta_{t^\prime, t_y - t_z} \nn \\ 
    & \cdot & e^{-i \vec p \cdot (\vec x - \vec z) + i \vec{p}^{\,\prime} \cdot 
    (\vec x - \vec y)} ~ ,
\ee
where $V_0(x) = \bar{u}(x) \gamma_0 u(x)$ and $O_\pi^\dagger(z) = \bar u(z) 
\gamma_5 d(z)$ is the operator interpolating the $\pi^+$ mesons.
Note that, since we want to use all-to-all propagators, in Eqs.~(\ref{eq:c2pt}) 
and (\ref{eq:c3pt}) there is an additional sum over the space-time lattice 
volume, which helps improving the signal quality with respect to the case of a 
fixed-point source ($z = 0$).

Using the completeness relation and taking $t$ and $(t^\prime - t)$ large enough, 
one gets
 \be
    \label{eq:c2ptexp}
    C^\pi(t, \vec p) & _{\overrightarrow{ {\mbox{\tiny $t \to \infty$}} }} & 
    \frac{Z_\pi}{2 E_\pi(\vec p)} ~ e^{- E_\pi(\vec p) t} ~ , \\[2mm]
    C_0^{\pi \pi}(t, t^\prime, \vec p, \vec{p}^{\,\prime}) ~ & 
    _{\overrightarrow{\stackrel{\mbox{\tiny $t \to \infty$}}{\mbox{\tiny $(t^\prime - t) 
    \to \infty$}}}} & \frac{Z_\pi} {2 E_\pi(\vec p) ~ 2 E_\pi(\vec{p}^{\,\prime})} 
    \frac{1}{Z_V} \langle \pi^+(p^\prime) | \widehat{V}_0 | \pi^+(p) \rangle 
    \nn \\[2mm] 
    & & e^{- E_\pi(\vec p) t} ~ e^{ - E_\pi(\vec{p}^{\,\prime}) (t^\prime - t)} ~ ,
    \label{eq:c3ptexp}
\ee
where, up to discretization effects, $E_\pi(\vec p) = \sqrt{M_{\pi}^2 + |\vec{p}|^2}$ 
and $\sqrt{Z_\pi} = \langle 0 | O_\pi(0) | \pi^+ \rangle$ is independent on the meson 
momentum $\vec p$.
Note that in tmLQCD at maximal twist the value of the coupling constant $Z_\pi$ 
determines the pion decay constant $f_\pi$ \cite{improvement} without the need of 
the knowledge of any renormalization constant, namely
 \be
    f_\pi = 2m \frac{\sqrt{Z_\pi}}{M_\pi^2} ~ ,
    \label{eq:fpi_tm}
 \ee
where $m$ is the bare twisted quark mass.

Taking advantage of the choice of the Breit frame where $\vec{p}^{\,\prime} = - 
\vec{p}$, it follows
 \be
    \frac{C_0^{\pi \pi}(t, t^\prime, \vec p, -\vec p)}{C^\pi(t^\prime, \vec p )} ~ 
    _{\overrightarrow{\stackrel{\mbox{\tiny $t \to \infty$}}{\mbox{\tiny 
    $(t^\prime - t) \to \infty$}}}} ~ \frac{1}{Z_V} \frac{\langle \pi^+(p^\prime) | 
    \widehat{V}_0 | \pi^+(p) \rangle }{2 E_\pi(\vec p)} = \frac{1}{Z_V} 
    F_\pi(q^2) ~ , 
    \label{eq:standard}
 \ee
where 
 \be
    q^2 \equiv \left[ E_\pi(\vec p) - E_\pi(\vec{p}^{\,\prime}) \right]^2 - 
    |\vec p - \vec{p}^{\,\prime}|^2 ~ _{\overrightarrow{\vec{p}^{\,\prime} = - 
    \vec p}} ~ -4 |\vec p|^2
    \label{eq:q2}
 \ee
is independent of the simulated pion mass.

The vector renormalization constant can be obtained from Eq.~(\ref{eq:standard}) 
by using the absolute normalization of the pion form factor at $q^2 = 0$, namely 
$F_\pi(q^2 = 0) = 1$, which implies
 \be
    Z_V ~ _{\overrightarrow{\stackrel{\mbox{\tiny $t \to \infty$}}{\mbox{\tiny 
    $(t^\prime - t) \to \infty$}}}} ~ \frac{C^\pi(t^\prime, \vec 0 )}{C_0^{\pi \pi}(t, 
    t^\prime, \vec 0, \vec 0)} ~ .
    \label{eq:ZV}
 \ee
Combining Eqs.~(\ref{eq:standard}) and (\ref{eq:ZV}) one gets
 \be
    R_0(t, t^\prime; q^2) \equiv \frac{C_0^{\pi \pi}(t, t^\prime, \vec p, 
    -\vec p)}{C_0^{\pi \pi}(t, t^\prime, \vec 0, \vec 0)} ~ \frac{C^\pi(t^\prime, 
    \vec 0 )}{C^\pi(t^\prime, \vec p )} ~ _{\overrightarrow{\stackrel{\mbox{\tiny 
    $t \to \infty$}}{\mbox{\tiny $(t^\prime - t) \to \infty$}}}} ~ F_\pi(q^2) 
    \label{eq:ffpion}
 \ee
which means that the pion form factor can be obtained directly from the plateau of 
the double ratio given by the l.h.s.~of Eq.~(\ref{eq:ffpion}) at large time 
distances. 
Note that in this way the normalization condition $F_\pi(q^2 = 0) = 1$ is fulfilled at 
all quark masses, lattice volumes and spacings.

The (mass-dependent) renormalization constant $Z_V$ can be obtained alternatively using 
the 3-point correlation function calculated in a frame in which the initial and final 
pions have the same momentum $\vec p$, i.e.~from the  plateau of the ratio $C^\pi(t^\prime, 
\vec p )~/$ $C_0^{\pi \pi}(t, t^\prime, \vec p, \vec p)$  at large time distances. 
In this way the pion form factor can be extracted from the plateau of a ratio of 
3-point correlation functions only, i.e.~from $C_0^{\pi \pi}(t, t^\prime, 
\vec p, -\vec p)~/$ $C_0^{\pi \pi}(t, t^\prime, \vec p, \vec p)$. 
Such an alternative approach has been tested in Ref.~\cite{DWF} and shown to have a 
statistical precision similar to the one based on Eq.~(\ref{eq:ffpion}).

In terms of the all-to-all quark propagators $S^{u(d)}(x, z)$, where the flavor labels 
$u$ and $d$ correspond to $\tau_3 = \pm 1$ in Eq.~(\ref{eq:Dirac}), the two-point 
function $C^\pi(t, \vec p)$ of the charged pion becomes
 \be
    C^\pi(t, \vec p) = \sum_{x, z} \langle Tr[ S_u(x, z) \gamma_5 S_d(z, x) 
    \gamma_5 ] \rangle ~ \delta_{t, t_x - t_z} ~ e^{-i \vec p \cdot (\vec x - 
    \vec z)} . 
    \label{eq:C2pt_PS}
 \ee
For the tmLQCD action the $\gamma_5$-hermiticity property 
 \be
    S_d(z, x) = \gamma_5 S_u^{\dagger}(x, z) \gamma_5
    \label{eq:hermiticity}
 \ee
holds with the dagger operator acting in the (suppressed) color and Dirac spaces.

As for the three-point correlation function $C_0^{\pi \pi}(t_x, t_y, \vec p, -\vec p)$, 
according to the discussion on the disconnected diagrams made before Eq.~(\ref{eq:c3pt}), 
up to discretization effects one gets
 \be
    C_0^{\pi \pi}(t, t^\prime, \vec p, -\vec p) & = & \sum_{x, z} \langle Tr[ S_u(x, z) 
    \gamma_5 \overline{\Sigma}_{du}(z, x; t^\prime; -\vec p) \gamma_0 ] \rangle \nn \\ 
    & \cdot & ~ \delta_{t, t_x - t_z} ~ e^{-2i \vec p \cdot (\vec x - \vec z)} ~ , 
    \label{eq:C3pt_PS}
 \ee
where $\overline{\Sigma}_{du}(z, x; t^\prime; -\vec p) = \gamma_5 [ \Sigma_{du}(x, z; 
t^\prime; -\vec p) ]^{\dagger} \gamma_5$ and
 \be
    \Sigma_{du}(x, z; t^\prime; \vec p) = \sum_y S_d(x, y) \gamma_5 S_u(y, z) ~ 
    e^{-i\vec p \cdot (\vec z - \vec y)} ~ \delta_{t^\prime, t_y - t_z} ~ .
    \label{eq:S1}
 \ee
The {\em sequential} propagator $\Sigma_{du}(x, z; t^\prime; \vec p)$ satisfies 
the equation 
 \be
    \sum_y D_d(x, y) ~ \Sigma_{du}(y, z; t^\prime; \vec p) = \gamma_5 S_u(x, z) ~ 
    \delta_{t^\prime, t_x - t_z} ~ e^{i \vec p \cdot (\vec x - \vec z)} ~ .
    \label{eq:Sigma}
 \ee

As it has been shown in Ref.~\cite{improvement}, the calculation of correlation 
functions of parity symmetric operators is automatically ${\cal{O}}(a)$ improved 
at maximal twist.
Thus for non-vanishing values of the spatial momenta the ${\cal{O}}(a)$ terms 
can be eliminated by appropriate averaging of the correlation functions over 
initial and final momenta of opposite sign.
Using the symmetry of correlation functions under the spatial inversion and the 
simultaneous exchange of u and d quarks (which is fulfilled only after gauge 
averaging at maximal twist in the physical basis) as well as the charge conjugation 
symmetry and the $\gamma_5$-hermiticity property, one gets that: ~ i) the 
correlators (\ref{eq:C2pt_PS}) and (\ref{eq:C3pt_PS}) are real, and ~ ii) 
$C^\pi(t^\prime, \vec p) = C^\pi(t^\prime, -\vec p)$ and $C_0^{\pi \pi}(t, 
t^\prime, \vec p, -\vec p) = C_0^{\pi \pi}(t, t^\prime, -\vec p, \vec p)$.
Thus discretization effects in both $C^\pi(t^\prime, \vec p)$ and $C_0^{\pi 
\pi}(t, t^\prime, \vec p, -\vec p)$ start automatically at order 
${\cal{O}}(a^2)$\footnote{This result holds as well in all reference frames and 
it explains the findings shown in Fig.~12 of Ref.~\cite{tm_q}, where the 
correlation functions with opposite momenta have been calculated 
explicitly and found to be identical within statistical errors.}.

Let us now consider the case of quark fields with twisted BC's. 
Equations~(\ref{eq:C2pt_PS})-(\ref{eq:Sigma}) hold as well by simply replacing 
the propagators S and $\Sigma$ with the corresponding twisted ones, $\widetilde{S}$ 
and $\widetilde{\Sigma}$, and by taking into account the change of the quantized 
momenta, namely $p_j \to \widetilde{p}_j$ [see Eq.~(\ref{eq:pj_twisted})]. 
The two- and three-point correlators can be expressed in terms of quark propagators 
satisfying periodic BC's, e.g.~ in terms of $S^{\widetilde{\theta}}$ [see 
Eq.~(\ref{eq:Stilde})]. 
We now write down the explicit formulae for sake of completeness. 

In order to work in the Breit frame we consider three choices of the twisting 
four-vector $\widetilde{\theta}$, namely $\widetilde{\theta} = \widetilde{\theta}_\pm 
= (L/2T, \pm \vec{\theta})$ and $\widetilde{\theta} = \widetilde{\theta}_0 = 
(L/2T, \vec 0)$ for various values of $\vec{\theta}$.
Writing $\vec{p}$ in the generic form $\vec{p} = 2 \pi \vec{\theta} / L$, we get 
 \be
    C^\pi(t, \frac{2 \pi}{L} \vec{\theta}) = \sum_{x, z} \langle 
    Tr[ S_u^{\widetilde{\theta}_+}(x, z) \gamma_5 S_d^{\widetilde{\theta}_0}(z, x) 
    \gamma_5 ] \rangle ~ \delta_{t, t_x - t_z} , 
    \label{eq:C2pt_twisted}
 \ee
 \be
    C_0^{\pi \pi} (t, t^\prime, \frac{2 \pi}{L} \vec{\theta}, -\frac{2 \pi}{L} 
    \vec{\theta}) & = & \sum_{x, z} \langle Tr[ S_u^{\widetilde{\theta}_+}(x, z) 
    \gamma_5 \overline{\Sigma}_{du}^{\widetilde{\theta}_0, {\widetilde{\theta}_-}}(z, 
    x; t^\prime) \gamma_0 ] \rangle ~ \delta_{t, t_x - t_z} ~ ,
    \label{eq:C3pt_twisted}
 \ee
where thanks to the $\gamma_5$-hermiticity property one has
 \be
    \overline{\Sigma}_{du}^{\widetilde{\theta}_0, {\widetilde{\theta}_-}}(z, x; t^\prime) 
    = \gamma_5 [ \Sigma_{du}^{\widetilde{\theta}_-, {\widetilde{\theta}_0}} ]^\dagger 
    (x, z; t^\prime) \gamma_5
    \label{eq:hermiticity_sigma}
 \ee
and the sequential propagator $\Sigma_{du}^{\widetilde{\theta}_-, 
{\widetilde{\theta}_0}}(x, z; t^\prime)$ satisfies the modified Dirac equation
 \be
    \sum_y D_d^{\widetilde{\theta}_-}(x, y) ~ \Sigma_{du}^{\widetilde{\theta}_-, 
    \widetilde{\theta}_0}(y, z; t^\prime) = \gamma_5 S_u^{\widetilde{\theta}_0}(x, z) 
    ~ \delta_{t^\prime, t_x - t_z} ~ .
    \label{eq:Sigma_twisted}
 \ee

Note that, because of Eq.~(\ref{eq:Stilde}), no exponential factors appear in the 
r.h.s.~of Eqs.~(\ref{eq:C2pt_twisted})-(\ref{eq:Sigma_twisted}) and the dependence 
on the vector $\vec{\theta}$ is totally embedded in the twisted quark propagators 
$S^{\widetilde{\theta_+}}$ and $\Sigma^{\widetilde{\theta}_-, 
{\widetilde{\theta}_0}}$.

\subsection{Stochastic procedures\label{subsec:stochastic}}

The next point to be addressed is the evaluation of the all-to-all propagator 
$S^{\widetilde{\theta}}(x, z)$ which is the solution of the modified Dirac 
equation (\ref{eq:Stheta}).
Restoring color and spin indices, denoted by Latin and Greek letters respectively, 
one has
 \be
    \sum_y [D^{\widetilde{\theta}}(x, y)]_{\alpha \beta}^{a b}~ 
    [S^{\widetilde{\theta}}(y, z)]_{\beta \gamma}^{b c} = \delta_{x, z} ~ 
    \delta_{a, c} ~ \delta_{\alpha, \gamma} ~ .
    \label{eq:Stheta_full}
 \ee
The computation of exact all-to-all quark propagators is a formidable task well beyond 
present computational capabilities, because it involves a huge number of inversions of 
the Dirac equation for all possible locations of the source in space and time. 
Consequently most of the lattice computations of connected 2- and 3-point correlation 
functions are till now carried out using the point-to-all propagator by fixing the 
source at some space-time point, referred to as the origin. 
To get the expressions of our 2- and 3-point correlators in terms of point-to-all 
propagators it is enough to limit the sum over the variable $z$ to $z = 0$ everywhere 
in Eqs.~(\ref{eq:C2pt_twisted})-(\ref{eq:Sigma_twisted}).
The basic advantage of the all-to-all propagator with respect to the point-to-all one 
relies in the fact that the former contains all the information on the gauge 
configuration, which in turn means that the calculation of 2- and 3-point 
functions using all-to-all propagators is expected to have much less 
gauge noise.

An efficient way to estimate the all-to-all propagator is based on stochastic 
techniques with the help of variance reduction methods to better separate the 
signal from the noise (see Ref.~\cite{Wilcox} and references therein).
In recent years new stochastic methods have been developed, like the dilution 
method of Ref.~\cite{dilution} and the so-called "one-end-trick" of Ref.~\cite{stoc}.
The latter, already applied by the ETM collaboration to the calculation of neutral 
meson masses (see Refs.~\cite{ETMC2} and \cite{ETMC3}), allows to achieve a great 
reduction of the noise-to-signal ratio and it will be applied in this work to the 
calculation of 3-point correlation functions (see also Refs.~\cite{ETMC_pion} 
and \cite{RBC_UKQCD}).

The starting point of all stochastic approaches is to consider random sources 
$\eta_r^a(x)$, which, for reasons that will become clear later on, we take 
independent of both the spin variable and the twisting vector $\widetilde{\theta}$ 
(i.e., of the quark momentum). 
The index $r$ ($r = 1, ... N$) enumerates the generated random sources, which 
must satisfy the following constraint
 \be
    \lim_{N \to \infty} \frac{1}{N} \sum_{r = 1}^N \eta_r^a(x) [\eta_r^b(y)]^* = 
    \delta_{a, b} ~ \delta_{x, y} ~ .
    \label{eq:etar}
  \ee
In this work we adopt for the sources a random choice of $\pm 1$ values.
Then one introduces the ``$\phi$-propagator"
 \be
    [\phi_r^{\widetilde{\theta}}(x)]_{\alpha \beta}^a = \sum_y
    [S^{\widetilde{\theta}}(x, y)]_{\alpha \beta}^{a b} ~ \eta_r^b(y) ~ ,
    \label{eq:phi}
 \ee
which is solution of the equation
 \be
    \sum_y [D^{\widetilde{\theta}}(x, y)]_{\alpha \beta}^{a b} ~ 
    [\phi_r^{\widetilde{\theta}}(y)]_{\beta \gamma}^b =  \eta_r^a(x) ~ 
    \delta_{\alpha, \gamma} ~ .
    \label{eq:Dphi}
 \ee
where the sum over repeated color or spin indices is understood. 
As explained in details in Ref.~\cite{ETMC2}, the quantity $(1/N) \sum_{r = 1}^N 
[\phi_r^{\widetilde{\theta}}(x)]_{\alpha \beta}^a [\eta_r^b(y)]^*$ is an unbiased 
estimator of the all-to-all propagator $[S^{\widetilde{\theta}}(x, y)]_{\alpha 
\beta}^{a b}$. However, while the signal is of order ${\cal{O}}(1)$, the noise is 
of the order $\sqrt{V / N}$ (where V is the space-time volume) and therefore a 
huge number of random sources and inversions of Eq.~(\ref{eq:Dphi}) would be 
required.

The ``one-end-trick" is based on the observation that the product of two
``$\phi$-propagators" is an unbiased estimator of the product of two all-to-all 
propagators summed over the intermediate space-time points.
In this case, however, the signal is of order V, while the noise is of order $V / 
\sqrt{N}$, so that it is even sufficient to employ one random source per gauge 
configuration, as we do in this work.

Choosing the random source $\eta_a^r(x)$ to be non-vanishing only for a 
randomly-chosen time slice, located at $t_r$ \footnote{The random choice of 
the time slice at $t_r$ is mainly motivated by the reduction of autocorrelations 
observed for fermionic quantities using the ETM gauge ensembles (see Ref.~\cite{ETMC2}).}, 
the 2-point correlation function (\ref{eq:C2pt_twisted}) can be estimated as 
 \be
    C^\pi(t, \frac{2 \pi}{L} \vec{\theta}) = \sum_{\vec x, t_x} 
    \langle [\phi_{u, r}^{\widetilde{\theta}_+}(\vec x, t_x)]_{\alpha \beta}^a
    ~ \{ [\phi_{u, r}^{\widetilde{\theta}_0}(\vec x, t_x)]_{\beta \alpha}^a \}^* ~ 
    \delta_{t, t_x - t_r} \rangle
    \label{eq:C2pt_stoc}
 \ee
where we notice that the two $\phi$'s have the same flavor.
Looking at the above equation the $\phi$-propagator 
$[\phi_r^{\widetilde{\theta}}(x)]_{\alpha \beta}^a$ plays a role quite similar to the 
one of the point-to-all propagator $[S^{\widetilde{\theta}}(x, 0)]_{\alpha \beta}^{a b}$ 
with only one color index, being the other one carried by the random source. 
This means that the time needed for the calculation of the $\phi$-propagator 
is $1/3$ of the one required for the point-to-all propagator.
Note also that both $\phi_r^{\widetilde{\theta}_+}(x)$ and 
$\phi_r^{\widetilde{\theta}_0}(x)$ are solutions of Eq.~(\ref{eq:Dphi}) with the 
same random source $\eta_r(x)$. 
This is essential to properly get the r.h.s.~of Eq.~(\ref{eq:C2pt_stoc}).
Moreover the independence of the random source from spin indices allows to 
evaluate 2-point correlation functions with interpolating fields of the form 
($\bar{q} \Gamma q^\prime$) for any Dirac matrix $\Gamma$.

The stochastic estimate of the 3-point correlation function (\ref{eq:C3pt_twisted}) 
requires the introduction of the sequential ``$\Phi$-propagator"
 \be
    [\Phi_{du, r}^{\widetilde{\theta}_-, \widetilde{\theta}_0}(x; 
    t^\prime)]_{\alpha \beta}^a = \sum_y [\Sigma_{du, r}^{\widetilde{\theta}_-, 
    \widetilde{\theta}_0}(x, y; t^\prime)]_{\alpha \beta}^{a b} ~ \eta_r^b(y) ,
    \label{eq:Phi}
 \ee
which is solution of the equation
 \be
    \sum_y [D_d^{\widetilde{\theta}_-}(x, y)]_{\alpha \beta}^{a b} ~ 
    [\Phi_{du, r}^{\widetilde{\theta}_-, \widetilde{\theta}_0}(y; 
    t^\prime)]_{\beta \rho}^b =  [\gamma_5]_{\alpha \gamma} ~ 
    [\phi_{u, r}^{\widetilde{\theta}_0}(x)]_{\gamma \rho}^a 
    ~ \delta_{t^\prime, t_x - t_r} ~ .
    \label{eq:DPhi}
 \ee
One gets
 \be
    C_0^{\pi \pi}(t, t^\prime, \frac{2 \pi}{L} \vec{\theta}, -\frac{2 \pi}{L} 
    \vec{\theta}) & = & \sum_{\vec x, t_x} \langle ~ [\phi_{u, r}^{\widetilde{\theta}_+}
    (\vec x, t_x)]_{\alpha \beta}^a ~ \{ [\Phi_{du, r}^{\widetilde{\theta}_-, 
    \widetilde{\theta}_0}(\vec x, t_x; t^\prime)]_{\beta \gamma}^a \}^* \nn \\ 
    & \cdot & [\gamma_5 \gamma_0]_{\gamma \alpha} ~ \delta_{t, t_x - t_r} \rangle ~ .
    \label{eq:C3pt_stoc}
 \ee
Note that: ~ i) the quark propagators required in Eqs.~(\ref{eq:C2pt_stoc}) and 
(\ref{eq:C3pt_stoc}) are those of one single flavor, while the other quark flavor 
appears only in the modified Dirac operator of Eq.~(\ref{eq:DPhi}), and ~ ii) for 
each value of the quark momentum injected via the twisted BC's a new inversion of 
the Dirac operator is required.

\section{The charged pion form factor\label{sec:results}}

As already mentioned in the Introduction, the ETM collaboration has started an 
intensive, systematic program of calculations of three-point correlation 
functions relevant for the determination of meson form factors at low, 
intermediate and heavy quark masses.
In this work we concentrate on the results obtained for the vector form factor of 
the pion. 

In Table \ref{tab:setup} we collect the simulation set-up for all the runs carried out 
at $\beta = 3.9$ and for the two runs performed at a finer lattice spacing ($\beta = 
4.05$). 
Approximate values of the (charged) pion mass $M_\pi$ in physical units as well as of 
the quantity $M_\pi L$, which governs finite volume effects in the so-called 
$p$-regime of ChPT, are reported for each run.

\begin{table}[!htb]

\begin{center}
\small{\begin{tabular}{||c|c||c|c||c|c|c|c|c|c||}
\hline
$\beta$ & $a$ & $Run$ & $Refs.$ & $a m_{sea}$ & $V \cdot T ~ / ~ a^4$ & $M_\pi$ & $M_\pi L$ & $No.~gauge$ \\ 
        & $(fm)$ & & $\cite{ETMC2,ETMC_tm}$ & & & $(\mev)$  & & $config.$ \\ \hline \hline
$3.9$  & $\simeq 0.09$ & $R_1$    & $B_7$        & $0.0030$ & $32^3 \cdot 64$ & $\simeq 260$ & $\simeq 3.7$ & $240$ \\ \hline
       &               & $R_{2a}$ & $B_6$        & $0.0040$ & $32^3 \cdot 64$ & $\simeq 300$ & $\simeq 4.2$ & $240$ \\ \hline
       &               & $R_{2b}$ & $B_{1a,b,c}$ & $0.0040$ & $24^3 \cdot 48$ & $\simeq 300$ & $\simeq 3.2$ & $480$ \\ \hline
       &               & $R_3$    & $B_2$        & $0.0064$ & $24^3 \cdot 48$ & $\simeq 380$ & $\simeq 4.0$ & $240$ \\ \hline
       &               & $R_4$    & $B_{3a,b}$   & $0.0085$ & $24^3 \cdot 48$ & $\simeq 440$ & $\simeq 4.7$ & $240$ \\ \hline
       &               & $R_{5a}$ & $B_4$        & $0.0100$ & $24^3 \cdot 48$ & $\simeq 480$ & $\simeq 5.1$ & $240$ \\ \hline
       &               & $R_6$    & $B_{5a,b}$   & $0.0150$ & $24^3 \cdot 48$ & $\simeq 580$ & $\simeq 6.1$ & $240$ \\ \hline \hline
$4.05$ & $\simeq 0.07$ & $R_{2c}$ & $C_1$        & $0.0030$ & $32^3 \cdot 64$ & $\simeq 300$ & $\simeq 3.4$ & $240$  \\ \hline
       &               & $R_{5b}$ & $C_3$        & $0.0080$ & $32^3 \cdot 64$ & $\simeq 480$ & $\simeq 5.4$ & $240$  \\ \hline

\end{tabular}
}

\caption{\it Set-up of the lattice simulations for the various runs considered in this 
work.
\label{tab:setup}}

\end{center}

\end{table}

The gauge configurations used for the measurements are selected from the trajectories 
produced by the ETM collaboration (see Refs.~\cite{ETMC1,ETMC2,ETMC_tm}) at various 
values of the sea (bare) quark mass, $a m_{sea}$. 
We have chosen 1 configuration out of at least 20 (equilibrated) trajectories in 
all cases except for the run at the lightest pion mass (1 out of 10).

Volume effects can be checked through the runs $R_{2a}$ and $R_{2b}$ (see later 
subsection \ref{subsec:volume}), while lattice artifacts can be studied by means 
of the runs $R_{2b}$ and $R_{2c}$ at a pion mass around $300~\mev$ and of the 
runs $R_{5a}$ and $R_{5b}$ for $M_\pi \simeq 480~\mev$ (see later subsection 
\ref{subsec:a2}).

In this work at each value of the (bare) quark mass, $a m = a m_{sea}$, the statistical 
errors are evaluated with the jackknife procedure, while a bootstrap sampling will 
be applied in order to combine the jackknives for different quark masses (see 
later Section \ref{sec:radius}).

\subsection{Vector renormalization constant $Z_V$\label{subsec:ZV}}

In tmLQCD tuned at maximal twist the constant $Z_V$ renormalizes both the isovector part 
of the (local) e.m.~current [see Eq.~(\ref{eq:vector})] and the isovector off-diagonal 
components of the (local) axial current, e.g.~$A_\mu(x) = \bar{d}(x) \gamma_\mu 
\gamma_5 u(x)$.
Therefore the renormalization constant $Z_V$ can be calculated in two ways. 
The first one is from Eq.~(\ref{eq:ZV}), which makes use of 2- and 3-point correlation 
functions and is equivalent to fix the absolute normalization of the pion form factor, 
$F_\pi(0) = 1$. 
The second way is from the non-singlet axial Ward Identity ($WI$), which, up to 
discretization effects, in tmLQCD at maximal twist reads as \cite{improvement}
 \be
    Z_V \partial_\mu A_\mu(x) = 2 a m P_5(x)
    \label{eq:WI}
 \ee
where $a m$ is the bare quark mass and $P_5(x) = \bar{d}(x) \gamma_5 u(x)$ is the bare 
pseudo-scalar density.
The presence of bare operators in the r.h.s.~of Eq.~(\ref{eq:WI}) is due to the fact that 
at maximal twist the mass renormalization constant is equal to the inverse of the 
pseudo-scalar renormalization constant, i.e.~$Z_m = Z_P^{-1}$.
At zero momentum it follows 
 \be
    Z_V = 2 a m ~ \frac{C^\pi(t, \vec 0)}{\partial_t A^\pi(t, \vec 0)}
    \label{eq:ZV_WI}
  \ee
with $A^\pi(t, \vec 0) = \sum_{x, z} \langle A_0(x) ~ P_5(z) \rangle ~ \delta_{t, 
t_x - t_z}$.

\begin{figure}[!hbt]

\centerline{\includegraphics[scale=0.80]{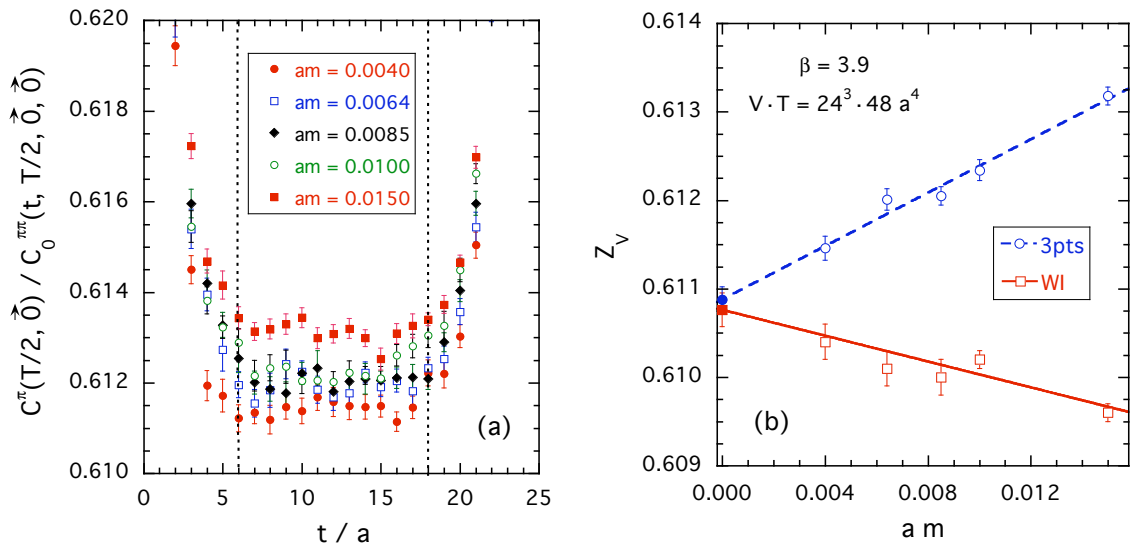}}

\caption{\it (a) Ratio of 2-point and 3-point correlation functions given by the r.h.s.~of 
Eq.~(\ref{eq:ZV}), evaluated for $t^\prime = T / 2$ at $\beta = 3.9$ and $V \cdot T = 24^3 
\cdot 48 ~ a^4$, versus the (Euclidean) time $t$ in lattice units. 
(b) The vector renormalization constant $Z_V$ as obtained at different values of the bare 
quark mass in lattice units. 
Open dots correspond to the values extracted from the plateau region denoted by the vertical 
dotted lines in (a).
Open squares are the results obtained from the WI using Eq.~(\ref{eq:ZV_WI}) in 
Ref.~\cite{renorm}.
The solid and dashed lines are simple linear interpolations of the lattice points and the 
full markers denote the corresponding values at the chiral point.}

\label{fig:ZV}

\end{figure}

The results obtained for the ratio given by the r.h.s.~of Eq.~(\ref{eq:ZV}), 
evaluated for all the runs at $\beta = 3.9$ and $V \cdot T = 24^3 \cdot 48 ~ a^4$, 
are shown in Fig.~\ref{fig:ZV}(a). 
The time distance $t^\prime$ between the time slices of the source and the sink 
is fixed at $t^\prime = T /2$, so that the 3-point correlation function 
(\ref{eq:c3pt}) becomes antisymmetric with respect to $t = T / 2$ and 
it can be appropriately averaged to reduce the statistical fluctuations. 
Moreover for finite time extension $T$ the 2-point correlation function 
$C^\pi(t, \vec p)$ is symmetric with respect to $t = T /2$, so that a 
second exponential $e^{-E_\pi(\vec p) (T - t)}$ appears in 
Eq.~(\ref{eq:c2ptexp}) and a factor $1/2$ has to be applied 
to the r.h.s~of Eq.~(\ref{eq:ZV}).

From the plateau region denoted by the vertical dotted lines in Fig.~\ref{fig:ZV}(a) 
an estimate of the renormalization constant $Z_V$ can be obtained at each value of 
the bare quark mass.
The results are reported in Fig.~\ref{fig:ZV}(b) and compared with the corresponding 
results obtained from the WI using Eq.~(\ref{eq:ZV_WI}) (see Ref.~\cite{renorm}).
Both methods exhibit an extremely high statistical precision of the order of $0.3~\%$.

The quark mass dependence visible in Fig.~\ref{fig:ZV}(b) is a pure discretization effect 
and is different between the two methods. 
It appears to be linear in both cases, which is not in contradiction with the ${\cal{O}}(a)$ 
improvement, since terms proportional to $a^2 m \Lambda_{QCD}$ may be dominant with respect 
to terms proportional to $a^2 m^2$. 

The extrapolations to the chiral limit should therefore coincide, providing the value 
of the renormalization constant $Z_V$, which is indeed defined in such a limit.
From Fig.~\ref{fig:ZV}(b) it can be seen that the values obtained by a simple linear 
fit at the chiral point coincide nicely within quite small statistical errors, namely 
$Z_V = 0.61088 (14)$ from Eq.~(\ref{eq:ZV}) and $Z_V = 0.61076 (19)$ from the WI 
[Eq.~(\ref{eq:ZV_WI})].

\subsection{Momentum dependence of the 2-point correlation function\label{subsec:2pts}}

\begin{figure}[!hbt]

\centerline{\includegraphics[scale=0.80]{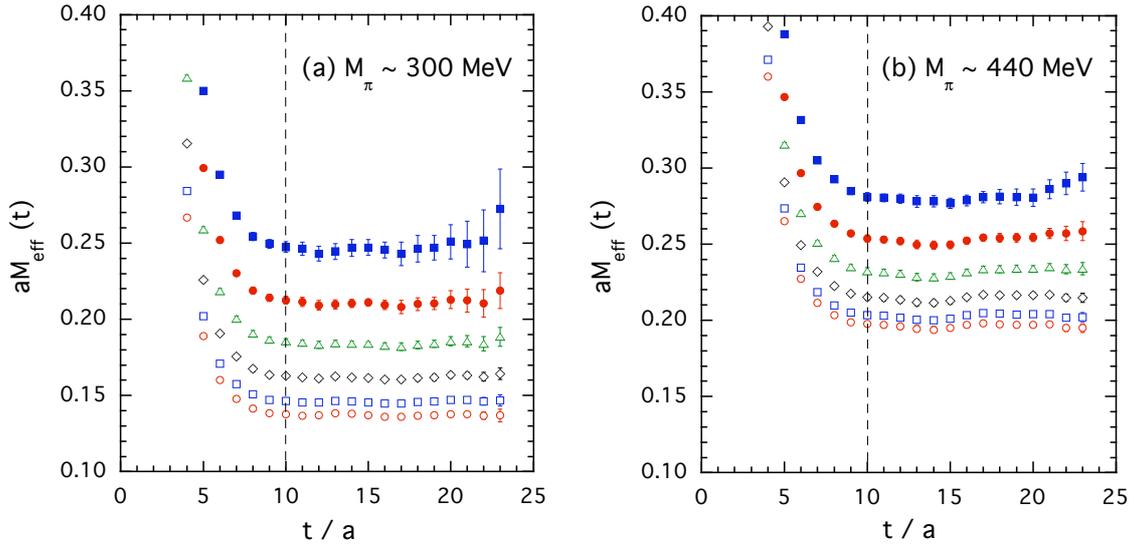}}

\caption{\it Effective mass of the pion (\ref{eq:M_eff}) versus the (Euclidean) time distance 
in lattice units for $M_\pi \simeq 300~\mev$ (a) and $M_\pi \simeq 440~\mev$ (b) at $\beta = 
3.9$ and $V \cdot T = 24^3 \cdot 48 ~ a^4$. 
The twisting angle $\vec{\theta}$ is chosen in the symmetric form $\vec{\theta} = ( \theta, 
\theta, \theta)$. 
The dots, squares, diamonds, triangles, the full dots and the full squares correspond to 
$\theta = \{ 0.0, 0.11, 0.19, 0.27, 0.35, 0.44 \}$, respectively.
The dashed vertical line is drawn at $t / a = 10$, where the ground state starts to dominate.}

\label{fig:Meff}

\end{figure}

The 2-point correlation function (\ref{eq:C2pt_stoc}) has been calculated for various values 
of the twisting angle $\vec{\theta}$ chosen always in the symmetric form $\vec{\theta} = 
( \theta, \theta, \theta)$ with $\theta = \{ 0.0, 0.11, 0.19, 0.27, 0.35, 0.44 \}$.
The time behavior of the effective mass (or logarithmic slope) $a M_{eff}(t)$, defined as
 \be
    a M_{eff}(t) \equiv \mbox{log} \left[ \frac{C^\pi(t, 2\pi \vec{\theta} / L)}{C^\pi(t + a, 
    2\pi \vec{\theta} / L)} \right] ~ , ~
    \label{eq:M_eff}
 \ee
is shown in Fig.~\ref{fig:Meff} for two (representative) values of $M_\pi$ at $\beta = 3.9$ 
and $V \cdot T = 24^3 \cdot 48 ~ a^4$.

It can be seen that the statistical precision is remarkably high and it allows to extract 
quite precisely the energy $E_\pi(\vec{p})$ [see Eq.~(\ref{eq:c2ptexp})] corresponding 
to the pion ground state, which starts to dominate from $t / a = 10$.

\begin{figure}[!hbt]

\centerline{\includegraphics[scale=0.80]{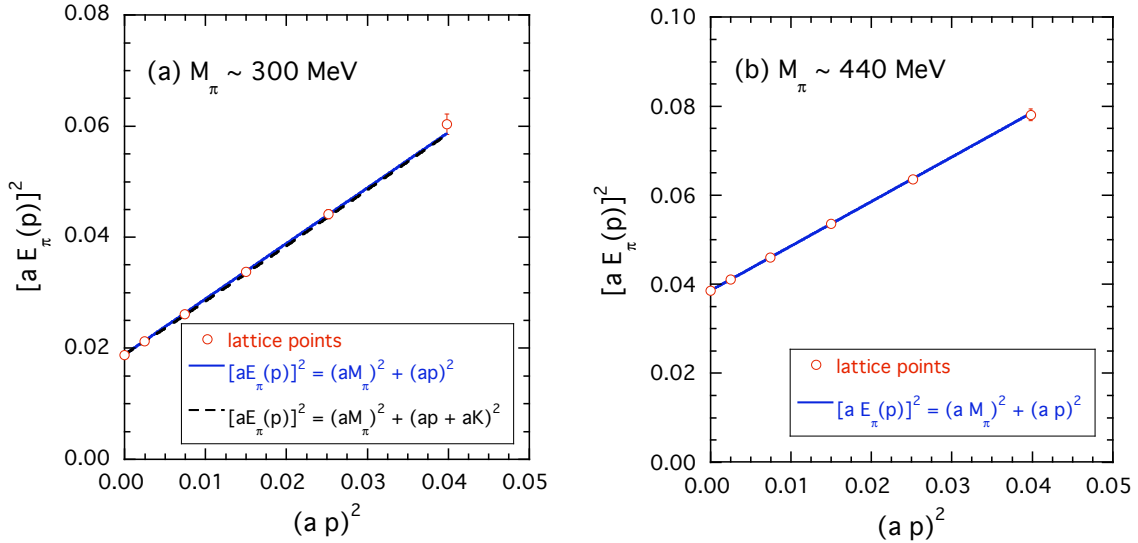}}

\caption{\it Squared pion energy $E_\pi^2(\vec p)$ in lattice units, obtained from the time 
plateaux of the effective mass shown in Fig.~\ref{fig:Meff} (by choosing the time interval 
$10 \leq t /a \leq 21$), versus the squared pion momentum $p^2 \equiv 3 (2 \pi \theta / L)^2$ 
in lattice units, for $M_\pi \simeq 300~\mev$ (a) and $M_\pi \simeq 440~\mev$ (b) at $\beta 
= 3.9$ and $V \cdot T = 24^3 \cdot 48 ~ a^4$. 
The solid line is the continuum-like dispersion relation $E_\pi^2(\vec p) = M_\pi^2(L) + |\vec p|^2$, 
while the dashed line in (a), which can be hardly distinguished from the solid one, represents 
the modified dispersion relation (\ref{eq:p_renorm}) predicted by partially twisted and 
partially quenched ChPT at NLO elaborated in Ref.~\cite{JT07}.}

\label{fig:Epion}

\end{figure}

The values obtained for the pion energy $E_\pi(\vec{p})$ are shown in Fig.~\ref{fig:Epion} 
as a function of the pion momentum given by $\vec p \equiv 2\pi \vec{\theta} / L$, always 
at $\beta = 3.9$ and $V \cdot T = 24^3 \cdot 48 ~ a^4$.
The lattice points appear to be in remarkable agreement with the continuum-like dispersion 
relation $E_\pi(\vec p) = \sqrt{ M_\pi^2(L) + |\vec p|^2}$, where $M_\pi(L)$ is the 
charged pion mass at finite volume.
We have also checked that, assuming the continuum dispersion relation for the energy, all the 
2-point correlation functions of moving pions can be simultaneously fitted very well together 
with the one at rest using a single momentum-independent matrix element $Z_\pi$ [see 
Eq.~(\ref{eq:c2ptexp})].
These findings clearly indicate that, at least for $t /a \geq 10$, where the ground state 
dominates, and for the pion momenta considered in this study, the discretization effects 
on the 2-point correlation functions $C^\pi(t, 2\pi \vec{\theta} / L)$ are almost the 
same as those affecting the correlator at rest $C^\pi(t, \vec{0})$, which were 
investigated in Ref.~\cite{ETMC_tm} and found to be small.

The use of twisted BC's is expected to produce finite volume corrections to the 
continuum-like dispersion relation. 
Such corrections have been investigated in Ref.~\cite{JT07} using partially 
quenched ChPT at NLO.
In the case of charged pion and adopting twisted BC's for one flavor only, the 
pion momentum $\vec p$ acquires an additive correction term $\vec K$, namely
 \be
    E_\pi^2(\vec p) = M_\pi^2(L) + (\vec p + \vec K)^2
    \label{eq:p_renorm}
 \ee
where the components of the vector $\vec K$ are given by ($i \neq j \neq k$)
 \be
    K_i = - \frac{1}{\sqrt{\pi} f_\pi^2 L^3} \int_0^\infty d\tau ~ \frac{1}{\sqrt{\tau} } ~ 
    e^{-\tau (\frac{M_\pi L}{2 \pi})^2} ~ \overline{\Theta}(\tau, \theta_i) ~ 
    \Theta(\tau, \theta_j) ~ \Theta(\tau, \theta_k)
    \label{eq:Ki}
 \ee
with $\Theta(\tau, \theta)$ and $\overline{\Theta}(\tau, \theta)$ being the elliptic 
Jacobi function and its derivative.
Explicitly one has $\Theta(\tau, \theta) \equiv \sum_{n = - \infty}^\infty e^{-\tau 
(n + \theta)^2}$ and $\overline{\Theta}(\tau, \theta) = \sum_{n = - \infty}^\infty 
(n + \theta) e^{-\tau (n + \theta)^2}$.

\begin{figure}[!hbt]

\parbox{9.5cm}{\centerline{\includegraphics[scale=0.55]{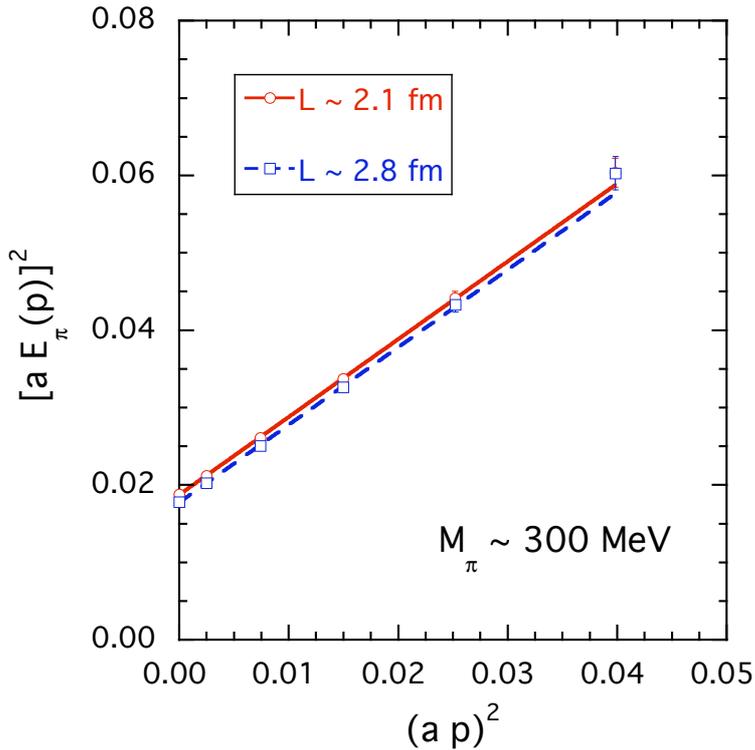}}} \ $~$ \
\parbox{4.5cm}{\vspace{-0.5cm} \caption{\it Squared pion energy $E_\pi^2(\vec p)$ in 
lattice units at $M_\pi \simeq 300~\mev$ and $\beta = 3.9$ for the two runs $R_{2a}$ 
and $R_{2b}$, performed at the volumes $V \cdot T = 24^3 \cdot 48 ~ a^4$ (dots) and 
$V \cdot T = 32^3 \cdot 64 ~ a^4$ (squares). 
The values of the twisting angle $\theta$ are chosen in such a way that $\theta / L$ has 
the same values in the two runs.
The solid and dashed lines represent the continuum-like dispersion relation $E_\pi^2(\vec p) 
= M_\pi^2(L) + |\vec p|^2$.
\label{fig:Epion_32}}}

\end{figure}

We have evaluated Eq.~(\ref{eq:Ki}) for the run $R_{2b}$, which has the smallest 
value of $M_\pi L$ (see Table \ref{tab:setup}).
The results are reported in Fig.~\ref{fig:Epion}(a) (dashed line) and they clearly 
indicate the smallness of the volume corrections to the pion momentum and therefore 
to the continuum dispersion relation expected at NLO.
Thus finite size effects may be limited mainly to the pion mass and thus expected 
to be small (see Refs.~\cite{ETMC1,ETMC2}). 
This is confirmed by the lattice results shown in Fig.~\ref{fig:Epion_32}, where 
the pion energies obtained in case of the runs $R_{2a}$ and $R_{2b}$, which differs 
only for the lattice size, are compared.

\subsection{Momentum dependence of the pion form factor\label{subsec:pionff}}

\begin{figure}[!hbt]

\centerline{\includegraphics[scale=0.80]{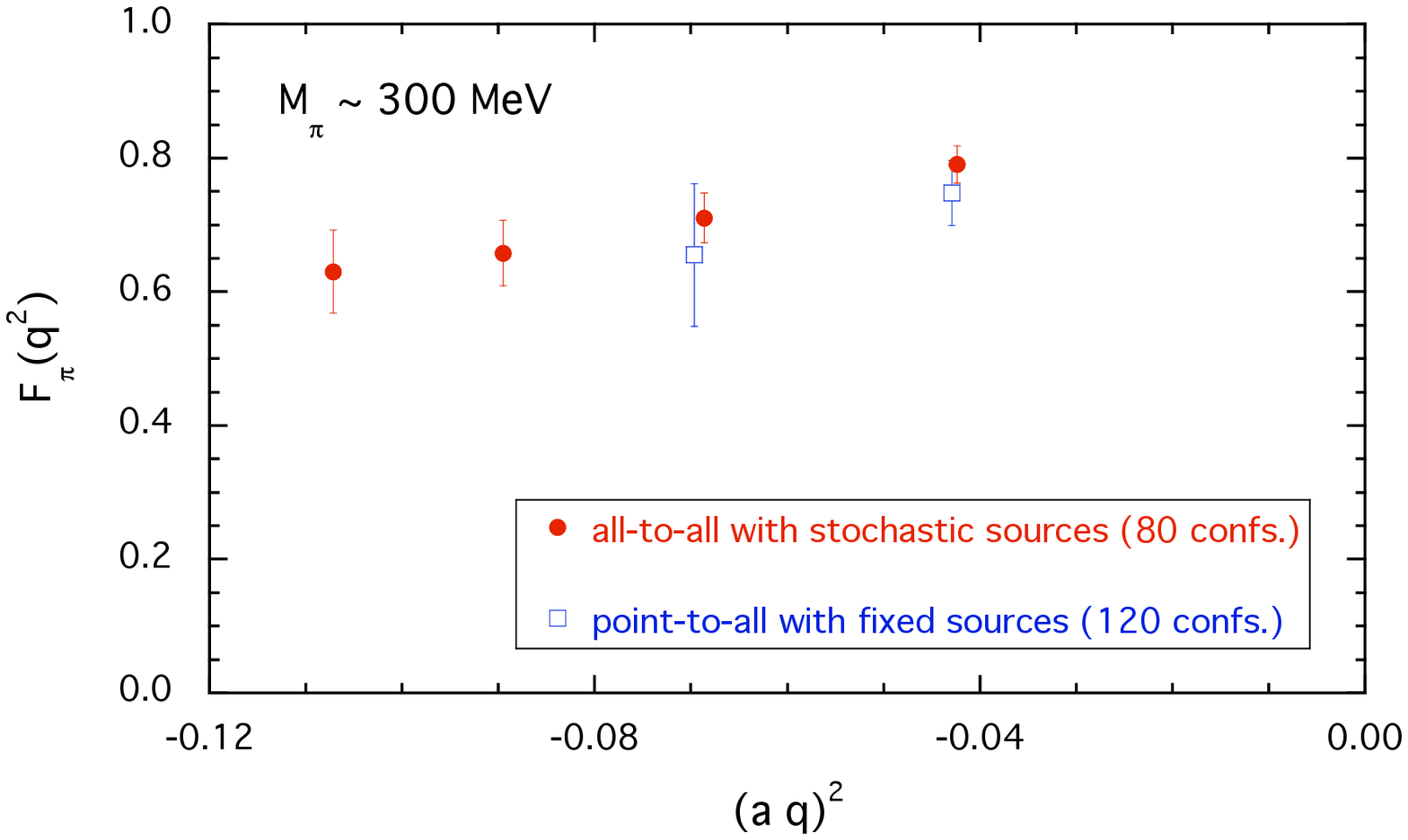}}

\caption{\it Pion form factor $F_\pi(q^2)$ versus $q^2$ in lattice units for a simulated 
pion mass of $\simeq 300~\mev$. 
The full dots are the results obtained using twisted BC's in the Breit frame and the 
one-end-trick procedure for calculating the all-to-all propagators for an ensemble 
of 80 gauge configurations taken from the run $R_{2b}$. 
The open squares correspond to the results of the standard procedure based on point-to-all 
propagators with fixed sources for 120 gauge configurations of the run $R_{2b}$. 
In this case spatially periodic BC's are applied in the frame where the final pion is at 
rest ($\vec p^{\,\prime} = 0$) and the momentum of the initial pion is given by $\vec p = 
2\pi/L $ $\{ (1,0,0), (1,1,0), (1,1,1), (2,0,0) \}$.
At the two smallest values of $q^2$ and for the ensemble of gauge configurations 
considered, only the stochastic procedure provides time plateaux of enough good 
quality to allow the extraction of the pion form factor.}

\label{fig:comparison}

\end{figure}

The advantage of calculating the pion form factor using all-to-all propagators, evaluated 
by the one-end-trick procedure with twisted BC's, with respect to the standard procedure 
based on point-to-all propagators with fixed sources and (spatially) periodic BC's is 
illustrated in Fig.~\ref{fig:comparison}. 
From the run $R_{2b}$ we choose a different number of gauge configurations for the 
stochastic and non-stochastic procedures in order to get the same total computational 
time\footnote{Let us remind that the one-end-trick requires less computational time for 
a single inversion of the Dirac operator (a factor of about $1/3$), but for each quark 
momentum a new inversion is needed by the use of twisted BC's.}.

Despite the more limited ensemble of gauge configurations the stochastic approach provides 
a much better precision at the two lowest values of $q^2$ (a factor between $\sim 2$ and 
$\sim 3$). 
It also allows a very good determination of the form factor at the two highest values of 
$q^2$ considered in this study, where the procedure based on point-to-all propagators 
fails to give reliable signals even in the presence of a larger ensemble of gauge 
configurations.

\begin{figure}[!hbt]

\centerline{\includegraphics[scale=0.80]{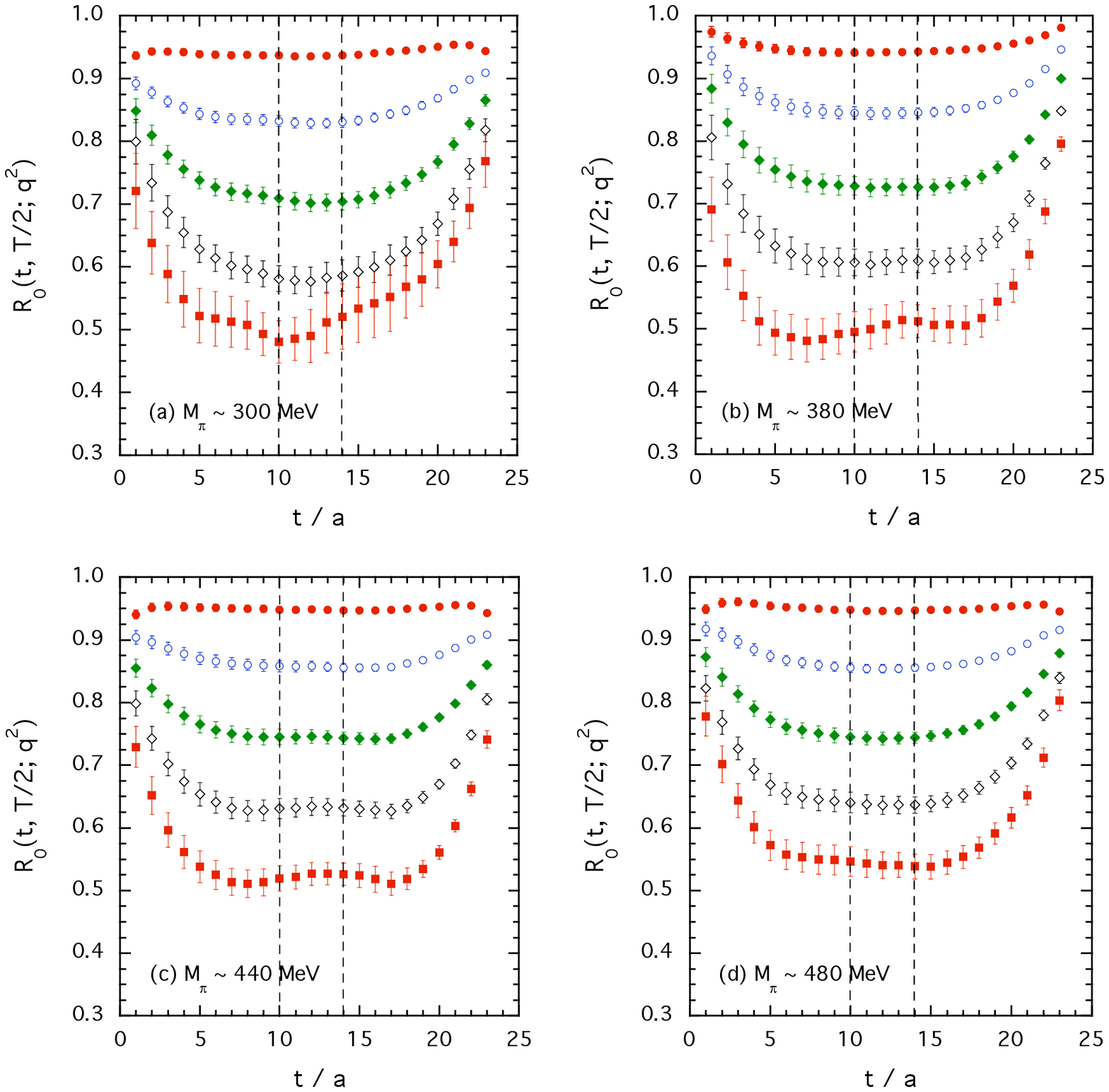}}

\caption{\it Ratio $R_0(t, t'; q^2)$, defined by Eq.~(\ref{eq:ffpion}), at $t' = T / 2$ versus
the time distance $t$ in lattice units, for $M_\pi \simeq 300~\mev$ (a), $M_\pi \simeq 380~\mev$ 
(b), $M_\pi \simeq 440~\mev$ (c), $M_\pi \simeq 480~\mev$ (d) at $\beta = 3.9$ and $V \cdot T = 
24^3 \cdot 48 ~ a^4$.
The full dots, open squares, full diamonds, open diamonds and full squares correspond to 
$a^2 q^2 = -0.01, -0.03, -0.06, -0.10$ and $-0.16$, respectively.
The dashed vertical lines identify the region $10 \leq t / a \leq 14$, where both the initial 
and the final pion ground states are isolated, so that the pion form factor $F_\pi(q^2)$ can 
be extracted.}

\label{fig:plateaux}

\end{figure}

\begin{figure}[!hbt]

\centerline{\includegraphics[scale=0.80]{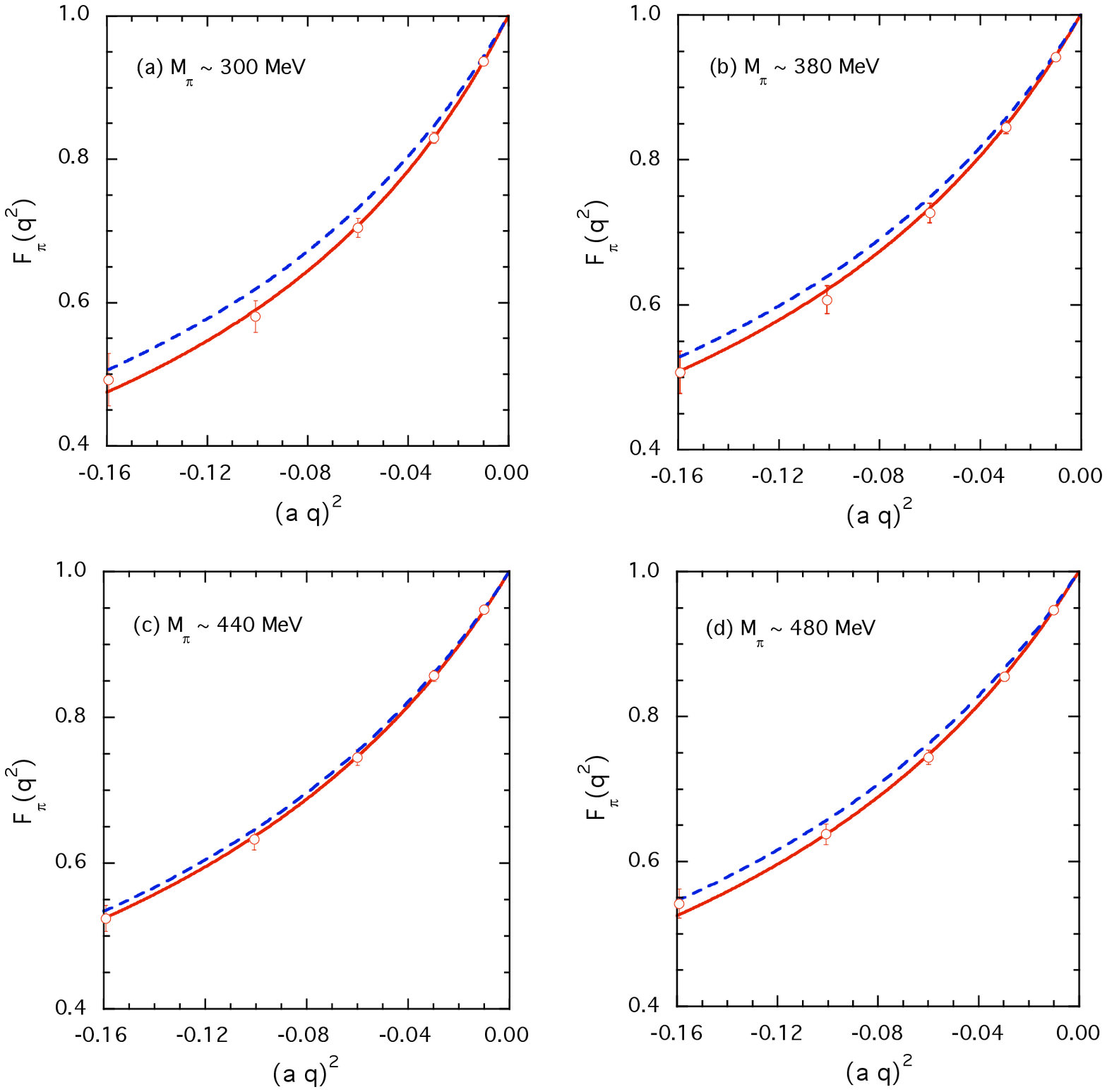}}

\caption{\it Pion form factor $F_\pi(q^2)$, extracted from the plateau region $10 \leq t / a 
\leq 14$ of the ratio $R_0(t, T / 2; q^2)$ (see Fig.~\ref{fig:plateaux}), versus the squared 
4-momentum transfer $q^2$ in lattice units, for $M_\pi \simeq 300~\mev$ (a), $M_\pi \simeq 
380~\mev$ (b), $M_\pi \simeq 440~\mev$ (c), $M_\pi \simeq 480~\mev$ (d) at $\beta = 3.9$ 
and $V \cdot T = 24^3 \cdot 48 ~ a^4$.
The solid line is the pole behavior (\ref{eq:pole}) with the parameter $M_{pole}$ fitted 
to the lattice points, while the dashed line is the VMD prediction with $M_{pole}$ fixed 
at the value of the lightest vector-meson mass taken from Ref.~\cite{vector}.}

\label{fig:Fpion}

\end{figure}

As discussed in the previous Section, the pion form factor $F_\pi(q^2)$ can be determined 
from the plateau of the ratio $R_0(t, t^\prime; q^2)$, defined by Eq.~(\ref{eq:ffpion}), 
at large time distances.
The quality of the time plateaux is illustrated in Fig.~\ref{fig:plateaux}, while the momentum 
dependence of the extracted pion form factor $F_\pi(q^2)$ is shown in Fig.~\ref{fig:Fpion} 
for various values of $M_\pi$ at $\beta = 3.9$ and $V \cdot T = 24^3 \cdot 48 ~ a^4$.
We have checked that different choices of the time interval for the plateau region lead 
to values of $F_\pi(q^2)$ which are largely consistent within the statistical precision.
The values of the pion form factor obtained for all the simulations of Table \ref{tab:setup} 
are reported in the Appendix.

In the whole range of values of both $q^2$ and the quark mass, considered in this 
work, our lattice data can be fitted very nicely using a simple pole ansatz
 \be
    F_\pi^{(pole)}(q^2) = \frac{1}{1 - q^2 / M_{pole}^2} ~ ,
    \label{eq:pole}
 \ee
as it is shown in Fig.~\ref{fig:Fpion}. 
For comparison we also show the predictions of the Vector Meson Dominance (VMD) 
model, in which the parameter $M_{pole}$ is fixed at the value of the lightest 
vector-meson mass ($M_{VMD}$) taken from Ref.~\cite{vector}.
The values obtained for $M_{pole}$ by fitting our lattice points at $\beta = 3.9$ 
are given in Table \ref{tab:Mpole}. 

\begin{table}[!htb]

\begin{center}
\begin{tabular}{||c||c|c|c||}
\hline
 $Run$ & $M_\pi(\mev)$ & $V \cdot T ~ / ~ a^4$ & $a M_{pole}$ \\ \hline
 $R_1$    & $\simeq 260$ & $32^3 \cdot 64$ & $0.359 \pm 0.016$ \\ \hline
 $R_{2a}$ & $\simeq 300$ & $32^3 \cdot 64$ & $0.363 \pm 0.011$ \\ \hline
 $R_{2b}$ & $\simeq 300$ & $24^3 \cdot 48$ & $0.379 \pm 0.010$ \\ \hline
 $R_3$    & $\simeq 380$ & $24^3 \cdot 48$ & $0.399 \pm 0.014$ \\ \hline
 $R_4$    & $\simeq 440$ & $24^3 \cdot 48$ & $0.420 \pm 0.012$ \\ \hline
 $R_{5a}$ & $\simeq 480$ & $24^3 \cdot 48$ & $0.419 \pm 0.011$ \\ \hline
 $R_6$    & $\simeq 580$ & $24^3 \cdot 48$ & $0.440 \pm 0.005$ \\ \hline \hline

\end{tabular}

\caption{\it Values of the fit parameter $M_{pole}$ appearing in Eq.~(\ref{eq:pole}) 
obtained at $\beta = 3.9$ in lattice units.
\label{tab:Mpole}}

\end{center}

\end{table}

From Fig.~\ref{fig:Fpion} it can be seen that the VMD prediction, which considers 
the contribution of the lowest vector resonance only, is not exactly fulfilled, 
since $M_{pole}$ turns out to be systematically lower than $M_{VMD}$.
However such a comparison might be plagued by systematic uncertainties 
affecting the lattice determination of the lightest vector-meson mass 
particularly at the lowest values of the pion mass (see Ref.~\cite{vector}).
Nevertheless, a simple extrapolation of $M_{pole}$ to the physical point, based 
on a polynomial fit in terms of quark masses (see later subsection \ref{subsec:pol}), 
yields the value $M_{pole}^{phys} = 0.713 \pm 0.044~\gev$, which is lower than 
the VMD prediction $M_{VMD}^{phys} = M_\rho = 0.776~\gev$ from PDG \cite{PDG}.

Note also that, defining the squared {\em pole} radius in terms of 
Eq.~(\ref{eq:pole}) as
 \be
    r_{pole}^2 \equiv 6 / M_{pole}^2 = 6 \left[ \frac{dF_\pi^{pole}(q^2)}{dq^2} 
    \right]_{q^2 = 0} ~ ,
    \label{eq:r2pole}
 \ee 
the VMD model leads at the physical point to $r_{pole}^2 = 6 / M_\rho^2 \simeq 
0.388~\mbox{fm}^2$, which underestimates by $\simeq 15 \%$ the (quite precise) 
experimental value of the squared pion charge radius, $\langle r^2 
\rangle^{exp.} = 0.452 \pm 0.011~\mbox{fm}^2$~\cite{PDG}.
On the contrary the value $M_{pole}^{phys} = 0.713 \pm 0.044~\gev$ implies 
$r_{pole}^2 = 0.459 \pm 0.057~\mbox{fm}^2$ in nice agreement with the 
experimental charge radius.

\subsection{Finite size effects\label{subsec:volume}}

We have investigated the effects of the finite spatial extension L of our lattice 
boxes by comparing the results of runs $R_{2a}$ and $R_{2b}$.
In our simulations the latter has the smallest value of the quantity $M_\pi L$, 
which governs finite size effects (FSE) in the p-regime.
The physical extension of the two boxes is $L \simeq 2.8$ fm and $L \simeq 2.1$ fm, 
respectively.
The values of the angle $\theta$ are chosen differently at the two volumes in order to 
keep the values of $q^2$ fixed.

\begin{figure}[!hbt]

\parbox{9.5cm}{\centerline{\includegraphics[scale=0.55]{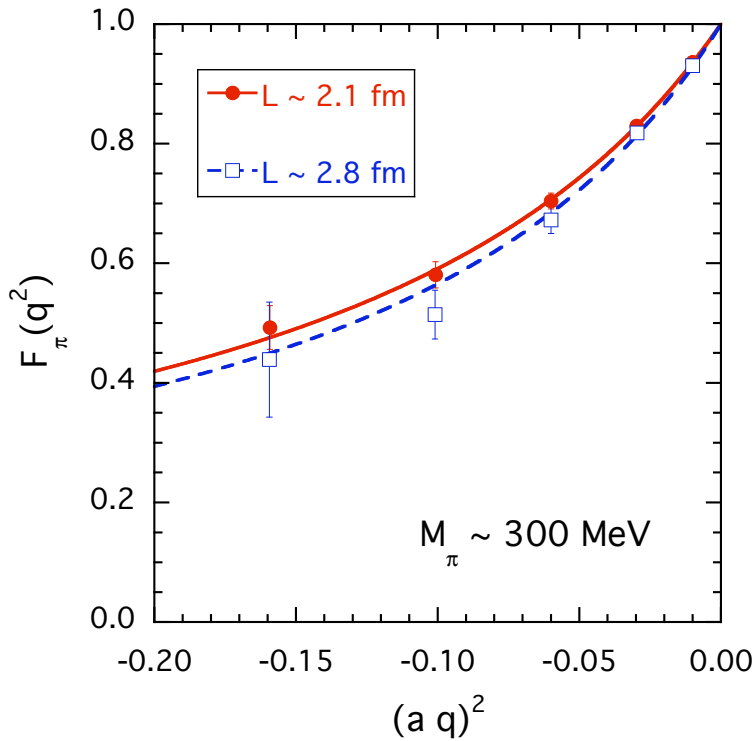}}} \ $~$ \
\parbox{4.5cm}{\vspace{-0.5cm} \caption{\it Pion form factor $F_\pi(q^2)$ obtained for 
the runs $R_{2a}$ (open squares) and $R_{2b}$ (full dots), which correspond to different 
lattice boxes of size $L \simeq 2.8$ fm and $L \simeq 2.1$ fm, respectively.
The solid and dashed lines are the results of the pole fit (\ref{eq:pole}).
\label{fig:volume}}}

\end{figure}

The results for the pion form factor are shown in Fig.~\ref{fig:volume}, while a direct 
comparison of the results for the pion mass and decay constant as well as for the squared 
pole radius, $r_{pole}^2$, is illustrated in Table \ref{tab:volume}.

\begin{table}[!htb]

\begin{center}
\begin{tabular}{||c|c||c|c|c||}
\hline
 $Run$ & $L~(fm)$ & $a M_\pi$ & $a f_\pi$ & $r_{pole}^2 / a^2$ \\ \hline
 $R_{2a}$ & $\simeq 2.8$ & $0.13377~(24)$ & $0.06625~(16)$ & $45.5 \pm 2.8$ \\ \hline
 $R_{2b}$ & $\simeq 2.1$ & $0.13623~(65)$ & $0.06459~(37)$ & $41.7 \pm 2.3$ \\ \hline \hline

\end{tabular}

\caption{\it Values of the pion mass and decay constant from the high-statistics 
work of Ref.~\cite{ETMC2} and of the squared pole radius [see Eq.~(\ref{eq:r2pole})] 
in lattice units for the runs $R_{2a}$ and $R_{2b}$.
\label{tab:volume}}

\end{center}

\end{table}

It can clearly be seen that FSE effects are larger on the pion form factor (or, equivalently, 
on the pole radius) with respect to the case of the pion mass and decay constant.
They indeed amount to $\simeq 8 \%$ on $r_{pole}^2$ in contrast to a $\approx 2 \%$ effect 
in the case of $M_\pi$ and $f_\pi$.
However we notice that FSE effects on $r_{pole}^2$ are comparable to our statistical precision 
($\simeq 6 \%$), while they are much larger in the case of $M_\pi$ and $f_\pi$ ($0.2 \div 0.6 
\%$).
Thus it is mandatory to include volume corrections to our results at least on the pion mass 
and decay constant.

On the theoretical side FSE on $M_\pi$ and $f_\pi$ have been studied with ChPT at NLO in 
Ref.~\cite{GL87} and using a resummed asymptotic formula in Ref.~\cite{CDH05}, where both 
leading and subleading exponential terms are taken into account and the chiral expansion 
is applied to the $\pi-\pi$ forward scattering amplitude. 
When the leading chiral representation of the latter is considered, the resummed approach 
coincides with the NLO result of Ref.~\cite{GL87}.
Viceversa at NNLO the resummation technique includes only a part of the two-loop effects 
as well as of higher-loop effects.
Recently the resummed approach has been positively checked against a full NNLO 
calculation of the pion mass in Ref.~\cite{CH06}, showing that the missing 
two-loop contributions are actually negligible.

The volume corrections predicted by the resummed approach have been already considered 
in the analysis of the ETMC results for $M_\pi$ and $f_\pi$ carried out in 
Refs.~\cite{ETMC1,ETMC2,ETMC_tm}.

On the contrary, till now, the theoretical investigation of FSE on the pion form 
factor is limited to the application of ChPT at NLO only.
The case of periodic BC's is considered in Ref.~\cite{FSE_periodic}, while twisted BC's 
are studied in Refs.~\cite{JT07,JT08} adopting two different reference frames, namely 
the rest frame of the final meson \cite{JT07} and the Breit one \cite{JT08}.

The sign of the volume effects on the pion form factor depends crucially on the 
absolute value and the spatial direction of the twisting vector $\vec{\theta}$.
The sign of FSE on the charge radius turns out to be opposite between the cases 
of periodic (Ref.~\cite{FSE_periodic}) and twisted (Refs.~\cite{JT07,JT08}) BC's.
When periodic BC's are used the extraction of the charge radius requires the use of the 
smallest available momentum, which is equal to $2 \pi / L$. 
Such a restriction is absent with twisted BC's and therefore volume effects are 
different.

Moreover the volume corrections depend on the reference frame: in the rest frame, 
besides the usual term related to the difference between the infinite volume loop 
integral and the sum over quantized momenta, there are two further contributions 
\cite{JT07} arising from isospin and hypercubic invariance breakings generated 
by flavor-dependent twisted BC's. 
Such two terms are vanishing in the Breit frame as shown in Ref.~\cite{JT08}.

Only the results of Ref.~\cite{JT08}, in which both the twisted BC's and the Breit 
reference frame are considered, can be directly applied to our data.
Thus one gets
 \be
    F_\pi(q^2; L) - F_\pi(q^2; \infty) & = & \frac{1}{f_\pi^2} \left\{ \int_0^1 dx ~ 
    I_{1/2}\left[ (1 - 2x) \frac{2\pi \vec{\theta}}{L}; M_\pi^2 - x (1 - x) q^2 \right] 
    - \right. \nn \\
    && \left. I_{1/2}\left( \frac{2\pi \vec{\theta}}{L}; M_\pi^2 \right) \right\}
    \label{eq:FVL}
 \ee
with $q^2 = - 4 (2 \pi \vec{\theta} / L)^2$ and
 \be
    I_{1/2}\left( \frac{2\pi \vec{\theta}}{L}; M_\pi^2 \right) = \frac{1}{2 \pi^{3/2} L^2} 
    \int_0^\infty d\tau ~ \frac{1}{\sqrt{\tau} } ~ e^{-\tau (\frac{M_\pi L}{2 \pi})^2} ~ 
    \left[ \prod_{i = 1}^3 \Theta(\tau, \theta_i) - \left( \frac{\pi}{\tau} \right)^{3/2} 
    \right] ~ 
    \label{eq:I1/2}
 \ee
where $\Theta(\tau, \theta)$ is defined after Eq.~(\ref{eq:Ki}).
The NLO volume corrections on the pion form factor expected for our run $R_{2b}$ do not 
exceed half of the statistical error, and they are even smaller in the case of the run 
$R_{2a}$ at the largest volume.
The FSE's predicted by Eqs.~(\ref{eq:FVL}-\ref{eq:I1/2}) are quite small and 
have the same sign for all the choices of the twisting angle $\vec{\theta}$ made 
in this work. 
The NLO corrections go to the right direction decreasing slightly the differences 
between the pion form factor obtained at the two box sizes.

As for the squared pole radius, the shift with the lattice volume reported in Table 
\ref{tab:volume} has the same sign expected from the volume correction (\ref{eq:FVL}).
However the FSE calculated at NLO for the run $R_{2b}$ corresponds to an increase of 
$\simeq 3 \%$ only, that is almost a factor 3 less than the observed FSE 
($\simeq 8 \%$).
This suggests that higher-order chiral effects might be relevant on the pion form 
factor still for $M_\pi L \simeq 3$, although our statistical precision ($\simeq 6 \%$) 
does not exclude FSE's on $r_{pole}^2$ as small as the ones predicted at NLO by 
Eq.~(\ref{eq:FVL}).

In the case of our runs $R_1$ and $R_{2a}$, which correspond to $M_\pi L \simeq 4$, 
the NLO volume corrections on $r_{pole}^2$ are expected to be $\simeq 1 \%$.
After multiplying such a value by a factor $\approx 3$ in order to take into account 
conservatively higher-loop effects, the expected FSE remains well below the 
statistical precision.

Therefore in this work we decide to analyze our form factor data using only simulations 
with $M_\pi L \gtrsim 4$, which means in practice that the run $R_{2b}$ is excluded 
from our analyses of the pion form factor. 
On the contrary in case of the pion mass and decay constant we keep the run 
$R_{2b}$ in the set of fitted data, but the FSE's, calculated through the 
resummed asymptotic formula of Ref.~\cite{CDH05} at the NNLO accuracy for 
the $\pi-\pi$ forward scattering amplitude, will be taken into account 
(see Sections \ref{sec:radius} and \ref{sec:ffpion}).

\subsection{Discretization effects\label{subsec:a2}}

\begin{figure}[!hbt]

\centerline{\includegraphics[scale=0.80]{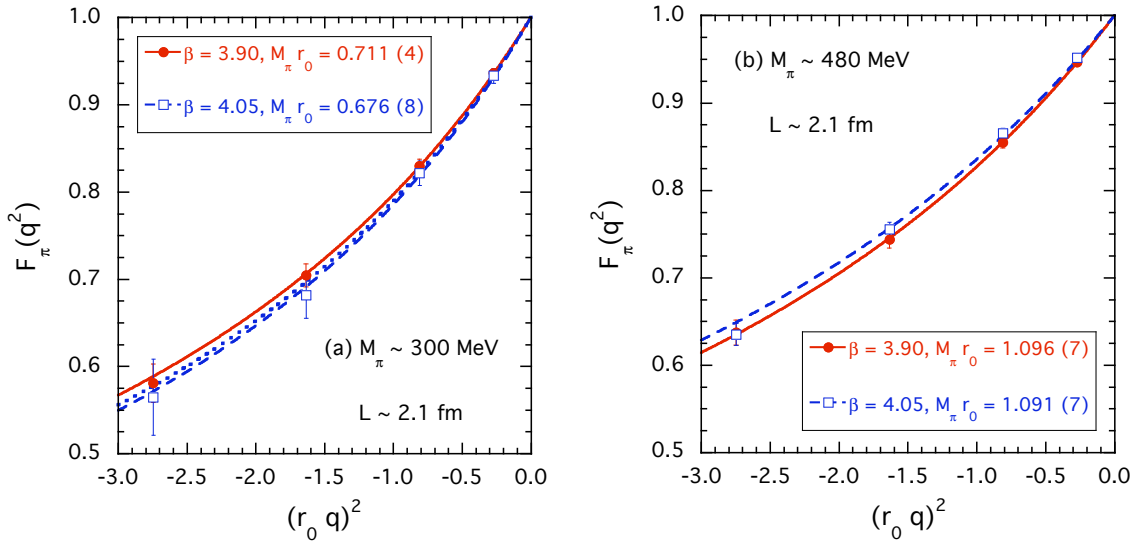}}

\caption{\it Results of the pion form factor $F_\pi(q^2)$ versus $q^2$ in units of the Sommer 
parameter $r_0$, obtained for the runs $R_{2b}$ (full dots) and $R_{2c}$ (open squares) at 
$M_\pi \simeq 300~\mev$ in (a), and for the runs $R_{5a}$ (full dots) and $R_{5b}$ (open 
squares) at $M_\pi \simeq 480~\mev$in (b).
The physical lattice size is the same in both runs ($L \simeq 2.1$ fm).
The solid and dashed lines are the results of the pole fit (\ref{eq:pole}), while the dotted 
line in (a) corrects the dashed one for the pion mass difference (see text).}

\label{fig:scaling}

\end{figure}

We have investigated the impact of lattice artifacts on the pion form factor by 
considering the runs $R_{2c}$ and $R_{5b}$ at the finer spacing $a \simeq 0.07$ 
fm (see Table~\ref{tab:setup}). 
These runs correspond to pion masses equal to $M_\pi \simeq 300~\mev$ and $M_\pi 
\simeq 480~\mev$, respectively, which are very similar to those of the runs $R_{2b}$ 
and $R_{5a}$ at $a \simeq 0.09$ fm, while the physical lattice size is almost kept 
fixed ($L \simeq 2.1$ fm).
Our results are shown in Fig.~\ref{fig:scaling} in terms of the Sommer parameter 
$r_0$ instead of the lattice spacing $a$. 
The ratio $r_0 / a$ has been determined in the chiral limit at the two lattice 
spacings in Ref.~\cite{ETMC_tm}, obtaining $r_0 / a = 5.22 \pm 0.02$ at $\beta = 
3.9$ and $r_0 / a = 6.61 \pm 0.03$ at $\beta = 4.05$.

It can clearly be seen that the size of discretization effects is comparable to the 
statistical error at both pion masses.
At the lowest pion mass there is a slight mismatch between the values of $M_\pi r_0$ 
corresponding to the runs $R_{2b}$ and $R_{2c}$.
Using the ChPT formulae at NNLO evaluated in Ref.~\cite{BCT}, which will be used in 
the next Sections, and adopting for the relevant LEC's the values given in Ref.~\cite{CGL}
we have estimated the correction due to the pion mass difference and applied it to the 
results of the run $R_{2c}$ (see dotted line in Fig.~\ref{fig:scaling}).
The correction is small, but reduces the impact of discretization effects, which now in 
terms of $r_{pole}^2$ do not exceed $\simeq 5 \%$ at both pion masses.

A more complete investigation of the scaling properties of the pion form factor, 
which requires the study of its mass dependence at two additional values of the 
lattice spacing, is needed and it is in progress.

In the next Sections continuum ChPT will be applied to the chiral extrapolation of the 
results of the runs $R_1$, $R_{2a}$, $R_3$, $R_4$, $R_{5a}$ and $R_6$, which correspond 
to a single lattice spacing ($a \simeq 0.09$ fm) and to a pion mass range between $\simeq 
260~\mev$ and $\simeq 580~\mev$ with $M_\pi L \gtrsim 4$.
The impact of lattice artifacts will be estimated by substituting the results of the runs 
$R_{2a}$ and $R_{5a}$ with those of the runs $R_{2c}$ and $R_{5b}$, respectively.

\section{Slope and curvature of the pion form factor\label{sec:radius}}

The slope $s$ and the curvature $c$ of the pion form factor are defined from the 
expansion in $q^2$
 \be
    F_\pi(q^2) = 1 + s ~ q^2 + c ~ q^4 + {\cal{O}}(q^6) ~ .
    \label{eq:expansion}
 \ee
In terms of the pole ansatz (\ref{eq:pole}) the slope is given by 
 \be
    s_{pole} = \frac{1}{M_{pole}^2} = \frac{r_{pole}^2}{6} ~ ,
    \label{eq:spole}
 \ee
while the curvature is constrained to be 
 \be
    c_{pole} = s_{pole}^2 = \frac{1}{M_{pole}^4} = \left( \frac{r_{pole}^2}{6} 
    \right)^2 ~ .
    \label{eq:cpole}
  \ee
We have therefore compared the slope and curvature obtained from the pole ansatz 
(\ref{eq:pole}) with those of a simple cubic fit in $q^2$
 \be
    F_\pi^{(cub)}(q^2) = 1 + s_{cub} ~ q^2 + c_{cub} ~ q^4 + d_{cub} ~ q^6 ~ .
    \label{eq:cubic}
 \ee

\begin{figure}[!hbt]

\parbox{9.5cm}{\centerline{\includegraphics[scale=0.55]{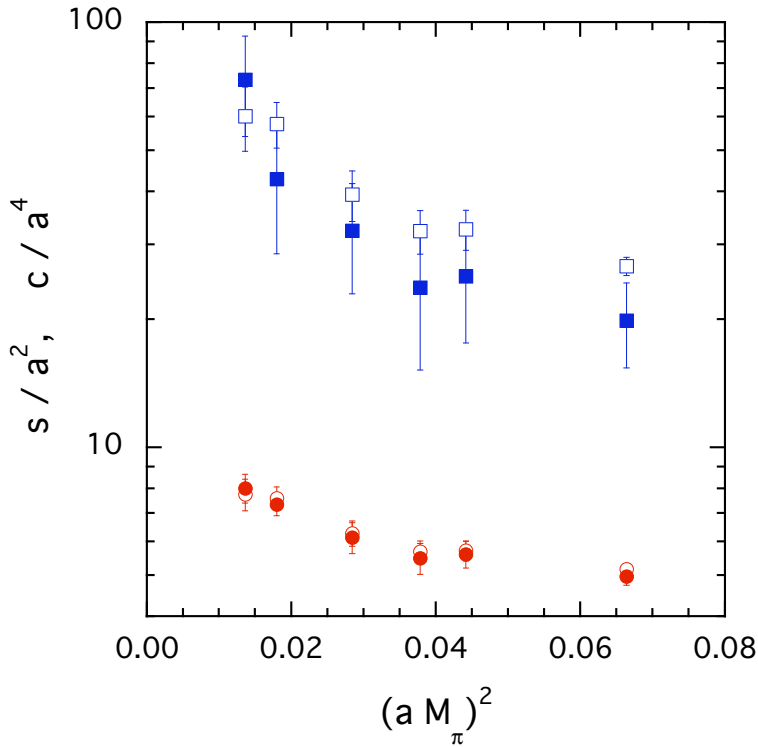}}} \ $~$ \
\parbox{4.5cm}{\vspace{-0.5cm} \caption{\it The slope $s$ (dots) and the curvature $c$ 
(squares) of the pion form factor [see Eq.~(\ref{eq:expansion})] versus the squared 
pion mass in lattice units, for the runs $R_1$, $R_{2a}$, $R_3$, $R_4$, $R_{5a}$ and 
$R_6$. 
Open dots and squares correspond to the results of the pole fit given by Eqs.~(\ref{eq:spole}) 
and (\ref{eq:cpole}), respectively.
Full markers are the results obtained with the cubic fit (\ref{eq:cubic}).
\label{fig:slopes}}}

\end{figure}

The results obtained by including in the fitting procedure the form factor corresponding 
to the four highest, negative values of $q^2$ are shown in Fig.~\ref{fig:slopes} (see 
also Table \ref{tab:radiusV_ETMC}) \footnote{The lattice points at the lowest, 
negative value of $q^2$ are the noisiest data (see Fig.~\ref{fig:Fpion} and 
also Fig.~\ref{fig:plateaux} for the corresponding time plateaux).
The inclusion of these data in the fitting procedure does not change significantly 
the determination of the various parameters appearing in Eqs.~(\ref{eq:spole}) and 
(\ref{eq:cubic}).}.
It can be seen that the two determinations of the slope are in very good agreement and 
the results for the curvature are consistent within the statistical errors, which 
turn out to be lower in the case of the pole fit.

\begin{figure}[!hbt]

\centerline{\includegraphics[scale=0.80]{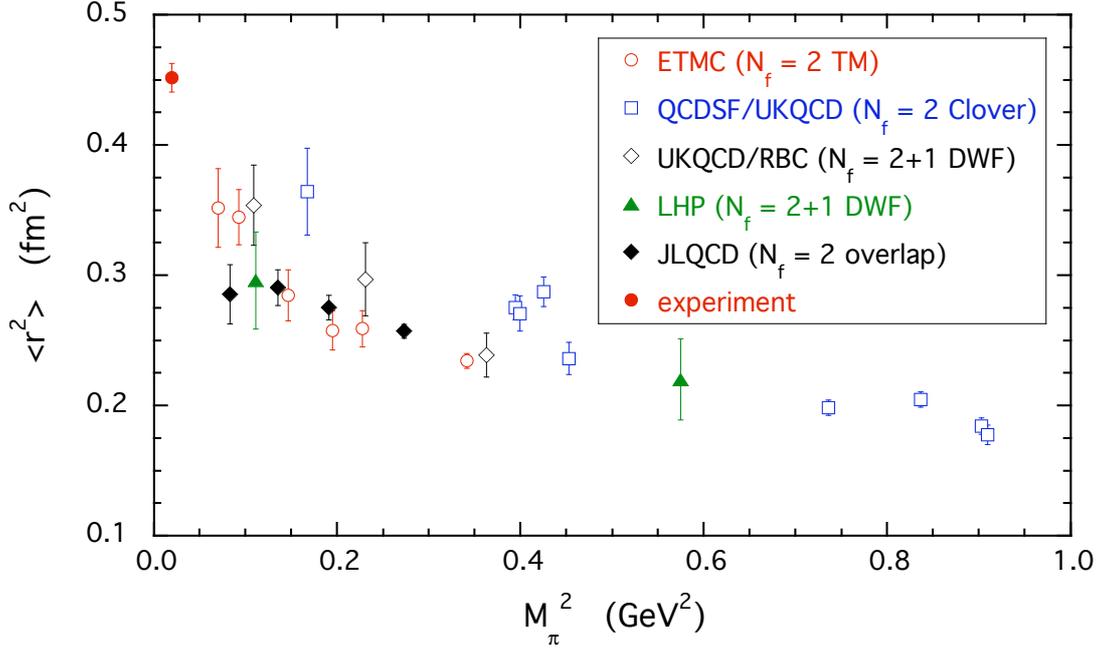}}

\caption{\it The squared pion charge radius versus the squared pion mass. 
Open dots: this work (see third column of Table \ref{tab:radiusV_ETMC}).
Open squares: results from Ref.~\cite{Clover2} corrected as explained in the text.
Open diamonds, full triangles, full diamonds correspond to Refs.~\cite{DWF,LHP,overlap}, 
respectively.
The full dot represents the experimental value of the squared pion charge radius 
$\langle r^2 \rangle^{exp.} = 0.452 \pm 0.011~\mbox{fm}^2$ from PDG \cite{PDG}.}

\label{fig:radiusV}

\end{figure}

In what follows we take as our best estimates the values of the slope and the curvature 
coming from the pole ansatz.
The former ones, expressed in physical units using the value $a = 0.087$ fm from 
Ref.~\cite{ETMC1}, are collected in the third column of Table \ref{tab:radiusV_ETMC} 
and shown in Fig.~\ref{fig:radiusV}, where they are compared with the available 
results of other lattice collaborations that employ O(a)-improved lattice 
actions and unquenched gauge configurations.
It can be seen that all the determinations of the pion charge radius exhibit a 
quite similar mass dependence, indicating that lattice artifacts are presumably 
under control.
The results labeled as ``QCDSF/UKQCD" in Fig.~\ref{fig:radiusV} do 
not correspond to the original ones reported in Ref.~\cite{Clover2}.
There the lattice spacing, instead of the Sommer parameter $r_0$, was assumed to 
depend on the sea quark mass and such a procedure reintroduces non-negligible 
lattice artifacts.
The ``QCDSF/UKQCD" results shown in Fig.~\ref{fig:radiusV} are obtained after 
properly extrapolating the ratio $r_0 / a$ to the chiral limit\footnote{We thank 
J.~Zanotti for providing us the extrapolated values of $r_0 / a$ at the chiral 
point.}.

\begin{table}[!htb]

\begin{center}
\begin{tabular}{||c||c||c|c||c|c||}
\hline
 $Run$ & $M_\pi$  & $r_{\rm pole}^2 \equiv 6 s_{\rm pole}$ & $c_{\rm pole}$ 
                  & $r_{\rm cub}^2 \equiv 6 s_{\rm cub}$   & $c_{\rm cub}$ \\
       & $(\mev)$ & $(\mbox{fm}^2)$                        & $(10^{-3}~\mbox{fm}^4)$ 
                  & $(\mbox{fm}^2)$                        & $(10^{-3}~\mbox{fm}^4)$ \\\hline
 $R_1$    & $265$ & $0.352 \pm 0.030$ & $3.44 \pm 0.59$ & $0.364 \pm 0.028$ & $4.20 \pm 1.12$ \\ \hline
 $R_{2a}$ & $304$ & $0.345 \pm 0.021$ & $3.30 \pm 0.40$ & $0.333 \pm 0.020$ & $2.45 \pm 0.82$ \\ \hline
 $R_3$    & $383$ & $0.285 \pm 0.019$ & $2.25 \pm 0.31$ & $0.278 \pm 0.023$ & $1.85 \pm 0.54$ \\ \hline
 $R_4$    & $441$ & $0.258 \pm 0.015$ & $1.85 \pm 0.22$ & $0.249 \pm 0.021$ & $1.36 \pm 0.49$ \\ \hline
 $R_{5a}$ & $477$ & $0.259 \pm 0.014$ & $1.87 \pm 0.20$ & $0.254 \pm 0.018$ & $1.45 \pm 0.44$ \\ \hline
 $R_6$    & $584$ & $0.234 \pm 0.006$ & $1.53 \pm 0.07$ & $0.225 \pm 0.010$ & $1.14 \pm 0.26$ \\ \hline \hline

\end{tabular}

\caption{\it Values of the pion mass, charge radius and curvature, determined from 
the pole (\ref{eq:spole}-\ref{eq:cpole}) and cubic (\ref{eq:cubic}) fits, for the 
various ETMC runs.
Physical units are used taking for the lattice spacing the value $a = 0.087$ fm 
from Ref.~\cite{ETMC1}.
The uncertainties are statistical (jackknife) errors.
\label{tab:radiusV_ETMC}}

\end{center}

\end{table}

\subsection{ChPT formulae at NNLO\label{subsec:ChPT_radius}}

In Ref.~\cite{BCT} the pion form factor, as well as the pion mass and decay 
constant, have been calculated in continuum SU(2) ChPT at NNLO in infinite 
volume using a modified minimal subtraction ($\overline{MS}$) scheme to 
regulate the infinities.
Using the quark mass $\hat{m}$ as the expansion parameter, one has
 \be
    \label{eq:MPi2}
    M_\pi^2 & = & 2B \hat{m} + \left[ M_\pi^2 \right]_{\mbox{NLO}} +
    \left[ M_\pi^2 \right]_{\mbox{NNLO}} + {\cal{O}}(\hat{m}^4) ~ , \\
    \label{eq:MPi2_1loop}
    \left[ M_\pi^2 \right]_{\mbox{NLO}} & = & 2B \hat{m} \cdot 2 x_2 
    \left[ 2\ell_3^r + \frac{1}{2} L(\mu) \right] ~ , \\ 
    \left[ M_\pi^2 \right]_{\mbox{NNLO}} & = & 2B \hat{m} \cdot 4 x_2^2 
    \left\{ \frac{1}{N} \left[ \ell_1^r + 2 \ell_2^r - \frac{13}{3} L(\mu) \right] 
    + \frac{163}{96} \frac{1}{N^2} \right. \nn \\
    & - & \left. \frac{7}{2} k_1 - 2 k_2 + 4 \ell_3^r \left( \ell_4^r - \ell_3^r 
    \right) - \frac{9}{4} k_3 + \frac{1}{4} k_4 \right. \nn \\
    & + & \left. r_M^r + \Delta_M \left( \Delta_M - \Delta_F + \frac{1}{2N} \right)
    \right\} ~ ,
    \label{eq:MPi2_2loops}
 \ee
 \be
    \label{eq:fPi}
    f_\pi & = & F + \left[ f_\pi \right]_{\mbox{NLO}} + 
    \left[ f_\pi \right]_{\mbox{NNLO}} + {\cal{O}}(\hat{m}^3) ~ , \\ 
    \left[ f_\pi \right]_{\mbox{NLO}} & = & 2 F x_2 \left[ \phantom{\frac{1}{1}} 
    \hspace{-0.25cm} \ell_4^r - L(\mu) \right] ~ , 
    \label{eq:fPi_1loop}
 \ee
 \be
    \left[ f_\pi \right]_{\mbox{NNLO}} & = & 4 F x_2^2 \left\{ \frac{1}{N} 
    \left[ -\frac{1}{2} \ell_1^r - \ell_2^r + \frac{29}{12} L(\mu) \right] - 
    \frac{13}{192} \frac{1}{N^2} \right. \nn \\
    & + & \left. \frac{7}{4} k_1 + k_2 + 2 \ell_4^r \left( \ell_4^r - \ell_3^r 
    \right) - \frac{5}{4} k_4 \right. \nn \\
    & + & \left. r_F^r + \frac{1}{2} \Delta_F (\Delta_M - \Delta_F) - \frac{1}{N} 
    \Delta_M \right\} ~ , 
    \label{eq:fPi_2loops}
 \ee
 \be
    \label{eq:r2}
    \langle r^2 \rangle & = & \left[ \langle r^2 \rangle \right]_{\mbox{NLO}} 
    + \left[ \langle r^2 \rangle \right]_{\mbox{NNLO}} + {\cal{O}}(\hat{m}^2) 
    ~ , \\[2mm]
    \left[ \langle r^2 \rangle \right]_{\mbox{NLO}} & = & - \frac{2}{F^2} \left( 
    6\ell_6^r + L(\mu) + \frac{1}{N} \right) ~ , 
    \label{eq:r2_1loop}
 \ee
 \be
    \left[ \langle r^2 \rangle \right]_{\mbox{NNLO}} & = & 4 \frac{x_2}{F^2} 
    \left\{ \frac{1}{N} \left[ - 2 \ell_4^r + \frac{31}{6} L(\mu) + \frac{13}{192} 
    - \frac{181}{48 N} \right] \right. \nn \\
    & - & \left. 3 k_1 + \frac{3}{2} k_2 - \frac{1}{2} k_4 + 3 k_6 - 12 \ell_4^r 
    \ell_6^r \right. \nn \\
    & + & \left. 6 r_1^r + \Delta_F \left( 6 \ell_6^r + L(\mu) + \frac{1}{N} 
    \right) - \frac{1}{N} \Delta_M \right\} ~ , 
    \label{eq:r2_2loops}
 \ee
 \be
    \label{eq:curv}
    c & = & \left[ c \right]_{\mbox{NLO}} + \left[ c \right]_{\mbox{NNLO}} 
    + {\cal{O}}(\hat{m}) ~ , \\[2mm]
    \label{eq:c_1loop}
    \left[ c \right]_{\mbox{NLO}} & = & \frac{2}{60 N F^4 ~ x_2} ~ , \\
    \left[ c \right]_{\mbox{NNLO}} & = & \frac{4}{F^4} \left\{ \frac{1}{N} \left[ 
    - \frac{13}{540} L(\mu) + \frac{1}{720} - \frac{8429}{25920 N} \right] \right. 
    \nn \\
    & + & \left. \frac{1}{12} k_1 - \frac{1}{24} k_2 + \frac{1}{24} k_6 + \frac{1}{3 N} 
    \left( \ell_1^r - \frac{1}{2} \ell_2^r + \frac{1}{10} \ell_4^r + \frac{1}{2} 
    \ell_6^r \right) \right. \nn \\
    & + & \left. r_2^r - \frac{1}{60 N} (\Delta_M + \Delta_F) \right\} ~ , 
    \label{eq:c_2loops}
 \ee
where $2B \hat{m}$ is the celebrated GMOR term, $F$ is the pion decay constant 
in the chiral limit ($f_\pi$ is normalized such that $f_\pi \approx 130~\mev$ 
at the physical point) and 
 \be
    N & \equiv & \left(4 \pi \right)^2 ~ , \nn \\
    x_2 & \equiv & \frac{2B \hat{m}}{F^2} ~ , \nn \\
    L(\mu) & \equiv & \frac{1}{N} \mbox{log}\left( \frac{2B \hat{m}}{\mu^2} 
    \right) ~ , \nn \\
    k_i & \equiv & [4 \ell_i^r - \gamma_i L(\mu)] ~ L(\mu) ~ , \nn \\
    \Delta_M & \equiv & 2\ell_3^r + \frac{1}{2} L(\mu) ~ , \nn \\
    \Delta_F & \equiv & 2 \left[ \ell_4^r - L(\mu) \right] ~ .
    \label{eq:notations}
 \ee

The constants $\ell_i^r$ are the finite part of the coupling constants appearing 
in the $O(p^4)$ Lagrangian after the application of the $\overline{MS}$ procedure 
and their values depend on the renormalization scale $\mu$ through the anomalous 
dimensions $\gamma_i$ as $\mu^2 d\ell_i^r / d\mu^2 = - \gamma_i / 2N$. 
The coefficients $\gamma_i$ are calculated in Ref.~\cite{GL} and those relevant 
in this work are given by: $\gamma_1 = 1/3, ~ \gamma_2 = 2/3, ~ \gamma_3 = -1/2, 
~ \gamma_4 = 2, ~ \gamma_6 = -1/3$. 
The four constants $r_M^r$, $r_F^r$, $r_1^r$, $r_2^r$ denote the contributions of 
the $O(p^6)$ Lagrangian after $\overline{MS}$ subtraction.
Though the values of all the above constants depend on $\mu$, at each order in the 
chiral expansion the physical observables are independent (as they should be) of 
the value of the renormalization scale $\mu$.

At LO only two chiral parameters appear, namely $B$ (related to the chiral 
condensate) and $F$.
At NLO three further LEC's, $\ell_3^r$, $\ell_4^r$ and $\ell_6^r$, are present. 
At NNLO the total number of LEC's increases up to 11 due to the inclusion of 
$\ell_1^r$, $\ell_2^r$, $r_M^r$, $r_F^r$, $r_1^r$ and $r_2^r$.

We notice that the NNLO terms for the charge radius (\ref{eq:r2_2loops}) and 
the curvature (\ref{eq:c_2loops}) do not depend upon the LEC's $\ell_1^r$ and 
$\ell_2^r$ separately, but only through the linear combination ($\ell_1^r - 
\ell_2^r / 2$).
However different linear combinations of $\ell_1^r$ and $\ell_2^r$ appear in the 
NNLO terms of both the pion mass (\ref{eq:MPi2_2loops}) and decay constant 
(\ref{eq:fPi_2loops}).
Therefore the LEC's $\ell_1^r$ and $\ell_2^r$ can be determined by a simultaneous 
analysis of the the charge radius (and/or the curvature) together with the pion 
mass and decay constant.

In what follows the $O(p^4)$ constants $\ell_i^r$ will be substituted by scale-invariant 
quantities, $\bar{\ell}_i$, defined via the relations
 \be
    \ell_i^r \equiv \frac{\gamma_i}{2N} \left[ \bar{\ell}_i + N L(\mu) \right] ~ .
    \label{eq:ellebar}
 \ee
The new quantities, which depend (logarithmically) on the quark mass, can be expressed 
as $\bar{\ell}_i = \mbox{log}(\Lambda_i^2 / 2B \hat{m})$ and their values are commonly 
given at the physical point.

We notice that in Ref.~\cite{BCT} the quark mass is not actually used as the 
expansion parameter.
Instead of it the physical pion mass and decay constant are adopted.
In order to recover the formulae of Ref.~\cite{BCT} it's enough to replace in 
Eqs.~(\ref{eq:MPi2})-(\ref{eq:notations}) $x_2$ with $M_\pi^2 / f_\pi^2$, 
$L(\mu)$ with $(1/N) \cdot \mbox{log}(M_\pi^2 / \mu^2)$ and to set 
$\Delta_M = \Delta_F = 0$ wherever they appear explicitly.

As explained in Sec.~\ref{subsec:volume} we apply to the pion mass and decay 
constant the corrections for FSE computed in Ref.~\cite{CDH05}.
Using again the quark mass $\hat{m}$ as the expansion parameter, one gets 
 \be
    \frac{M_\pi(L) - M_\pi}{M_\pi} & = & \frac{2 x_2}{N} \sum_{n = 1}^{\infty} 
    \frac{m(n)}{\lambda_n} \left\{ \phantom{\frac{1}{1}} \hspace{-0.25cm} 
    K_1(\lambda_n) \right. \nn \\
    & - & \left. \frac{2 x_2}{N} \left[ K_1(\lambda_n) \left(-\frac{55}{18} + 
    4 \bar{\ell}_1 + \frac{8}{3} \bar{\ell}_2 - \frac{5}{2} \bar{\ell}_3 - 
    2 \bar{\ell}_4 \right) \right. \right. \nn \\
    & + & \left. \left. \frac{K_2(\lambda_n)}{\lambda_n} \left( \frac{112}{9} 
    - \frac{8}{3} \bar{\ell}_1 - \frac{32}{3} \bar{\ell}_2 \right) + 
    \frac{13}{3} g_0 K_1(\lambda_n) \right. \right. \nn \\
    & - & \left. \left. \frac{1}{3} \left( 40 g_0 + 32 g_1 + 26 g_2 \right) 
    \frac{K_2(\lambda_n)}{\lambda_n} \right. \right. \nn \\
    & + & \left. \left. \frac{N}{2} \left[ \Delta_M \lambda_n K_0(\lambda_n) + 
    2 \Delta_F K_1(\lambda_n) \right] \phantom{\frac{1}{1}} \hspace{-0.25cm} 
    \right] \right\} + {\cal{O}}(\hat{m}^3) ~ ,
    \label{eq:MPiL}
 \ee
 \be
    \frac{f_\pi(L) - f_\pi}{f_\pi} & = & - 2 ~ \frac{2 x_2}{N} \sum_{n = 1}^{\infty} 
    \frac{m(n)}{\lambda_n} \left\{ \phantom{\frac{1}{1}} \hspace{-0.25cm} 
    2 K_1(\lambda_n) \right. \nn \\
    & - & \left. \frac{2 x_2}{N} \left[ K_1(\lambda_n) \left(-\frac{7}{9} + 
    2 \bar{\ell}_1 + \frac{4}{3} \bar{\ell}_2 - 3 \bar{\ell}_4 \right) 
    \right. \right. \nn \\
    & + & \left. \left. \frac{K_2(\lambda_n)}{\lambda_n} \left( \frac{112}{9} 
    - \frac{8}{3} \bar{\ell}_1 - \frac{32}{3} \bar{\ell}_2 \right) + \frac{1}{6} 
    \left( 8 g_0 - 13 g_1 \right) K_1(\lambda_n) \right. \right. \nn \\
    & - & \left. \left. \frac{1}{3} \left( 40 g_0 - 12 g_1 - 8 g_2 -13 g_3 
    \right) \frac{K_2(\lambda_n)}{\lambda_n} \right. \right. \nn \\
    & + & \left. \left. N \left[ \Delta_M \lambda_n K_0(\lambda_n) + 2 \Delta_F 
    K_1(\lambda_n) \right] \phantom{\frac{1}{1}} \hspace{-0.25cm} \right] \right\} 
    + {\cal{O}}(\hat{m}^3) ~ ,
    \label{eq:fPiL}
 \ee
where $K_{0,1,2}$ are modified Bessel functions, the values of the multiplicities 
$m(n)$ are given in Ref.~\cite{CDH05} and
 \be
    \lambda_n & \equiv & \sqrt{n} ~ \sqrt{2B \hat{m}} ~ L ~ , \nn \\
    g_0 & \equiv & 2 - \frac{\pi}{2} ~ , \nn \\
    g_1 & \equiv & \frac{\pi}{4} - \frac{1}{2} ~ , \nn \\
    g_2 & \equiv & \frac{1}{2} - \frac{\pi}{8} ~ , \nn \\
    g_3 & \equiv & \frac{3 \pi}{16} - \frac{1}{2} ~ .
    \label{eq:notationsV}
 \ee
Notice that in Eqs.~(\ref{eq:MPiL})-(\ref{eq:fPiL}) no further LEC is introduced 
with respect to Eqs.~(\ref{eq:MPi2})-(\ref{eq:fPi_2loops}).

In order to recover the formulae of Ref.~\cite{CDH05}, where the physical pion 
mass and decay constant are adopted as expansion parameters, it is enough to 
replace in Eqs.~(\ref{eq:MPiL})-(\ref{eq:fPiL}) $x_2$ with $M_\pi^2 / f_\pi^2$, 
$\lambda_n$ with $\sqrt{n} ~ M_\pi L$ and to set $\Delta_M = \Delta_F = 0$.

\subsection{Chiral fits\label{subsec:fit1}}

Let us now apply Eqs.~(\ref{eq:MPi2})-(\ref{eq:c_2loops}) to the analyses of the 
quark mass dependence of our results.
As explained in the previous Section the set of lattice data chosen for the 
fitting procedure is given by the results of the runs $R_1$, $R_{2a}$, $R_3$, 
$R_4$, $R_{5a}$ and $R_6$ for four quantities: the pion mass and decay constant, 
the charge radius and the curvature of the pion form factor.
In case of the pion mass and decay constant also the results of the run $R_{2b}$ 
are considered and the FSE corrections given by Eqs.~(\ref{eq:MPiL})-(\ref{eq:fPiL}) 
are applied.

Since each run corresponds to an independent ensemble of gauge configurations a 
bootstrap procedure is applied in order to combine all the jackknives in different 
ways (1000 samples are used in practice).
The statistical uncertainties, which are reported here after, are therefore 
bootstrap errors.

In order to fix the lattice spacing and the up/down quark mass the experimental 
values of the pion mass and decay constant ($M_\pi^{phys.} = 134.98~\mev$ and 
$f_\pi^{phys.} = 130.7 \pm 0.4~\mev$ from Ref.~\cite{PDG}) are used\footnote{In 
order to account for the e.m.~isospin breaking effects which are not introduced 
in the lattice simulations, we use the experimental value of the neutral pion 
mass in accord with Refs.~\cite{Pi0,ETMC4}.}.
We determine firstly the value of the bare quark mass $a m_\pi$, at which the 
pion assumes its physical mass, by requiring that the ratio $M_\pi / f_\pi$ 
from Eqs.~(\ref{eq:MPi2}) and (\ref{eq:fPi}) takes the experimental value 
$134.98 / 130.7 \simeq 1.033$.
Secondly, using the physical value $f_\pi^{phys.}$ the lattice spacing $a$ is 
determined.
The value of the renormalized light quark mass in the $\overline{MS}$ scheme, 
$m^{\overline{MS}}(2~\gev)$, is obtained from $a m_\pi$ by considering the 
determination of the non-perturbative (multiplicative) renormalization 
constant $Z_m = 1 / Z_P$, evaluated using the RI-MOM scheme in 
Ref.~\cite{renorm}, and the matching factor with the $\overline{MS}$ 
scheme, which is known up to four loops \cite{matching}.

We start with a ChPT analysis at NLO including our lattice data only for the pion 
mass and decay constant up to $M_\pi \simeq 500~\mev$ in order to compare with the 
ETMC NLO analyses of Refs.~\cite{ETMC1,ETMC2,ETMC_scaling}.
The main differences are: 
~ i) the use of a single lattice spacing ($a \simeq 0.09$ fm) both in the present 
work and in Refs.~\cite{ETMC1,ETMC2}, while the results obtained from two lattice 
spacings ($a \simeq 0.07$ and $0.09$ fm) are taken into account in 
Ref.~\cite{ETMC_scaling};
~ ii) a better statistical accuracy of the data for $M_\pi$ and $f_\pi$ in 
Refs.~\cite{ETMC1,ETMC2,ETMC_scaling} due to the use of all the (correlated) 
trajectories produced by the ETM collaboration with respect to the present 
work in which only a subset of 240 (uncorrelated) trajectories are employed;
~ iii) the presence of the results of the run $R_1$ at $M_\pi \simeq 260~\mev$ 
in the present work and in Ref.~\cite{ETMC_scaling} at variance with 
Refs.~\cite{ETMC1,ETMC2}.

\begin{figure}[!hbt]

\centerline{\includegraphics[scale=0.80]{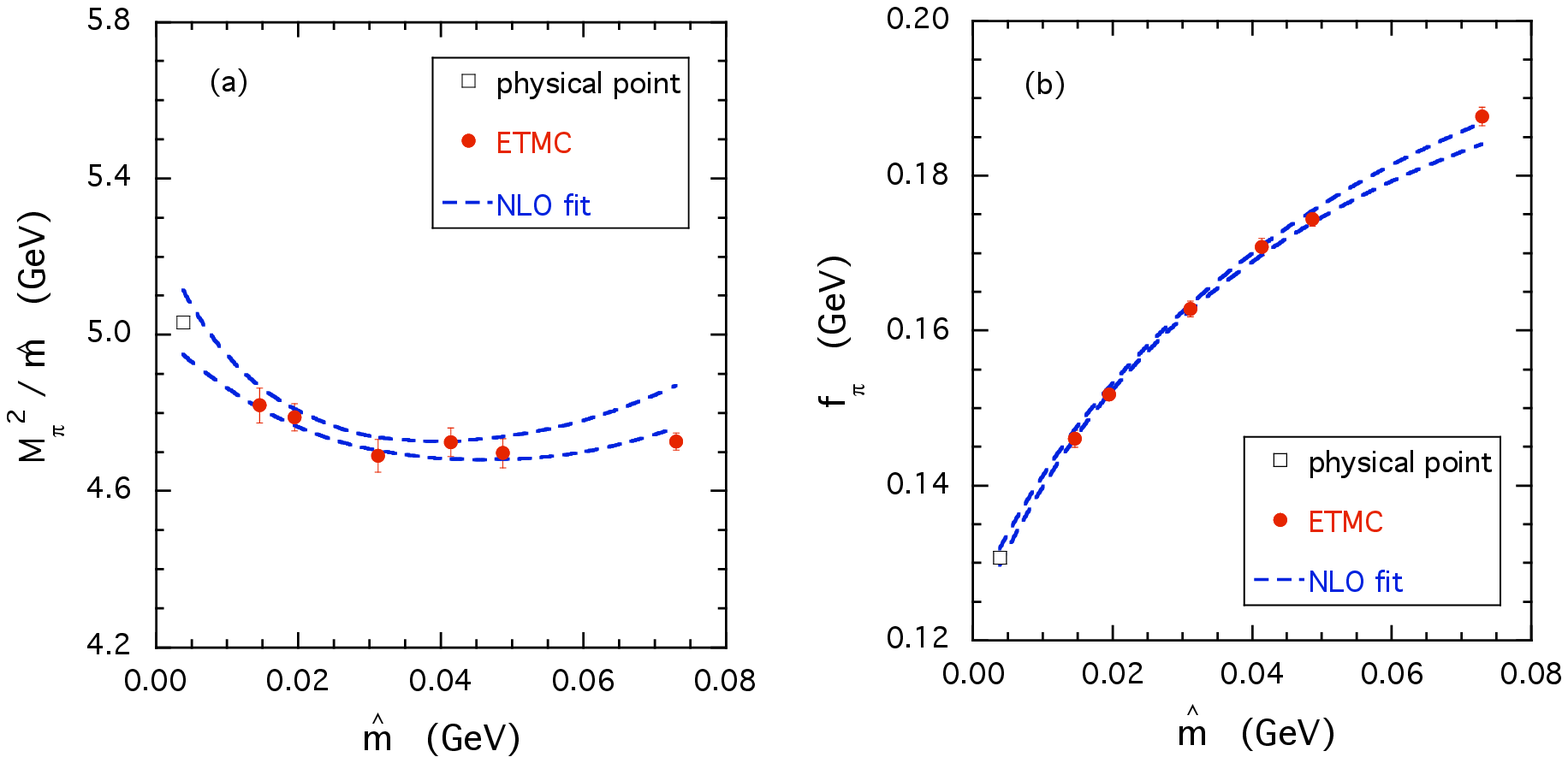}}

\caption{\it The ratio of the squared pion mass to the renormalized quark mass 
$\hat{m} = m^{\overline{MS}}(2~\gev)$ (a) and the pion decay constant (b) versus 
$\hat{m}$ in physical units.
The points at the highest value of $\hat{m}$ are not included in the analysis.
The dots are the ETMC results, corrected by the FSE effects given by 
Eqs.~(\ref{eq:MPiL})-(\ref{eq:fPiL}), and the squares represent the 
experimental value for each quantity from Ref.~\cite{PDG}. 
The dashed lines correspond to the region selected at $1 \sigma$ level by the 
ChPT analysis at NLO [see Eqs.~(\ref{eq:MPi2})-(\ref{eq:MPi2_1loop}) and 
(\ref{eq:fPi})-(\ref{eq:fPi_1loop})].
The values of the fitting parameters are listed in the second column of Table 
\ref{tab:chiral1}.}

\label{fig:MF_1loop}

\end{figure}

Following Ref.~\cite{ETMC2} the FSE correction can be evaluated beyond NLO using 
Eqs.~(\ref{eq:MPiL})-(\ref{eq:fPiL}) and adopting for the unknown LEC's 
$\bar{\ell}_1$ and $\bar{\ell}_2$ the central values given in Ref.~\cite{CGL}.
The values obtained for the fitting parameters (the LEC's of the chiral Lagrangian) 
are given in the second column of Table \ref{tab:chiral1}, while the best fit at 
NLO, including the FSE corrections given by Eqs.~(\ref{eq:MPiL})-(\ref{eq:fPiL}), 
is shown in Fig.~\ref{fig:MF_1loop} by the dashed lines.

\begin{table}[!htb]

\begin{center}
\begin{tabular}{||c||c||c|c||}
\hline
 $parameter$ & $NLO$ & $NNLO$ & $NNLO~+~<r^2>_S^{exp.}$ \\ \hline \hline
 $2 B~(\gev)$ & $5.21 \pm 0.05$ & $5.19 \pm 0.42$ & $4.89 \pm 0.08$\\ \hline 
 $F~(\mev)$ & $121.7 \pm 1.1$ & $121 \pm 10$ & $122.5 \pm 0.9$ \\ \hline \hline
 $\bar{\ell}_3$ & $3.48 \pm 0.12$ & $5.0 \pm 4.0$ & $3.1 \pm 0.7$ \\ \hline 
 $\bar{\ell}_4$ & $4.67 \pm 0.06$ & $5.2 \pm 1.8$ & $4.39 \pm 0.14$ \\ \hline 
 $\bar{\ell}_6$ & $14.59 \pm 0.03$ & $14.9 \pm 2.2$ & $15.9 \pm 1.3$ \\ \hline \hline
 $\bar{\ell}_1$ & $-0.4$~\cite{CGL} & $0.3 \pm 3.4$ & $-1.3 \pm 1.4$ \\ \hline 
 $\bar{\ell}_2$ & $\hskip 0.3cm 4.3$~\cite{CGL} & $5.3 \pm 1.1$ & $\hskip 0.3cm 5.1 \pm 1.1$ \\ \hline 
 $r_M^r \cdot 10^4$ & $--$ & $-0.1 \pm 1.1$ & $-0.60 \pm 0.29$ \\ \hline 
 $r_F^r \cdot 10^4$ & $--$ & $-0.5 \pm 1.5$ & $-0.15 \pm 0.14$ \\ \hline 
 $r_1^r \cdot 10^4$ & $--$ & $-0.95 \pm 0.15$ & $-0.92 \pm 0.15$ \\ \hline 
 $r_2^r \cdot 10^4$ & $--$ & $\hskip 0.3cm 0.70 \pm 0.17$ & $\hskip 0.3cm 0.77 \pm 0.04$ \\ \hline \hline 
 $a~(fm)$ & $0.0861 \pm 0.0007 $ & $0.0863 \pm 0.0047$ & $0.0884 \pm 0.0006$ \\ \hline
 $\hat{m}^{phys.}~(\mev)$ & $3.60 \pm 0.08$ & $3.6 \pm 0.6$ & $3.81 \pm 0.08$ \\ \hline \hline
 $\chi^2~/~d.o.f.$ & $0.72$ & $0.91$ & $0.88$ \\ \hline \hline

\end{tabular}

\caption{\it Values of the chiral parameters, the lattice spacing and the renormalized 
quark mass $\hat{m} = m^{\overline{MS}}(2~\gev)$ at the physical point for various ChPT 
analyses (see text).
For consistency with $\hat{m}$, the parameter $B$ is given in the $\overline{MS}$ scheme 
at a scale equal to $2~\gev$.
The values of the parameters $r_M^r$, $r_F^r$, $r_1^r$ and $r_2^r$ are given at the 
$\rho$-meson mass scale.
In the case of the NLO analysis the parameters $\bar{\ell}_1$ and $\bar{\ell}_2$ are 
used only for evaluating FSE's, while the parameter $\bar{\ell}_6$ is fixed by the 
experimental value of the pion charge radius. 
The latter is not included in the NNLO analyses.
The uncertainties are statistical (bootstrap) errors only.
\label{tab:chiral1}}

\end{center}

\end{table}

It can be seen that the values of all the chiral parameters, in particular of 
the LEC's $\bar{\ell}_3$ and $\bar{\ell}_4$, as well as the values of the lattice 
spacing $a$ and the light quark mass $\hat{m}^{phys.}$, are consistent with the 
findings of Refs.~\cite{ETMC1,ETMC2,ETMC_scaling}.
Note also that the statistical precision of the extracted values of $\bar{\ell}_3$ 
and $\bar{\ell}_4$ are very similar in this work and in 
Refs.~\cite{ETMC1,ETMC2,ETMC_scaling}.

The theoretical evaluation of FSE effects given by Eqs.~(\ref{eq:MPiL})-(\ref{eq:fPiL}) 
appears to work quite well.
Indeed after applying the FSE corrections, the pion masses and decay constants 
corresponding to the runs $R_{2a}$ (at $L = 32a$) and $R_{2b}$ (at $L = 24a$) 
become consistent within one standard deviation, as shown in Table 
\ref{tab:FSE}.

\begin{table}[!htb]

\begin{center}
\begin{tabular}{||c||c|c||}
\hline
 & $lattice~(\%)$ & $theoretical~(\%)$ \\ \hline
 $aM_\pi(L=24a) / aM_\pi(L=32a) - 1$ & $+1.8~(5)$ & $+1.2$ \\ \hline
 $af_\pi(L=24a) / af_\pi(L=32a) - 1$ & $-2.5~(6)$ & $-2.5$ \\ \hline \hline

\end{tabular}

\caption{\it Values of the quantities $[aM_\pi(L=24a) /aM_\pi(L=32a) - 1]$ and 
$[af_\pi(L=24a) / af_\pi(L=32a) - 1]$ from the runs $R_{2a}$ and $R_{2b}$ at a 
pion mass of $\simeq 300~\mev$.
The lattice results correspond to the values given in Table \ref{tab:volume}.
The theoretical results are those corresponding to Eqs.~(\ref{eq:MPiL})-(\ref{eq:fPiL}) 
using for the relevant LEC's the values reported in the second column of Table 
\ref{tab:chiral1}.
\label{tab:FSE}}

\end{center}

\end{table}

We have checked that the full exclusion of the run $R_{2b}$ from our analyses 
does not have any significant impact on the chiral fits as well as on the values 
obtained for the chiral parameters. 
Therefore, since the runs $R_{2a}$ and $R_{2b}$ are compatible once theoretical 
FSE's are included through Eqs.~(\ref{eq:MPiL})-(\ref{eq:fPiL}), in what follows 
we shall not show the results of the run $R_{2b}$ in the figures (i.e.~we show 
only lattice data having $M_\pi L \gtrsim 4$) and we will always apply to the 
pion mass and decay constant the FSE corrections given by 
Eqs.~(\ref{eq:MPiL})-(\ref{eq:fPiL}).

The quality of the NLO fit shown in Fig.~\ref{fig:MF_1loop} is quite remarkable, 
leaving apparently little room for higher-order corrections even at the highest 
pion mass ($\simeq 580~\mev$), though the latter point is not included in the 
fitting procedure.
However we now show that the same does not hold for the charge radius and the 
curvature of the pion form factor.

The NLO prediction for the charge radius [see Eq.~(\ref{eq:r2_1loop})] depends 
in practice only on one LEC, $\bar{\ell}_6$, being F fixed by the analysis of 
the pion decay constant with a precision of the level of $\simeq 1\%$ (see 
the second column of Table \ref{tab:chiral1}). 
Note also that both the derivative of $[ \langle r^2 \rangle ]_{\mbox{NLO}}$ with respect 
to the quark mass and the curvature $[ c ]_{\mbox{NLO}}$ (see Eq.~\ref{eq:c_1loop}) 
are independent of $\bar{\ell}_6$ and therefore basically parameter-free.

\begin{figure}[!hbt]

\centerline{\includegraphics[scale=0.80]{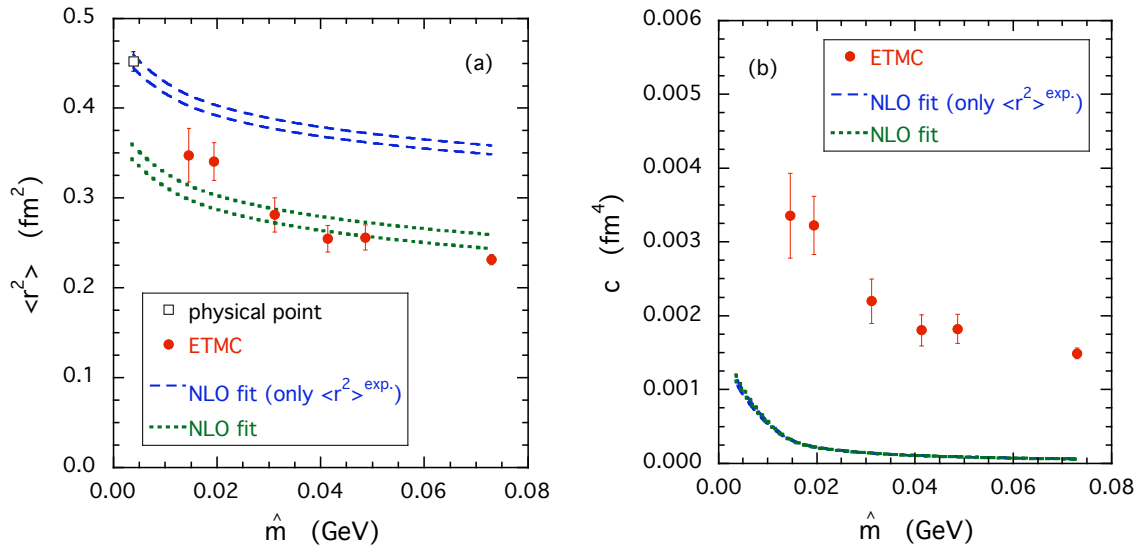}}

\caption{\it The squared charge radius (a) and the curvature (b) of the pion 
form factor versus the renormalized quark mass $\hat{m}$ in physical units.
The dots are our lattice results and the square represents the experimental 
value of the squared charge radius \cite{PDG}.
The dashed and dotted lines represent the region selected at $1 \sigma$ level by 
the ChPT predictions at NLO given by Eqs.~(\ref{eq:r2_1loop}) and 
(\ref{eq:c_1loop}).
In the case of the dashed lines the value of the LEC $\bar{\ell}_6$ is fixed by 
the experimental charge radius, while in the case of the dotted lines it is 
obtained by including in the fitting procedure our lattice data of the 
charge radius for pion masses up to $\simeq 500~\mev$ (i.e.~for $\hat{m} 
\lesssim 0.05~\gev$).}

\label{fig:RC_1loop}

\end{figure}

The value of the LEC $\bar{\ell}_6$ can be determined from the experimental value 
of the squared charge radius, $\langle r^2 \rangle^{exp.} = 0.452 \pm 
0.011~\mbox{fm}^2$ \cite{PDG}, since the latter is expected to be 
dominated by the NLO term (\ref{eq:r2_1loop}).
This leads to $\bar{\ell}_6 = 14.59 \pm 0.03$ (see the second column of 
Table~\ref{tab:chiral1}). 
The corresponding NLO predictions for the charge radius and the curvature are 
shown in Fig.~\ref{fig:RC_1loop} by the dashed lines.
They significantly overestimate our lattice data for the charge radius and largely 
underestimate those for the curvature.

Alternatively we have excluded the experimental value of the charge radius and 
included in the fitting procedure the lattice data of the charge radius for 
pion masses up to $\simeq 500~\mev$ (i.e.~for $\hat{m} \lesssim 0.05~\gev$) 
obtaining $\bar{\ell}_6 = 11.6 \pm 0.3$ (with $\chi^2 / 
\mbox{d.o.f.} \simeq 1.2$).
The corresponding NLO predictions are shown in Fig.~\ref{fig:RC_1loop} by the 
dotted lines. 
At the physical point the charge radius is $\langle r^2 \rangle^{phys.} = 0.352 \pm 
0.008~\mbox{fm}^2$ in clear contradiction with the experimental value\footnote{If 
only the lattice data of the charge radius for the two lowest pion masses ($M_\pi 
\lesssim 300~\mev$) are considered, the value of $\bar{\ell}_6$ becomes $12.8 
\pm 0.5$ (with $\chi^2 / \mbox{d.o.f.} \simeq 0.65$) and the predicted 
charge radius at the physical point is $\langle r^2 \rangle^{phys.} = 
0.393 \pm 0.017~\mbox{fm}^2$, which still deviates from the 
experimental value by three standard deviations.}.

Moreover, since the curvature is independent on $\bar{\ell}_6$, any NLO fit is 
unable to provide enough large values of the curvature consistent with the relation 
(\ref{eq:cpole}), i.e.~with the pole behavior of the pion form factor observed 
in subsection \ref{subsec:pionff}.
Such a finding suggests that a NLO analysis of the pion form factor is applicable 
only to quite low values of $|q^2|$ (see later Section \ref{sec:ffpion}). 

Thus, both the NLO results shown in Fig.~\ref{fig:RC_1loop} and the smallness of 
the systematic effects due to finite volumes and lattice spacings, estimated in 
subsections \ref{subsec:volume} and \ref{subsec:a2}, indicate that the quark 
mass dependences of our lattice data for both the charge radius and the 
curvature require to take into account chiral effects beyond the NLO.

\begin{figure}[!hbt]

\centerline{\includegraphics[scale=0.80]{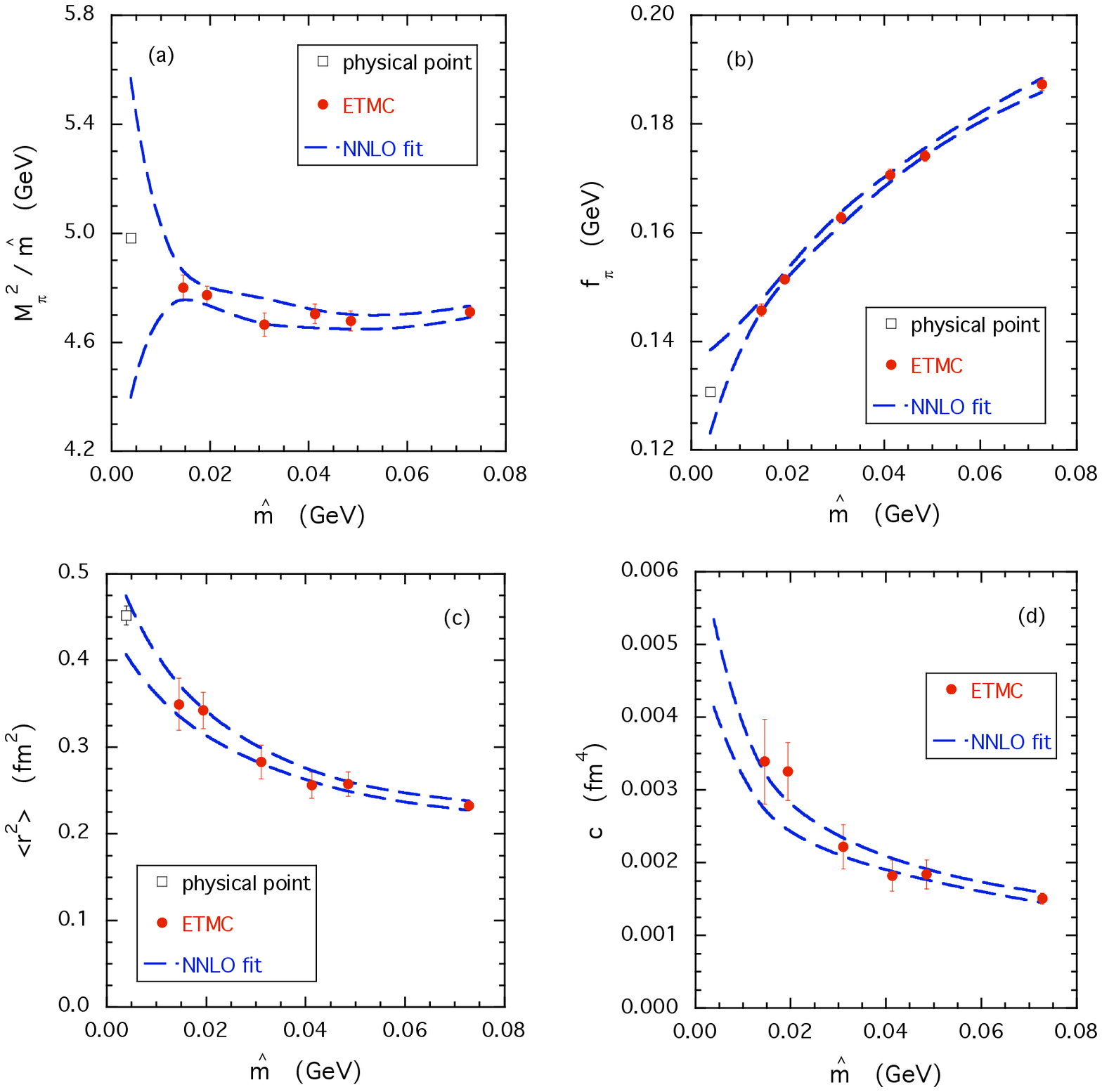}}

\caption{\it The ratio $M_\pi^2 / \hat{m}$ (a), the pion decay constant (b), the 
charge radius (c) and the curvature (d) of the pion form factor versus the 
renormalized quark mass $\hat{m}$ in physical units.
The dots are our lattice results and the squares represent the corresponding 
experimental values from PDG \cite{PDG}. 
The dashed lines correspond to the region selected at $1 \sigma$ level by the 
ChPT fit at NNLO based on Eqs.~(\ref{eq:MPi2})-(\ref{eq:c_2loops}).
The values of the fitting parameters are listed in the third column of Table 
\ref{tab:chiral1}.
The experimental value of the pion charge radius is not included in the fitting 
procedure.}

\label{fig:MFRC_2loop}

\end{figure}

The results of the fit performed using ChPT at NNLO [i.e.~based on 
Eqs.~(\ref{eq:MPi2})-(\ref{eq:c_2loops})] are shown in Fig.~\ref{fig:MFRC_2loop}, 
while the values of the fitting parameters are listed in the third column of 
Table \ref{tab:chiral1} with the renormalization scale $\mu$ fixed at the 
physical $\rho$-meson mass.
Notice that the experimental value of the pion charge radius is not included in 
the fitting procedure.

As already found in Ref.~\cite{ETMC2}, the inclusion of NNLO effects leads to 
quite large uncertainties in the values of all the LEC's, in particular both 
for the LEC's $\bar{\ell}_3$ and $\bar{\ell}_4$ appearing at NLO and NNLO, 
and for the LEC's $\bar{\ell}_1$ and $\bar{\ell}_2$ appearing only at NNLO.
Nevertheless the uncertainties in the chiral fits shown in Fig.~\ref{fig:MFRC_2loop} 
are of the order of the statistical errors in the mass range of the lattice points. 
This means that the large uncertainties reported in the third column of Table 
\ref{tab:chiral1} are strongly correlated; the effect of the variation of one 
fitting parameter can be always compensated by those generated by the variations 
of the other parameters.
However when we extrapolate the chiral predictions for the pion mass and decay 
constant outside the mass range of the lattice data towards the chiral point 
we end up with rather large uncertainties as shown in Fig.~\ref{fig:MFRC_2loop}.

Such a situation is clearly unsatisfactory both for a precise extraction of the 
LEC's and for the extrapolation to the physical point.
The inclusion of the experimental values of the pion charge radius and curvature 
in the set of fitted data can obviously reduce the uncertainties in the extraction 
of the chiral parameters, but in this way the predictive power of the chiral fits 
is lost.

Thus we look for an observable which should be: i) unrelated to the vector form 
factor of the pion, ii) known experimentally and iii) whose chiral expansion at 
NLO contains one of the LEC's, let's say $\bar{\ell}_3$ or $\bar{\ell}_4$.
In this way the experimental value of such an observable, expected to be dominated 
by the NLO contribution, can constrain sufficiently the range of the variability 
of one of the LEC's. 
In turn this could be beneficial to reduce the uncertainties of all our fitting 
parameters.

A possible, appropriate choice is the squared radius $\langle r^2 \rangle_S$ of 
the pion ``scalar'' form factor, defined as
 \be
    \label{eq:r2_S}
    \langle r^2 \rangle_S \equiv & = & \frac{6}{F_\pi^S(0)} \left[ \frac{dF_\pi^S(q^2)}{dq^2} 
    \right]_{q^2 = 0} ~ , 
  \ee
where
  \be
    \label{eq:ffS}
    F_\pi^S(q^2) & = & \langle \pi^+(p^\prime) | \bar{u} u + \bar{d} d | \pi^+(p) 
    \rangle ~ .
 \ee
Indeed, on one hand side the experimental value of the pion scalar radius is 
known quite accurately from the analysis of $\pi-\pi$ scattering data (see 
Ref.~\cite{CGL}), which gives
 \be
    \langle r^2 \rangle_S^{exp.} = 0.61 \pm 0.04~\mbox{fm}^2 ~ .
    \label{eq:r2S_exp}
 \ee
On the other hand side the chiral expansion of $\langle r^2 \rangle_S$, calculated 
at NNLO in Ref.~\cite{BCT}, reads as
 \be
    \label{eq:r2S}
    \langle r^2 \rangle_S & = & \left[ \langle r^2 \rangle_S \right]_{\mbox{NLO}} 
    + \left[ \langle r^2 \rangle_S \right]_{\mbox{NNLO}} + {\cal{O}}(\hat{m}^2) ~ ,
 \ee
 \be
    \label{eq:r2S_1loop}
    \left[ \langle r^2 \rangle_S \right]_{\mbox{NLO}} & = & \frac{2}{F^2} 
    \left( 6 \ell_4^r - 6 L(\mu) - \frac{13}{2N} \right) ~ ,
 \ee
 \be
    \left[ \langle r^2 \rangle_S \right]_{\mbox{NNLO}} & = & 4 \frac{x_2}{F^2} 
    \left\{ \frac{1}{N} \left[ 88 \ell_1^r + 36 \ell_2^r + 5 \ell_3^r - 13 
    \ell_4^r + \frac{145}{36} L(\mu) \right. \right. \nn \\
    & - & \left. \left. \frac{23}{192} + \frac{869}{108N} \right] + 31 k_1 + 17 
    k_2 - 6 k_4 + 12 \ell_4^r ( \ell_4^r - 2 \ell_3^r ) \right. \nn \\
    & + & \left. 6 r_S^r - \Delta_F \left( 3 \Delta_F - \frac{13}{2N} \right) - 
    \frac{6}{N} \Delta_M \right\} ~ , 
    \label{eq:r2S_2loops}
 \ee
It can be seen that the LEC $\bar{\ell}_4$, which also governs the NLO correction 
to the pion decay constant [see Eq.~(\ref{eq:fPi_1loop})], appears in 
Eq.~(\ref{eq:r2S_1loop}).

As already stressed, the experimental value $\langle r^2 \rangle_S^{exp.}$ is 
expected to be dominated by the NLO contribution (\ref{eq:r2S_1loop}). 
Using the values of the relevant LEC's of the second column of Table 
\ref{tab:chiral1} one gets $[\langle r^2 \rangle_S]_{NLO} = 0.716 \pm 
0.014~\mbox{fm}^2$, which overestimates the experimental value 
(\ref{eq:r2S_exp}) by almost three standard deviations.
Thus we want to use the NNLO calculation of $\langle r^2 \rangle_S$, also 
for consistency with the use of Eqs.~(\ref{eq:MPi2})-(\ref{eq:c_2loops}) 
for the other observables.
In order to do that we need to set the value of the parameter $r_S^r$ appearing 
in Eq.~(\ref{eq:r2S_2loops}).
In Ref.~\cite{BCT} an estimate of $r_S^r$ at the $\rho$-meson mass scale has 
been obtained using a resonance model, namely $r_S^r \approx -0.3 \cdot 10^{-4}$.
We have checked that using the above value or putting the parameter $r_S^r$ equal 
to zero does not produce any significant difference in our chiral fits.
This is not surprising since the effects of a non-vanishing value of $r_S^r$ 
are expected to be relevant at large quark masses only.
Thus in what follows the value $r_S^r = 0$ is assumed. 

\begin{figure}[!hbt]

\centerline{\includegraphics[scale=0.80]{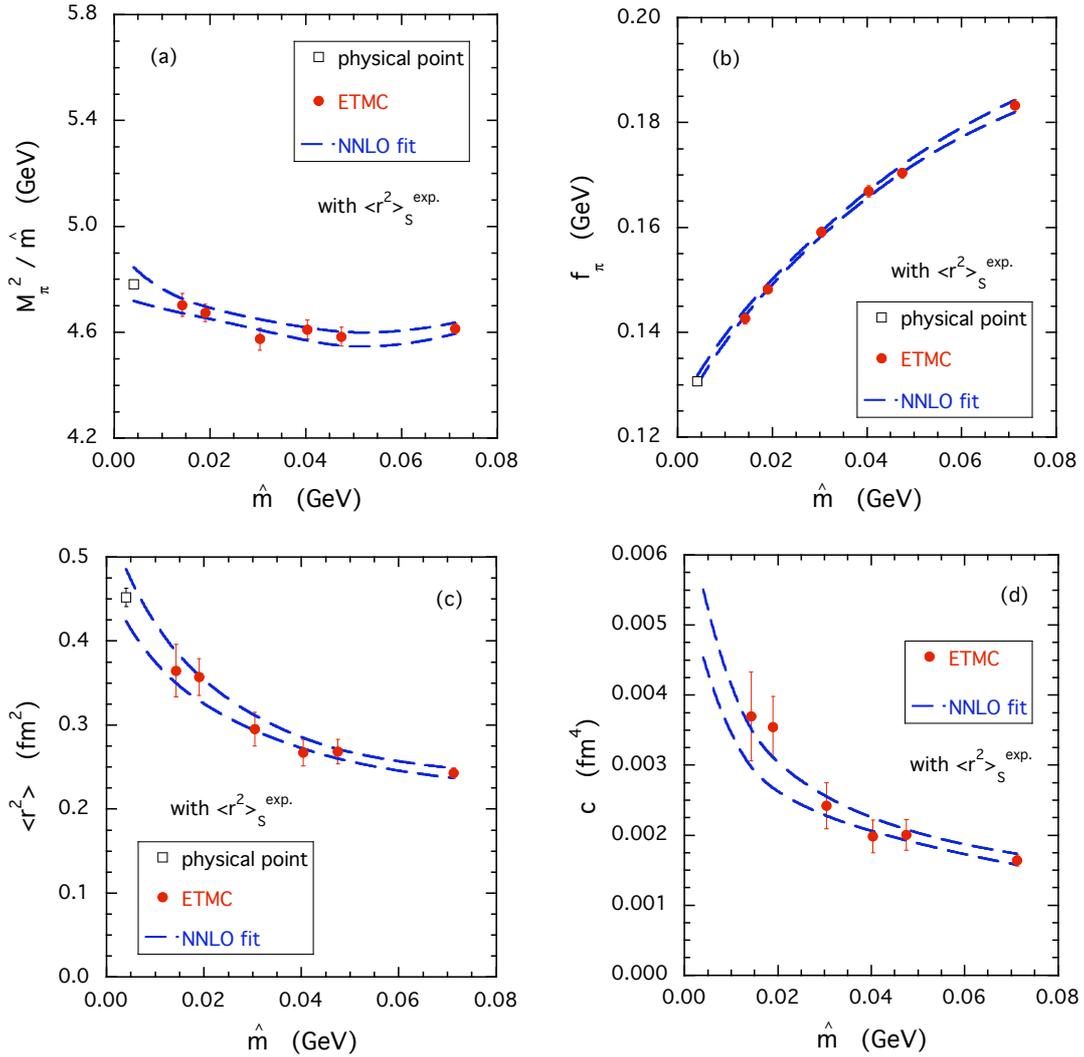}}

\caption{\it As in Fig.~\ref{fig:MFRC_2loop}, but including the experimental value 
of the pion scalar radius from Ref.~\cite{CGL} in the fitting procedure.
The resulting values of the fitting parameters are listed in the fourth column of 
Table \ref{tab:chiral1}.}

\label{fig:MFRCrs2_2loop}

\end{figure}

Including the experimental value $\langle r^2 \rangle_S^{exp.}$ in the ensemble 
of fitted data and Eqs.~(\ref{eq:r2S})-(\ref{eq:r2S_2loops}) in the fitting 
procedure, we obtain the results shown in Fig.~\ref{fig:MFRCrs2_2loop} 
with the values of the fitting parameters reported in the fourth 
column of Table \ref{tab:chiral1}.
Let us remind that the experimental value of the pion charge radius is not 
included in the fitting procedure.

Our expectation about the reduction of the uncertainties of the fitting parameters 
is fully confirmed.
Thanks to the introduction of the experimental value $\langle r^2 \rangle_S^{exp.}$ 
the value of $\bar{\ell}_4$ is determined quite accurately and this is beneficial 
for reducing the uncertainties of all the other LEC's (compare the 
third and the fourth columns of Table \ref{tab:chiral1}).

Note that with respect to the NLO analysis the values of both the parameter $B$ and 
the lattice spacing, obtained in the NNLO analysis which includes $\langle r^2 
\rangle_S^{exp.}$, change beyond the corresponding statistical errors. 
On the contrary the values of the LEC's $F$, $\bar{\ell}_3$, $\bar{\ell}_4$ and 
$\bar{\ell}_6$ do not change significantly.

Since our data used for the curvature rely on the assumption of the monopole 
behavior (see Eq.~(\ref{eq:cpole})), we have checked that the values of the LEC's, 
extracted by including the curvature data obtained from the cubic fit (\ref{eq:cubic}), 
change only slightly within the statistical errors with respect to the ones reported 
in the fourth column of Table \ref{tab:chiral1}.

The values of the charge radius and the curvature predicted at the physical point 
by the chiral fit shown in Fig.~\ref{fig:MFRCrs2_2loop}, are $\langle r^2 
\rangle^{phys.} = 0.456 \pm 0.030~\mbox{fm}^2$ and $c^{phys.} = (5.11 
\pm 0.47) \cdot 10^{-3}~\mbox{fm}^4$.
We perform a rough estimate of the systematic errors due to finite volume and 
discretization effects. 
Firstly we substitute the run $R_{2a}$ with the run $R_{2b}$ and fit the new set
of data; the changes in the central values of the chiral parameters provide an 
estimate of the finite volume effects. 
Secondly we further substitute the runs $R_{2b}$ and $R_{5a}$ with the runs 
$R_{2c}$ and $R_{5b}$ at the finer lattice spacing, respectively, obtaining 
an estimate of discretization effects.
Adding all the systematic uncertainties in quadrature, our final results are 
\be
   \label{eq:r2_phys}
   \langle r^2 \rangle^{phys.} & = & 0.456 \pm 0.030 \pm 0.024~\mbox{fm}^2 
   ~ , \\[2mm]
   \label{eq:curv_phys}
   c^{phys.} & = & (5.11 \pm 0.47 \pm 0.41) \cdot 10^{-3}~\mbox{fm}^4 ~ .
 \ee
where the first error is statistical and the second one systematic. 
We remind that the findings (\ref{eq:r2_phys}) and (\ref{eq:curv_phys}) are not 
completely independent, because they are based on an ensemble of fitted data 
which satisfy Eq.~(\ref{eq:cpole}).

\begin{figure}[!hbt]

\parbox{9.5cm}{\centerline{\includegraphics[scale=0.55]{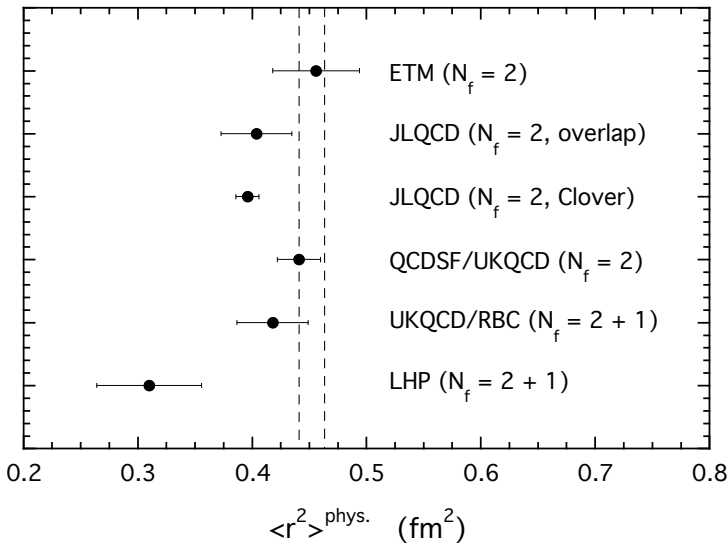}}} \ $~$ \
\parbox{4.5cm}{\vspace{-0.5cm} \caption{\it Results for the pion charge radius 
extrapolated at the physical point by various lattice collaborations performing 
unquenched calculations (see Table \ref{tab:r2phys}).
The vertical dashed lines show the experimental value $\langle r^2 \rangle^{exp.} 
= 0.452 \pm 0.011~\mbox{fm}^2$~\cite{PDG}.
\label{fig:r2Vphys}}}

\end{figure}

The results for the pion charge radius at the physical point obtained by various 
lattice collaborations performing unquenched calculations are compared in Table 
\ref{tab:r2phys} and in Fig.~\ref{fig:r2Vphys}.

\begin{table}[!htb]

\begin{center}
\scriptsize{
\begin{tabular}{||c||c|c|c|c|c||c||}
\hline
 $collaboration$ & $N_f$ & $Action$ & $V \cdot T ~ / ~ a^4$ & $a~(fm)$ & $M_\pi~(\mev)$ & 
 $\langle r^2 \rangle^{phys.}~(fm^2)$ \\ \hline \hline
 $ETM~\mbox{[this work]}$     & $2$     & $tlSym + tmW$    & $32^3 \cdot 64$ & $\sim 0.09$ & $\geq 260$ & $0.456 \pm 0.038$ \\ \hline 
 $JLQCD$~\cite{overlap}       & $2$     & $Iw. + overlap$  & $16^3 \cdot 32$ & $\sim 0.12$ & $\geq 290$ & $0.404 \pm 0.031$ \\ \hline 
 $JLQCD$~\cite{Clover1}       & $2$     & $plaq. + Clover$ & $20^3 \cdot 48$ & $\sim 0.09$ & $\geq 550$ & $0.396 \pm 0.010$ \\ \hline 
 $QCDSF/UKQCD$~\cite{Clover2} & $2$     & $plaq. + Clover$ & $24^3 \cdot 48$ & $\sim 0.08$ & $\geq 400$ & $0.441 \pm 0.019$ \\ \hline \hline 
 $UKQCD/RBC$~\cite{DWF}       & $2 + 1$ & $DWF$            & $16^3 \cdot 32$ & $\sim 0.12$ & $\geq 330$ & $0.418 \pm 0.031$  \\ \hline 
 $LHP$~\cite{LHP}             & $2 + 1$ & $Asqtad + DWF$   & $20^3 \cdot 64$ & $\sim 0.12$ & $\geq 320$ & $0.310 \pm 0.046$ \\ \hline \hline 

\end{tabular}
}

\caption{\it Results for the pion charge radius extrapolated at the physical point 
by various lattice collaborations performing unquenched calculations.
The uncertainty of the ETM result corresponds to the statistical and the systematic 
errors, given by Eq.~(\ref{eq:r2_phys}), added in quadrature.
\label{tab:r2phys}}

\end{center}

\end{table}

Our finding (\ref{eq:r2_phys}) agrees very well with the experimental value 
$\langle r^2 \rangle^{exp.} = 0.452 \pm 0.011~\mbox{fm}^2$~\cite{PDG}. 
It is also consistent within the errors with the results of JLQCD ($N_f = 2$), 
QCDSF/UKQCD ($N_f = 2$) and UKQCD/RBC ($N_f = 2 + 1$) collaborations, while the 
difference with the result of LHP ($N_f = 2 + 1$) collaboration is equal to 
$\approx 2$ standard deviations.

We stress that the use of twisted BC's and the inclusion of the NNLO terms 
in the ChPT analyses are two important features considered in this work. 
Note that twisted BC's are used only in Ref.~\cite{DWF} and a NNLO ChPT 
analysis is carried out only in Ref.~\cite{overlap}.

\begin{figure}[!hbt]

\parbox{9.5cm}{\centerline{\includegraphics[scale=0.55]{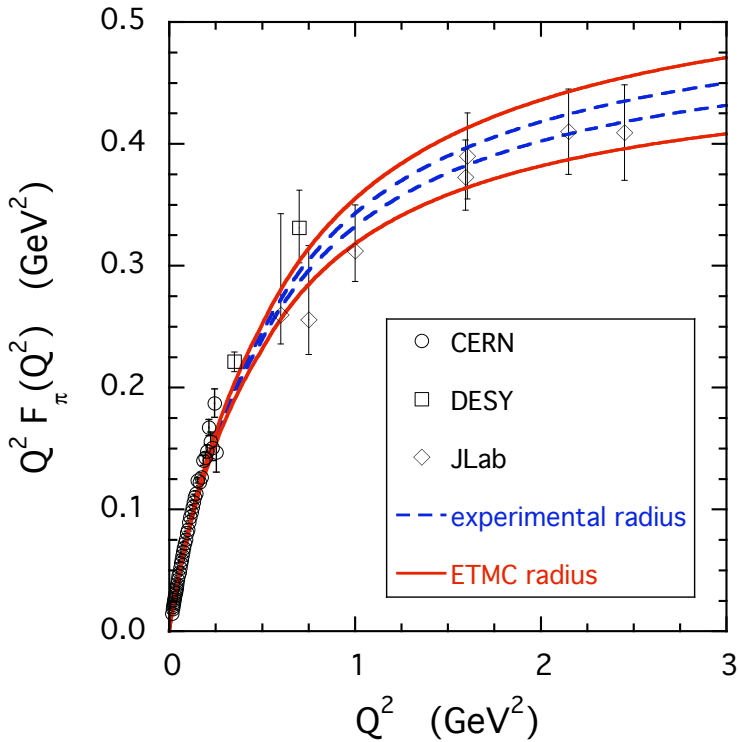}}} \ $~$ \
\parbox{4.5cm}{\vspace{-0.5cm} \caption{\it Pion form factor times $Q^2 = - q^2$,
$Q^2 F_\pi(Q^2)$, versus $Q^2$ in physical units.
The dots, squares and diamonds are experimental data from Refs.~\cite{CERN}, 
\cite{DESY1,DESY2} and \cite{JLAB1,JLAB2,JLAB3,JLAB4}, respectively.
The dashed and solid lines correspond to the regions selected at $1 \sigma$ level 
by the predictions of the monopole form (\ref{eq:monopole}) using $\langle r^2 
\rangle = \langle r^2 \rangle^{exp.} = 0.452 \pm 0.011~\mbox{fm}^2$ \cite{PDG} 
and $\langle r^2 \rangle = \langle r^2 \rangle^{phys.} = 0.456 \pm 
0.038~\mbox{fm}^2$, respectively.
\label{fig:monopole}}}

\end{figure}

The ETMC result (\ref{eq:r2_phys}) has been obtained by fitting the lattice data 
for values of the (squared) four-momentum transfer $Q^2 = - q^2$ up to $0.5~\gev^2$ 
using the pole ansatz (\ref{eq:pole}) or equivalently the monopole functional 
form
 \be
    F_\pi^{(monopole)}(Q^2) = \frac{1}{1 + \frac{\langle r^2 \rangle}{6} Q^2 } ~ .
    \label{eq:monopole}
 \ee
An interesting question is at which value of $Q^2$ the predictions based on 
Eq.~(\ref{eq:monopole}) start to deviate from the experimental data. 
Recently, thanks to the CEBAF facility at JLab, the pion form factor has been 
measured quite accurately up to few $\gev^2$. 
The comparison with the monopole prediction (\ref{eq:monopole}), using for the 
(squared) pion charge radius either the ETMC result (\ref{eq:r2_phys}) or its 
experimental value from PDG \cite{PDG}, is illustrated in Fig.~\ref{fig:monopole}.

It can clearly be seen that there is no hint of a deviation of the experimental 
data from the monopole ansatz up to $Q^2 \approx 2 \div 3~\gev^2$.

New experimental data up to $Q^2 \simeq 6~\gev^2$, expected to be taken after the 
completion of the JLab upgrade to $12~\gev$ \cite{upgrade}, may shed light on 
the range of validity of the monopole ansatz.

\subsection{Polynomial fit\label{subsec:pol}}

We want to discuss briefly an alternative fit to our lattice data based on a simple 
polynomial form, which does not have any logarithmic term; namely
 \be
    \label{eq:Mpi_pol}
    M_\pi^2 & = & 2 \bar{B} \hat{m} \cdot \left[ 1 + C_1 \hat{m} + C_2 \hat{m}^2 
    \right] ~ , \\[2mm]
    \label{eq:fpi_pol}
    f_\pi & = & \bar{F} \cdot \left[ 1 + D_1 \hat{m} + D_2 \hat{m}^2 \right] 
    ~ , \\[2mm]
    \label{eq:r2V_pol}
    \langle r^2 \rangle & = & 6 / (M_0 + E_1 \hat{m} + E_2 \hat{m}^2)^2 ~ .
 \ee

\begin{figure}[!hbt]

\centerline{\includegraphics[scale=0.80]{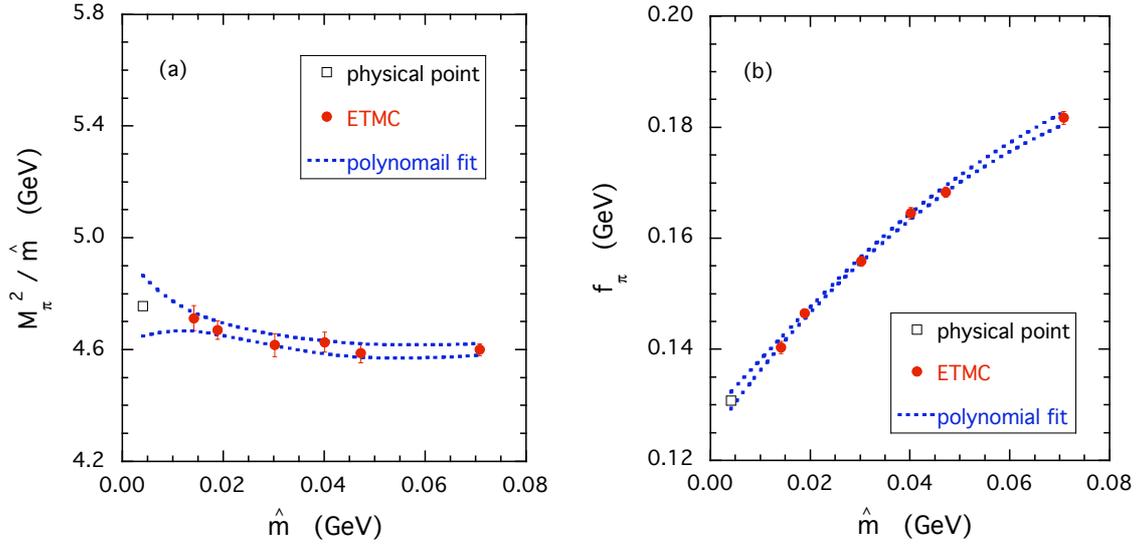}}

\caption{\it The ratio $M_\pi^2 / \hat{m}$ (a) and the pion decay constant (b) 
versus renormalized quark mass $\hat{m}$ in physical units.
The dots are the ETMC results, uncorrected for FSE effects, and the squares 
represent the experimental value for each quantity from PDG \cite{PDG}. 
The dotted lines correspond to the region selected at $1 \sigma$ level by the 
polynomial fits of the squared pion mass (\ref{eq:Mpi_pol}) and decay 
constant (\ref{eq:fpi_pol}).}

\label{fig:MFRC_pol}

\end{figure}

\begin{figure}[!hbt]

\parbox{9.5cm}{\centerline{\includegraphics[scale=0.55]{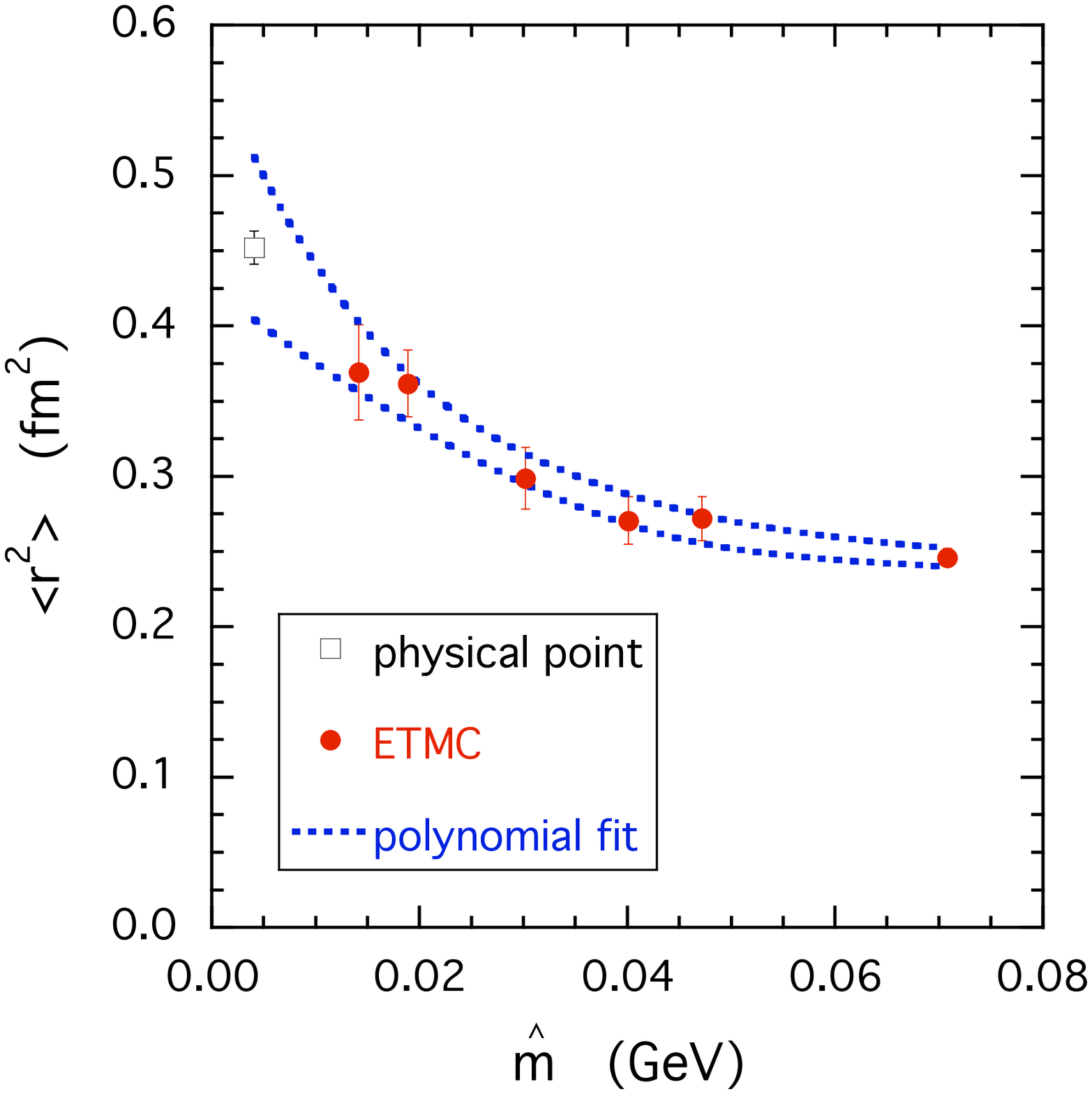}}} \ $~$ \
\parbox{4.5cm}{\vspace{-0.5cm} \caption{\it The charge radius of the pion form 
factor versus the renormalized quark mass $\hat{m}$ in physical units.
The dots are the ETMC results and the squares represent the experimental value 
of the pion charge radius \cite{PDG}, which is not included in the fitting 
procedure.
The dotted lines correspond to the region selected at $1 \sigma$ level by the 
polynomial fit (\ref{eq:r2V_pol}).
\label{fig:radiusV_pol}}}

\end{figure}

The results of such a fit, applied to our lattice data having $M_\pi L \gtrsim 4$ 
without applying any FSE correction, are illustrated in Figs.~\ref{fig:MFRC_pol} 
and \ref{fig:radiusV_pol}.

Among the fitting parameters we obtain $2 \bar{B} = 4.79 \pm 0.08~\gev$ and 
$\bar{F} = 126 \pm 2~\mev$, while the lattice spacing turns out to be $a = 
0.0889 \pm 0.0010~\mbox{fm}$ and the (renormalized) up/down quark mass is 
$\hat{m}^{phys.} = 3.8 \pm 0.1~\mev$.

It can clearly be seen that the quality of the polynomial fit is quite similar 
to the one of the ChPT fit shown in Fig.~\ref{fig:MFRCrs2_2loop} and therefore 
we have to conclude that our lattice results do not show a clear-cut evidence 
of chiral logs.

\section{Chiral fits of the pion form factor\label{sec:ffpion}}

In this Section we present the ChPT analysis of our results for the pion form 
factor including its momentum dependence. 
The ChPT expansion of $F_\pi(q^2)$ has been calculated in Ref.~\cite{BCT} at NNLO, 
obtaining
 \be
    \label{eq:FVpion}
    F_\pi(q^2) & = & 1 + \left[ F_\pi(q^2) \right]_{\mbox{1-loop}} 
    + \left[ F_\pi(q^2) \right]_{\mbox{2-loops}} + {\cal{O}}(\hat{m}^3) 
    ~ , \\[2mm]
    \label{eq:FVpion_1loop}
    \left[ F_\pi(q^2) \right]_{\mbox{1-loop}} & = & 2 x_2 \left[ \frac{1}{6}(w - 4) 
    \bar{J}(w) - w \left( \ell_6^r + \frac{1}{6} L(\mu) + \frac{1}{18N} \right) 
    \right] ~ , \\[2mm]
    \left[ F_\pi(q^2) \right]_{\mbox{2-loops}} & = & 4 x_2^2 ~ \left\{ \phantom{\frac{1}{1}} 
    \hskip -0.3cm P_V(w) + U_V(w) - \Delta_M \bar{J}(w) \right. \nn \\ 
    & - & \left. \Delta_F \left[ \frac{1}{6}(w - 4) \bar{J}(w) - w \left( \ell_6^r 
    + \frac{1}{6} L(\mu) + \frac{1}{18N} \right) \right] \right\} ~ , 
    \label{eq:FVpion_2loops}
 \ee
where $w \equiv q^2 / 2B \hat{m}$. 
The polynomial part $P_V(w)$ is given by
 \be
    P_V(w) & = & w \left[ -\frac{1}{2} k_1 + \frac{1}{4} k_2 - \frac{1}{12} 
    k_4 + \frac{1}{2} k_6 + r_1^r \right. \nn \\
    & + & \left. \frac{1}{N} \left( \frac{23}{36} L(\mu) + \frac{5}{576} + 
    \frac{37}{864N} \right) - \ell_4^r \left( 2 \ell_6^r + \frac{1}{9N} \right) 
    \right] \nn \\
    & + & w^2 \left[ \frac{1}{12} k_1 - \frac{1}{24} k_2 + \frac{1}{24} 
    k_6 + r_2^r \right. \nn \\
    & + & \left. \frac{1}{9N} \left( \ell_1^r - \frac{1}{2} \ell_2^r + \frac{1}{2} 
    \ell_6^r - \frac{1}{12} L(\mu) - \frac{1}{384} - \frac{47}{192N} \right) 
    \right] ~ ,
    \label{eq:PV}
 \ee
while the dispersive part $U_V(w)$ reads as
 \be
    U_V(w) & = & \bar{J}(w) \left[ - \frac{1}{3} w (w - 4) \left( \ell_1^r - 
    \frac{1}{2} \ell_2^r + \frac{1}{2} \ell_6^r \right) + \frac{1}{3} \ell_4^r 
    (w - 4) \right. \nn \\
    & - & \left. \frac{1}{36} L(\mu) (w^2 + 8w - 48) + \frac{1}{108N} (7 w^2 - 
    97 w + 81) \right] +  \frac{1}{9} H_1(w) \nn \\ 
    & + & \frac{1}{9} H_2(w) \left( \frac{1}{8} w^2 - w + 4 \right) + \frac{1}{6} 
    H_3(w) \left( w - \frac{1}{3} \right) - \frac{5}{3} H_4(w) ~ ,
    \label{eq:UV}
 \ee
where
 \be
     \bar{J}(w) & = & z ~ h(z) + \frac{2}{N} ~ , \\
     H_1(w) & = & z ~ h^2(z) ~ , \\
     H_2(w) & = & z^2 ~ h^2(z)- \frac{4}{N^2} ~ , \\
     H_3(w) & = & N \frac{z}{w} h^3(z)  + \frac{\pi^2}{Nw} h(z) - 
     \frac{\pi^2}{2N^2} ~ , \\
     H_4(w) & = & \frac{1}{wz} \left( \frac{1}{2} H_1(w) + \frac{1}{3} H_3(w) +
     \frac{1}{N} \bar{J}(w) + \frac{\pi^2 - 6}{12N^2} w \right) ~ ,
 \ee
with $z \equiv 1 - 4/w$ and 
 \be
    h(z) = \frac{1}{N \sqrt{z}} \mbox{log}\frac{\sqrt{z} - 1}{\sqrt{z}+1} ~ .
 \ee

Using the above formulae it is possible to test the momentum dependence of the 
pion form factor predicted by ChPT at NNLO.
Such a dependence is analytical up to the inelastic threshold $q_{thr}^2 = 4 
M_\pi^2$. 
Thus, in the chiral limit, terms of the form $q^2 ~ \mbox{log}(-q^2)$ appear in 
the pion form factor, which becomes a non-analytic function of $q^2$.
This is the origin of the divergency of both the charge radius and the curvature 
in the chiral limit [see Eqs.~(\ref{eq:r2})-(\ref{eq:c_2loops})].

It is easy to check that an expansion of Eqs.~(\ref{eq:FVpion})-(\ref{eq:FVpion_2loops}) 
in powers of $q^2$ leads to the result: $F_\pi(q^2) = 1 + \langle r^2 \rangle ~ 
q^2 / 6 + c ~ q^4 + {\cal{O}}(q^6)$, with $\langle r^2 \rangle$ and $c$ 
given by Eqs.~(\ref{eq:r2})-(\ref{eq:r2_2loops}) and 
Eqs.~(\ref{eq:curv})-(\ref{eq:c_2loops}), respectively.
Thus by using Eqs.~(\ref{eq:FVpion})-(\ref{eq:FVpion_2loops}) it is possible to 
take into account (at least partially) the effects of order ${\cal{O}}(q^6)$ 
in the momentum dependence of the pion form factor.

Note that the NNLO terms (\ref{eq:FVpion_2loops}-\ref{eq:UV}) do not depend 
upon the LEC's $\ell_1^r$ and $\ell_2^r$ separately, but only through the linear 
combination ($\ell_1^r - \ell_2^r / 2$).
Since different linear combinations of $\ell_1^r$ and $\ell_2^r$ appear in the 
NNLO terms of both the pion mass (\ref{eq:MPi2_2loops}) and decay constant 
(\ref{eq:fPi_2loops}), the LEC's $\ell_1^r$ and $\ell_2^r$ can be determined 
by a simultaneous analysis of the form factor together with the pion mass 
and decay constant.

\begin{figure}[!hbt]

\centerline{\includegraphics[scale=0.80]{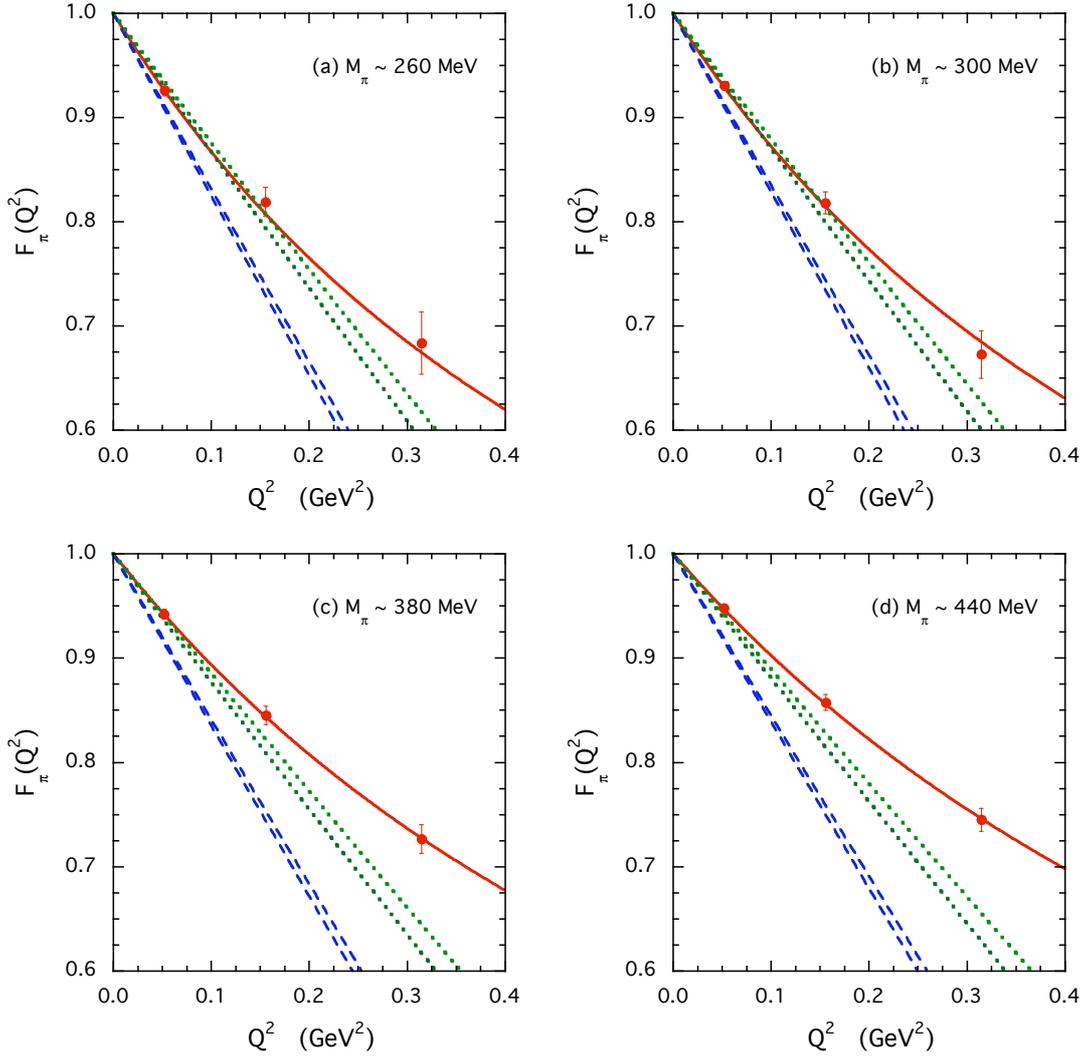}}

\caption{\it Pion form factor $F_\pi(Q^2)$ versus the squared 4-momentum transfer 
$Q^2 \equiv -q^2$ in physical units, for the run $R_1$ at $M_\pi \simeq 260~\mev$ 
(a), the run $R_{2a}$ at $M_\pi \simeq 300~\mev$ (b), the run $R_3$ at $M_\pi 
\simeq 380~\mev$ (c) and the run $R_4$ at $M_\pi \simeq 440~\mev$ (d).
For the lattice spacing the value $a = 0.0861~\mbox{fm}$ is adopted from the second 
column of Table \ref{tab:chiral1}.
The solid line is the pole behavior (\ref{eq:pole}) with the parameter $M_{pole}$ 
fitted to the lattice points.
The dashed and dotted lines are the regions selected at $1\sigma$ level by the ChPT 
predictions at NLO corresponding respectively to $\bar{\ell}_6 = 14.59 \pm 0.03$ 
and $\bar{\ell}_6 = 11.6 \pm 0.3$ with $2B = 5.21 \pm 0.05~\gev$ and $F = 121.7 
\pm 1.1~\mev$ (see text).}

\label{fig:Fpion_NLO}

\end{figure}

As already discussed in subsection \ref{subsec:fit1}, at NLO the pion form 
factor depends on one LEC, $\bar{\ell}_6$, which governs only the linear term 
in $q^2$.
The ChPT predictions at NLO corresponding respectively to $\bar{\ell}_6 = 14.59 \pm 
0.03$ and $\bar{\ell}_6 = 11.6 \pm 0.3$ (with $2B = 5.21 \pm 0.05~\gev$ and $F = 121.7 
\pm 1.1~\mev$), reported already in Fig.~\ref{fig:RC_1loop} in the case of the 
pion charge radius and curvature, are shown in Fig.~\ref{fig:Fpion_NLO} 
for various values of the pion mass.
We remind that the value $\bar{\ell}_6 = 14.59 \pm 0.03$ is fixed by the reproduction 
of the experimental charge radius, while the value $\bar{\ell}_6 = 11.6 \pm 0.3$ is 
obtained by fitting our lattice data of the charge radius for pion masses up to 
$\simeq 500~\mev$.
It can be seen that: 
\begin{itemize}

\item the momentum dependence predicted by ChPT at NLO is almost linear at variance 
with the pole behavior (\ref{eq:pole}) observed in our lattice data [see also 
Fig.~\ref{fig:RC_1loop}(b)] ; 

\item  using $\bar{\ell}_6 = 11.6 \pm 0.3$ the NLO approximation appears to work 
up to $Q^2 \equiv -q^2 \approx 0.15~\gev^2$ and for pion masses below $\approx 
300~\mev$. 
With such a value of $\bar{\ell}_6$ the NLO formula (\ref{eq:r2_1loop}) yields 
$\langle r^2 \rangle^{phys.} = 0.352 \pm 0.008~\mbox{fm}^2$, which underestimates 
significantly the experimental charge radius (see subsection \ref{subsec:fit1}).
A slight improvement can be achieved by using directly the NLO formula (\ref{eq:FVpion_1loop}) 
to fit our lattice data for the pion form factor at the lowest $Q^2$-value ($\simeq 
0.05~\gev^2$) and for the two lowest pion masses ($M_\pi \simeq 260$ and $\simeq 
300~\mev$).
We obtain $\bar{\ell}_6 = 12.2 \pm 0.5$ corresponding to $\langle r^2 \rangle^{phys.} 
= 0.373 \pm 0.017~\mbox{fm}^2$, which still deviates from the experimental value by 
four standard deviations;

\item using $\bar{\ell}_6 = 14.59 \pm 0.03$, which instead reproduces the 
experimental value of the pion charge radius, the range of applicability of the 
NLO approximation reduces to values of $Q^2$ at least not larger than $\approx 
0.03~\gev^2$ and to pion masses below $\approx 300~\mev$, which are not covered 
by our present lattice data ($Q^2 \gtrsim 0.05~\gev^2$). 
We notice that within the above restricted range of values of $Q^2$ and pion masses 
the deviation of the pion form factor from unity becomes smaller than few percent 
and therefore a particular attention should be paid to the statistical precision 
as well as to the systematic uncertainties related to cut-off and finite size 
effects.

\end{itemize}

From Fig.~\ref{fig:Fpion_NLO} it is clear that the description of our lattice 
data requires the inclusion of higher-order ChPT effects, which should be much 
larger than the finite volume corrections and the scaling violations observed 
in Figs.~\ref{fig:volume} and \ref{fig:scaling} for the values of $Q^2$ and 
$M_\pi$ considered in our calculations.

\begin{figure}[!hbt]

\centerline{\includegraphics[scale=0.80]{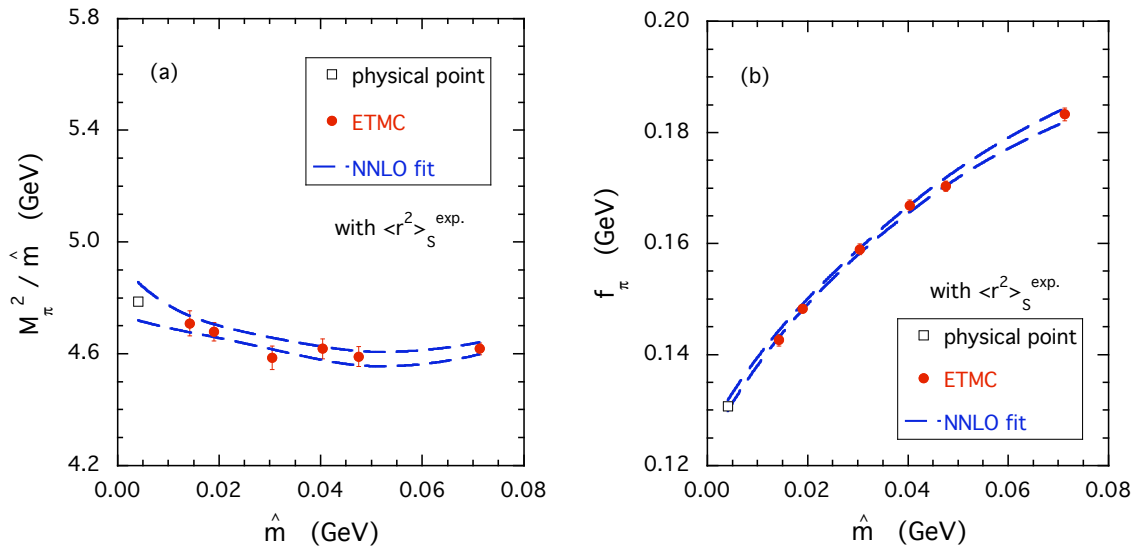}}

\caption{\it The ratio of the squared pion mass to the renormalized quark mass 
$\hat{m}$ (a) and the pion decay constant (b) versus $\hat{m}$ in physical 
units.
The dots are the ETMC results and the squares represent the experimental value 
for each quantity from Ref.~\cite{PDG}. 
The dashed lines correspond to the region selected at $1 \sigma$ level by the 
NNLO ChPT analysis of the ETMC results for the pion mass, decay constant 
and e.m.~form factor.
The experimental value of the pion scalar radius (\ref{eq:r2S_exp}) is added to 
the fitting procedure employing Eqs.~(\ref{eq:r2S})-(\ref{eq:r2S_2loops}).
The values of the fitting parameters are listed in the second column of Table 
\ref{tab:chiral2}.}

\label{fig:MF_FVrs2_2loops}

\end{figure}

\begin{figure}[!hbt]

\parbox{9.5cm}{\centerline{\includegraphics[scale=0.55]{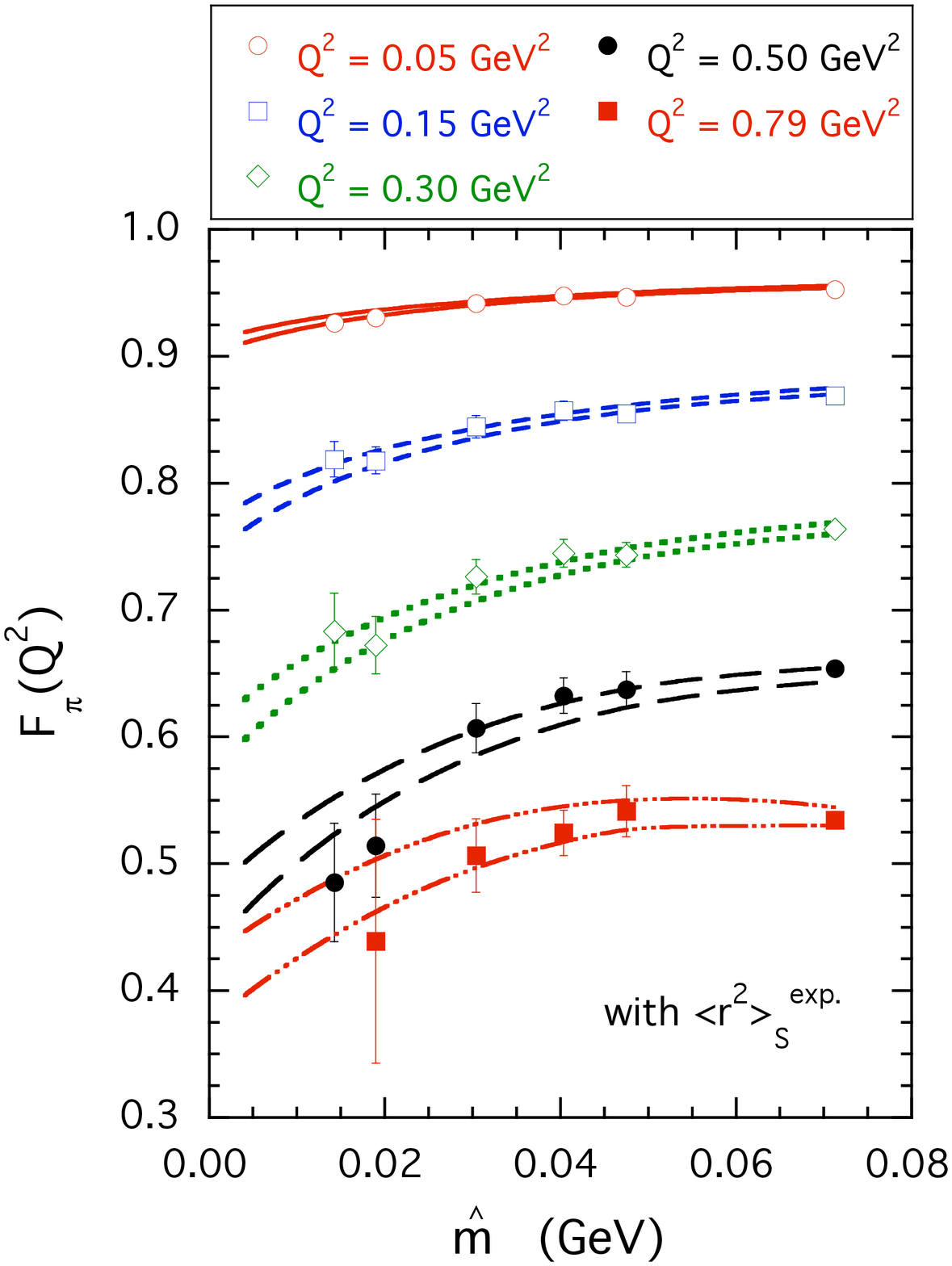}}} \ $~$ \
\parbox{4.5cm}{\vspace{-0.5cm} \caption{\it ETMC results for the pion e.m.~form 
factor versus the renormalized quark mass $\hat{m}$ at various values of $Q^2$ 
($\equiv - q^2$).
The various lines correspond to the regions selected at $1 \sigma$ level by the 
ChPT fit at NNLO based on Eqs.~(\ref{eq:FVpion})-(\ref{eq:FVpion_2loops}) with 
the fitting parameters given in the second column of Table \ref{tab:chiral2}.
The experimental value of the pion scalar radius (\ref{eq:r2S_exp}) is added to 
the fitting procedure using Eqs.~(\ref{eq:r2S})-(\ref{eq:r2S_2loops}).
\label{fig:FVrs2_2loops}}}

\end{figure}

We have therefore performed a simultaneous NNLO fit of the lattice data of the runs 
$R_1$, $R_{2a}$, $R_3$, $R_4$, $R_{5a}$ and $R_6$ for the quantities $M_\pi$, 
$f_\pi$ and $F_\pi(Q^2)$, including all values of $Q^2 \equiv -q^2$ from 
$\simeq 0.05~\gev^2$ up to $\simeq 0.8~\gev^2$.
As in the previous Section, the constraint (\ref{eq:r2S_exp}) on the pion scalar 
radius is included in the fitting procedure in order to reduce the uncertainties 
in the extracted values of the chiral parameters.
The latter are given in the second column of Table \ref{tab:chiral2}.
The nice quality of the NNLO fit is illustrated in Figs.~\ref{fig:MF_FVrs2_2loops} 
and \ref{fig:FVrs2_2loops}.
The corresponding value of the pion charge radius, calculated at the physical 
point using Eqs.~(\ref{eq:r2}-\ref{eq:r2_2loops}), is $\langle r^2 \rangle^{phys.} 
= 0.438 \pm 0.029~\mbox{fm}^2$, in nice agreement with the finding (\ref{eq:r2_phys}), 
which, we remind, is based on the use of the pole ansatz (\ref{eq:pole}) that 
describes very well the momentum dependence of our lattice data (see 
Figs.~\ref{fig:Fpion} and \ref{fig:Fpion_NLO}).

\begin{table}[!htb]

\begin{center}
\begin{tabular}{||c||c|c|c||}
\hline
 $parameter$ & $Q^2 \leq 0.8~\gev^2$ & $Q^2 \leq 0.5~\gev^2$ & $Q^2 \leq 0.3~\gev^2$ \\ \hline \hline
 $2 B~(\gev)$ & $4.89 \pm 0.10$ & $4.91 \pm 0.07$ & $4.90 \pm 0.09$ \\ \hline 
 $F~(\mev)$ & $122.6 \pm 1.1$ & $122.5 \pm 0.8$ & $122.5 \pm 1.0$ \\ \hline \hline
 $\bar{\ell}_3$ & $3.14 \pm 1.03$ & $3.24 \pm 0.53$ & $3.19 \pm 0.81$ \\ \hline 
 $\bar{\ell}_4$ & $4.37 \pm 0.27$ & $4.41 \pm 0.10$ & $4.39 \pm 0.19$ \\ \hline 
 $\bar{\ell}_6$ & $15.0 \pm 0.9$ & $14.8 \pm 1.1$ & $14.8 \pm 1.5$ \\ \hline \hline
 $\bar{\ell}_1$ & $-0.54 \pm 1.01$ & $-0.31 \pm 1.12$ & $-0.28 \pm 1.91$ \\ \hline 
 $\bar{\ell}_2$ & $\hskip 0.3cm 4.40 \pm 0.78$ & $\hskip 0.3cm 4.29 \pm 1.00$ & $\hskip 0.3cm 4.23 \pm 1.52$ \\ \hline 
 $r_M^r \cdot 10^4$ & $-0.48 \pm 0.30$ & $-0.43 \pm 0.23$ & $-0.44 \pm 0.37$ \\ \hline 
 $r_F^r \cdot 10^4$ & $\hskip 0.3cm 0.11 \pm 0.19$ & $\hskip 0.3cm 0.08 \pm 0.10$ & $\hskip 0.3cm 0.08 \pm 0.20$ \\ \hline 
 $r_1^r \cdot 10^4$ & $-0.98 \pm 0.12$ & $-0.94 \pm 0.13$ & $-0.90 \pm 0.14$ \\ \hline 
 $r_2^r \cdot 10^4$ & $\hskip 0.3cm 0.43 \pm 0.03$ & $\hskip 0.3cm 0.46 \pm 0.03$ & $\hskip 0.3cm 0.52 \pm 0.05$ \\ \hline \hline 
 $a~(fm)$ & $0.0883 \pm 0.0006 $ & $0.0883 \pm 0.0006$ & $0.0883 \pm 0.0007$ \\ \hline
 $\hat{m}^{phys.}~(\mev)$ & $3.80 \pm 0.09$ & $3.79 \pm 0.07$ & $3.79 \pm 0.09$ \\ \hline \hline
 $\chi^2~/~d.o.f.$ & $29~/~34$ & $19~/~28$ & $13~/~22$ \\ \hline \hline
 
\end{tabular}

\caption{\it Values of the chiral parameters, the lattice spacing and the renormalized 
light quark mass $\hat{m} = m^{\overline{MS}}(2~\gev)$ at the physical point, obtained 
from the simultaneous ChPT analysis of the pion mass, decay constant and form factor 
made at NNLO using Eqs.~(\ref{eq:MPi2})-(\ref{eq:fPi_2loops}) and 
(\ref{eq:FVpion})-(\ref{eq:FVpion_2loops}), including 
also the constraint (\ref{eq:r2S_exp}) on the pion 
scalar radius.
The second, third and fourth columns correspond to different $Q^2$-ranges of the 
lattice data of the form factor considered in the fitting procedure.
The values of the parameters $r_M^r$, $r_F^r$, $r_1^r$ and $r_2^r$ are given at the 
$\rho$-meson mass scale.
The uncertainties are statistical (bootstrap) errors only.
\label{tab:chiral2}}

\end{center}

\end{table}

The comparison of the results shown in the fourth column of Table \ref{tab:chiral1} 
and those in the second column of Table \ref{tab:chiral2} clearly indicates that 
within the statistical uncertainties the extracted values of the chiral parameters 
are quite stable against chiral effects of order ${\cal{O}}(q^6)$ in the pion form 
factor with the only exception of the parameter $r_2^r$, which instead exhibit a 
rather large variation.
The latter will be included in the systematic error (see later Table 
\ref{tab:chiral_final}), providing the dominant source of uncertainty 
for the parameter $r_2^r$.

We remind that, at variance with the results reported in Table \ref{tab:chiral1}, 
the ones shown in Table \ref{tab:chiral2} are obtained without the assumption of the 
pole ansatz (\ref{eq:pole}) for the momentum dependence of the form factor, but 
using only the functional forms (\ref{eq:FVpion}-\ref{eq:UV}) predicted by 
ChPT at NNLO.

Effects from higher orders in the chiral expansion are expected to become more 
and more important as the value of $Q^2$ increases.
In order to check their relevance in our analysis we repeat the NNLO fit by 
limiting the range of values of $Q^2$, i.e.~by including only lattice data 
with $Q^2 \leq 0.5~\gev^2$ (see third column of Table \ref{tab:chiral2}) and 
$Q^2 \leq 0.3~\gev^2$ (see fourth column of Table \ref{tab:chiral2}).
It can clearly be seen that, within the statistical precision, the extracted 
values of all the chiral parameters are only slightly sensitive to the 
$Q^2$-range used and therefore to higher-order effects.

Before closing this Section we mention that in Refs.~\cite{ETMC2,CGL,ETMC_scaling} 
a different definition of the LEC's at NNLO is adopted, namely the constants 
$r_M^r$ and $r_F^r$ are replaced by the constants $k_M$ and $k_F$.
Using the results of the second column of Table \ref{tab:chiral2} we obtain 
$k_M = -0.6 \pm 1.6$ and $k_F = 1.2 \pm 1.5$.

\section{Final results for the LEC's\label{sec:LECs}}

In this Section we provide the final estimates of the LEC's from the present work.
Our results, including both the statistical and the systematic uncertainties, 
are collected in the second column of Table \ref{tab:chiral_final}.
They have been evaluated by averaging the three central values reported 
in Table 8, using the quoted errors as the weights.

\begin{table}[!htb]

\begin{center}
\begin{tabular}{||c||c||c||}
\hline
 $LEC$ & $ETMC~\mbox{(NNLO)}$ & $\mbox{non-lattice}$\\ 
 $$    & $\mbox{(this work)}$      & $\mbox{estimates}$\\ \hline \hline
 $2 B~(\gev)$ & $4.90 \pm 0.09 \pm 0.20$ & $--$ \\ \hline 
 $F~(\mev)$ & $122.5 \pm 1.0 \pm 1.0$ & $--$ \\ \hline \hline
 $\bar{\ell}_1$ & $-0.4 \pm 1.3 \pm 0.6$ & $-0.4 \pm 0.6$~\cite{CGL} \\ \hline 
 $\bar{\ell}_2$ & $\hskip 0.3cm 4.3 \pm 1.1 \pm 0.4$ & $\hskip 0.3cm 4.3 \pm 0.1$~\cite{CGL} \\ \hline 
 $\bar{\ell}_3$ & $\hskip 0.2cm 3.2 \pm 0.8 \pm 0.2$ & $\hskip 0.35cm 2.9 \pm 2.4~\cite{CGL}$ \\ \hline 
 $\bar{\ell}_4$ & $\hskip 0.2cm 4.4 \pm 0.2 \pm 0.1$ & $\hskip 0.35cm 4.4 \pm 0.2$~\cite{CGL} \\ \hline 
 $\bar{\ell}_6$ & $14.9 \pm 1.2 \pm 0.7$ & $--$ \\ \hline \hline
 $r_M^r \cdot 10^4$ & $-0.45 \pm 0.30 \pm 0.10$ & $--$ \\ \hline 
 $r_F^r \cdot 10^4$ & $\hskip 0.3cm 0.08 \pm 0.16 \pm 0.05$ & $--$ \\ \hline 
 $r_1^r \cdot 10^4$ & $-0.94 \pm 0.13 \pm 0.10$ & $-2.0$~\cite{BCT} \\ \hline 
 $r_2^r \cdot 10^4$ & $\hskip 0.3cm 0.46 \pm 0.03 \pm 0.31$ & $\hskip 0.3cm 2.1$~\cite{BCT} \\ \hline \hline 

\end{tabular}

\caption{\it Values of the LEC's obtained from the NNLO ChPT analyses of the 
previous Section and compared with available estimates arising either from 
NNLO ChPT analyses of $\pi - \pi$ scattering data \cite{CGL} or from 
VMD models \cite{BCT}.
The values of the parameters $r_M^r$, $r_F^r$, $r_1^r$ and $r_2^r$ are given at the 
$\rho$-meson mass scale.
In the second column the first error is statistical and the second one systematic.
\label{tab:chiral_final}}

\end{center}

\end{table}

As in the case of the charge radius and curvature, discussed in the previous Section,
we estimate the systematic errors due to both finite volume and discretization 
effects. 
Firstly we substitute the run $R_{2a}$ with the run $R_{2b}$ and fit the new set
of data; the changes in the central values of the chiral parameters provide an 
estimate of the finite volume effects. 
Secondly we further substitute the runs $R_{2b}$ and $R_{5a}$ with the runs 
$R_{2c}$ and $R_{5b}$ at the finer lattice spacing, respectively, obtaining 
an estimate of discretization effects.
All the systematic errors, which include also the spread of the central values 
of Table \ref{tab:chiral2}, are finally added in quadrature.

In Table \ref{tab:chiral_final} our estimates of the chiral parameters are 
compared with available results from NNLO ChPT analyses of $\pi - \pi$ 
scattering data from Ref.~\cite{CGL}, and with estimates obtained using VMD 
models in Ref.~\cite{BCT}.
Our values for the LEC's $\bar{\ell}_1$, $\bar{\ell}_2$, $\bar{\ell}_3$ and 
$\bar{\ell}_4$ agree nicely with those extracted in Ref.~\cite{CGL}.
The uncertainties obtained in this work for the LEC $\bar{\ell}_4$ is quite similar 
to the one from Ref.~\cite{CGL}, while $\bar{\ell}_1$ and $\bar{\ell}_2$ are 
determined more precisely in Ref.~\cite{CGL} and $\bar{\ell}_3$ in the 
present work.

On the contrary the estimates of the counter-terms $r_1^r$ and $r_2^r$ obtained 
in Ref.~\cite{BCT} adopting VMD models turn out to be much larger than our values 
by a factor $\approx 2 \div 3$.

The results obtained for the lattice spacing, $a = 0.0883 \pm 0.0006~\mbox{fm}$, 
and the renormalized up/down quark mass, $\hat{m}^{phys.} = 3.79 \pm 0.08 \pm 
0.15~\mev$, are consistent within the errors with the findings of 
Refs.~\cite{ETMC1} and \cite{ETMC4} obtained at the same value 
of $\beta$ ($= 3.9$). 
The values $2B = 4.90 \pm 0.09 \pm 0.20~\gev$ and $F = 122.5 \pm 1.0 \pm 
1.0~\mev$ correspond to a light-quark condensate equal to 
 \be
   \langle q\bar{q} \rangle^{\overline{MS}}(2~\gev) = (- 264 \pm 3 \pm 5~\mev)^3 ~ ;
   \label{eq:condensate}
 \ee
moreover, the ratio $f_\pi^{phys.} / F$ is equal to
 \be
   f_\pi^{phys.} / F = 1.067 \pm 0.009 \pm 0.009 ~ .
   \label{eq:fpiF}
 \ee
The findings (\ref{eq:condensate}) and (\ref{eq:fpiF}) are in agreement with the 
corresponding values obtained by the scaling analysis of Ref.~\cite{ETMC_scaling}.

\begin{figure}[!hbt]

\centerline{\includegraphics[scale=0.80]{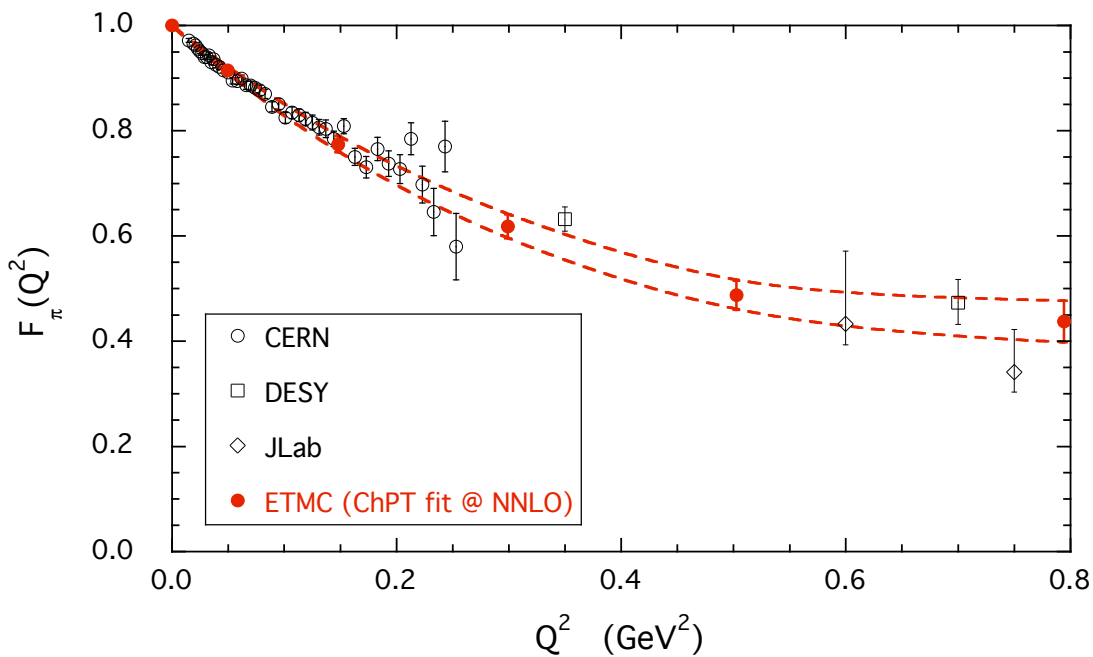}}

\caption{\it Pion form factor $F_\pi(Q^2)$ versus $Q^2 = -q^2$ in physical units. 
The full dots are the NNLO ChPT results of Table \ref{tab:ffpion_phys}, obtained 
at the physical point using for the LEC's the values given in 
Table \ref{tab:chiral_final}. 
The uncertainties of the ETMC results, illustrated also by the dashed lines, 
represent the statistical and the systematic errors of Table 
\ref{tab:ffpion_phys} added in quadrature. 
The open dots, squares and diamonds are experimental data from Refs.~\cite{CERN}, 
\cite{DESY1,DESY2} and \cite{JLAB1,JLAB2,JLAB3,JLAB4}, respectively.}

\label{fig:FVpion_phys}

\end{figure}

Using for the LEC's the values given in Table \ref{tab:chiral_final} we have calculated 
the values of the pion form factor at the physical point for the various values of 
$Q^2$ considered in this work.
Our results, including both statistical and systematic uncertainties, are collected 
in Table \ref{tab:ffpion_phys} and shown in Fig.~\ref{fig:FVpion_phys}, where they 
are also compared with available experimental data from 
Refs.~\cite{CERN,DESY1,DESY2,JLAB1,JLAB2,JLAB3,JLAB4}.

\begin{table}[!htb]

\begin{center}
\begin{tabular}{||c||c||}
\hline
 $Q^2~(\gev^2)$ & $F_\pi^{phys.}(Q^2)$ \\ \hline \hline
 $0.050$ & $0.914 \pm 0.005 \pm 0.003$ \\ \hline 
 $0.148$ & $0.774 \pm 0.013 \pm 0.008$ \\ \hline 
 $0.299$ & $0.618 \pm 0.019 \pm 0.013$ \\ \hline 
 $0.503$ & $0.487 \pm 0.022 \pm 0.017$ \\ \hline 
 $0.794$ & $0.437 \pm 0.030 \pm 0.026$ \\ \hline \hline

\end{tabular}

\caption{\it Values of the pion form factor $F_\pi^{phys.}(Q^2)$, extrapolated to 
the physical point using for the LEC's the results of Table \ref{tab:chiral_final}, 
for various values of $Q^2$.
The first error is statistical and the second one systematic.
\label{tab:ffpion_phys}}

\end{center}

\end{table}

It can be seen that our values are fully consistent with the experimental data 
in the whole range of values of $Q^2$ considered in this study. 
The agreement is particularly remarkable at low values of $Q^2$ ($Q^2 \lesssim 
0.15~\gev^2$), where the experimental data are very precise, as well as at larger 
values of $Q^2$ ($Q^2 \gtrsim 0.3~\gev^2$), where the uncertainties of our 
results become competitive with the experimental errors.

\section{Conclusions\label{sec:conclusions}}

We have presented a lattice calculation of the electromagnetic form factor of the 
pion obtained using the tree-level Symanzik improved gauge action with two flavors 
of dynamical twisted Wilson quarks.

The simulated pion masses range from $\simeq 260$ to $\simeq 580~\mev$ and the 
lattice box sizes are chosen in order to guarantee that $M_\pi L \gtrsim 4$. 

Accurate results for the form factor are obtained using all-to-all quark propagators 
evaluated with the stochastic procedure of Ref.~\cite{stoc}. 

The momentum dependence of the pion form factor is investigated up to values of the 
squared four-momentum transfer $Q^2 \simeq 0.8~\gev^2$ and, thanks to the use of 
twisted boundary conditions, down to $Q^2 \simeq 0.05~\gev^2$.
The $Q^2$-dependence at the simulated pion masses is well reproduced by a single 
monopole ansatz with a pole mass lighter by $\approx 10\% \div 15\%$ than the 
lightest vector meson mass.

Volume and discretization effects on the form factor have been directly evaluated 
performing simulations at different volumes and lattice spacings, and they turn 
out to be within the statistical errors.
A more complete investigation of the scaling properties of the pion form factor, 
based on the study of its mass dependence at two additional values of the lattice 
spacing is however desirable. 
The corresponding measurements are in progress.

The extrapolation of our results for the pion mass, decay constant and form factor 
to the physical point has been carried out using (continuum) ChPT at NNLO \cite{BCT}. 
The extrapolated value of the (squared) pion charge radius is $\langle r^2 \rangle^{phys} 
= 0.456 \pm 0.030_{\mbox{stat.}} \pm 0.024_{\mbox{syst.}}$ in nice agreement with 
the experimental result $\langle r^2 \rangle^{exp.} = 0.452 \pm 0.011~\mbox{fm}^2$ 
\cite{PDG}.
The extrapolated values of the pion form factor agree very well with the experimental 
data up to $Q^2 \simeq 0.8~\gev^2$ within uncertainties which become competitive 
with the experimental errors for $Q^2 \gtrsim 0.3~\gev^2$.

The relevant low-energy constants appearing in the chiral expansion of the pion 
form factor are extracted from our lattice data adding only the experimental 
value of the pion scalar radius \cite{BCT} in the fitting procedure. 
We get: $\bar{\ell}_1 = -0.4 \pm 1.3 \pm 0.6$, $\bar{\ell}_2 = 4.3 \pm 1.1 
\pm 0.4$, $\bar{\ell}_3 = 3.2 \pm 0.8 \pm 0.2$, $\bar{\ell}_4 = 4.4 \pm 0.2 
\pm 0.1$, $\bar{\ell}_6 = 14.9 \pm 1.2 \pm 0.7$, where the first error is 
statistical and the second one systematic.
Our findings are in nice agreement with the results of the NNLO ChPT analysis 
of $\pi - \pi$ scattering data of Ref.~\cite{CGL}.
The values found for the LEC's $\bar{\ell}_3$ and $\bar{\ell}_4$ are consistent 
with the corresponding results of the ETMC analysis of 
Ref.~\cite{ETMC_scaling}.
This is quite reassuring because different kinds of systematic uncertainties may 
affect the two analyses: the present one being a NNLO analysis limited mainly 
to data from a single lattice spacing, and that of Ref.~\cite{ETMC_scaling} 
having two values of the lattice spacing but limited mainly to a NLO 
analysis.

It is the aim of our collaboration to reduce as much as possible all the 
uncertainties of the extracted low-energy constants in the next future. 
To this end, data at more values of the lattice spacing and calculations of other 
physical quantities, like e.g.~the pion scattering lengths, will be considered.
This may allow to avoid any input from experiments obtaining a first principle 
computation of the low-energy constants.

In this respect a very interesting strategy is to include lattice data for the 
scalar form factor of the pion, because almost the same low-energy constants 
enter the chiral expansion of both vector and scalar form factors \cite{BCT}.
In this way the use of the experimental value of the pion scalar radius in the 
fitting procedure can be avoided.
However the lattice calculation of the scalar form factor requires the evaluation 
of both connected and disconnected diagrams.
While the former have been already calculated on the ETMC gauge configurations, 
a precise evaluation of the latter is in progress. 
The results will be reported elsewhere.

\section*{Acknowledgments} 

We thank all the members of the ETM collaboration for very fruitful discussions 
and for a very enjoyable collaboration. 
We gratefully acknowledge also several discussions with H.~Leutwyler and G.~Colangelo. 
The computer time for this project was made available to us by the apeNEXT systems in 
Rome and Zeuthen. 
We thank these computer centres and their staff for the invaluable technical advice 
and help.

\section*{Appendix}

In this Appendix we provide the values of the pion form factor $F_\pi(Q^2)$ 
obtained for all the simulations (see Table \ref{tab:setup}) and for the 
various values of the squared four-momentum transfer $Q^2 \equiv -q^2$ 
considered in this work.

\begin{table}[!htb]

\begin{center}
\begin{tabular}{||c||c|c||}
\hline
$Q^2~(\gev^2)$ & $R_1$         & $R_{2a}$      \\ \hline \hline
$0.050$        & $0.926~~~(5)$ & $0.930~~~(4)$ \\ \hline
$0.148$        & $0.819~(14)$  & $0.818~(11)$  \\ \hline
$0.299$        & $0.683~(30)$  & $0.672~(23)$  \\ \hline
$0.503$        & $0.485~(47)$  & $0.514~(41)$  \\ \hline
$0.794$        & $0.242~(95)$  & $0.439~(96)$  \\ \hline \hline

\end{tabular}

\caption{\it Values of the pion form factor $F_\pi(Q^2)$ for various values of 
$Q^2 \equiv -q^2$ (in physical units) in the case of the runs $R_1$ and $R_{2a}$ 
performed at $\beta = 3.9$ and at the lattice volume $V \cdot T = 32^3 \cdot 
64 ~ a^4$.
The uncertainties are statistical (jacknife) errors.
\label{tab:FVdata1}}

\end{center}

\end{table}

\begin{table}[!htb]

\begin{center}
\begin{tabular}{||c||c|c|c|c|c||}
\hline
$Q^2~(\gev^2)$ & $R_{2b}$      & $R_3$         & $R_4$         & $R_{5a}$      & $R_6$         \\ \hline \hline
$0.050$        & $0.936~~~(5)$ & $0.942~~~(5)$ & $0.948~~~(4)$ & $0.947~~~(4)$ & $0.953~(2)$ \\ \hline
$0.148$        & $0.830~~~(8)$ & $0.845~~~(9)$ & $0.857~~~(8)$ & $0.855~~~(6)$ & $0.869~(3)$  \\ \hline
$0.299$        & $0.704~(13)$  & $0.726~(14)$  & $0.745~(11)$  & $0.743~(10)$  & $0.764~(4)$  \\ \hline
$0.503$        & $0.581~(22)$  & $0.607~(19)$  & $0.632~(14)$  & $0.637~(14)$  & $0.654~(5)$  \\ \hline
$0.794$        & $0.492~(37)$  & $0.506~(29)$  & $0.524~(18)$  & $0.541~(20)$  & $0.534~(7)$  \\ \hline \hline

\end{tabular}

\caption{\it The same as in Table \ref{tab:FVdata1} but for the runs $R_{2b}$, $R_3$, 
$R_4$, $R_{5a}$ and $R_6$ performed at $\beta = 3.9$ and at the lattice volume 
$V \cdot T = 24^3 \cdot 48 ~ a^4$.
\label{tab:FVdata2}}

\end{center}

\end{table}

\begin{table}[!htb]

\begin{center}
\begin{tabular}{||c||c|c||}
\hline
$Q^2~(\gev^2)$ & $R_{2c}$      & $R_{5b}$      \\ \hline \hline
$0.050$        & $0.933~~~(9)$ & $0.952~~~(3)$ \\ \hline
$0.148$        & $0.821~(14)$  & $0.865~~~(6)$ \\ \hline
$0.299$        & $0.681~(26)$  & $0.756~~~(8)$ \\ \hline
$0.503$        & $0.565~(44)$  & $0.635~(12)$  \\ \hline
$0.794$        & $0.526~(93)$  & $0.495~(22)$  \\ \hline \hline

\end{tabular}

\caption{\it The same as in Table \ref{tab:FVdata1} but for the runs $R_{2c}$ and 
$R_{5b}$ performed at $\beta = 4.05$ and at the lattice volume $V \cdot T = 32^3 
\cdot 64 ~ a^4$.
\label{tab:FVdata3}}

\end{center}

\end{table}

\end{document}